\documentclass[12pt]{article}
\usepackage{amssymb,latexsym,amsmath,epsfig}

\newcommand{\comment}[1]{\vspace{5mm}\par
\framebox{\begin{minipage}[c]{.9 \textwidth}
\rm #1 \end{minipage}}\vspace{5 mm}\par}

\newcommand{\rem}[1]{}
\newcommand{\remfigure}[1]{#1}

\newtheorem{theorem}{Theorem}[section]
\newtheorem{definition}[theorem]{Definition}

\newtheorem{remark}[theorem]{Remark}
\newtheorem{proposition}[theorem]{Proposition}
\newtheorem{corollary}[theorem]{Corollary}

\rem{
\textheight6in
\textwidth9in}

\begin{document}
\title{Wave Structures and Nonlinear Balances in a Family of 1+1
Evolutionary PDEs}
\author{ Darryl D. Holm and Martin F. Staley
\\Theoretical Division and Center for Nonlinear Studies
\\Los Alamos National Laboratory, MS B284
\\Los Alamos, NM 87545
\\{\footnotesize email: dholm@lanl.gov} }
\date{February 26, 2002}
\maketitle

\begin{abstract} We introduce the following family of evolutionary 1+1 PDEs
that describe the balance between convection and stretching for
small viscosity in the dynamics of 1D nonlinear waves in fluids:
\[
m_t\
+\
\underbrace{\ \ um_x\ \
}
_{\hspace{-2mm}\hbox{convection}\hspace{-2mm}}\
+\
\underbrace{\ \ b\,u_xm\ \
}
_{\hspace{-2mm}\hbox{stretching}\hspace{-2mm}}\
=\
\underbrace{\ \
\nu\,m_{xx}\
}
_{\hspace{-2mm}\hbox{viscosity}
}
\,, \quad\hbox{with}\quad
u=g*m
\,.
\]
Here $u=g*m$ denotes
$
u(x)=\int_{-\infty}^\infty g(x-y)m(y)\,dy
\,.
$
This convolution (or filtering) relates velocity $u$ to momentum
density $m$ by integration against the kernel $g(x)$.  We shall
choose $g(x)$ to be an even function, so that $u$ and $m$ have
the same parity under spatial reflection. When $\nu=0$, this equation is
both reversible in time and parity invariant. We shall study the effects of
the balance parameter $b$ and the kernel $g(x)$ on the solitary wave
structures, and investigate their interactions analytically for $\nu=0$ and
numerically for small viscosity, $\nu\ne0$.

This family of equations admits the classic Burgers ``ramps and cliffs''
solutions which are stable for $-1<{b}<1$ with small viscosity.

For $b<-1$, the Burgers ramps and cliffs are
unstable. The stable solution for $b<-1$ moves leftward instead of
rightward and tends to a stationary profile.
When $m=u-\alpha^2u_{xx}$ and $\nu=0$, this profile is given by
$u(x)\simeq{\rm sech}^2(x/(2\alpha))$ for $b=-2$, and
by $u(x)\simeq{\rm sech}(x/\alpha)$ for $b=-3$.

For $b>1$, the Burgers ramps and cliffs are again unstable. The stable
solitary traveling wave for $b>1$ and $\nu=0$ is the ``pulson''
$u(x,t)=cg(x-ct)$, which restricts to the ``peakon'' solution in the
special case $g(x)=e^{-|x|/\alpha}$ when $m=u-\alpha^2u_{xx}$. Nonlinear
interactions among these pulsons or peakons are governed by the
superposition of solutions for $b>1$ and $\nu=0$,
\[u(x,t)=\sum_{i=1}^N p_i(t)\,g(x-q_i(t))\,.\]
These solutions obey a finite dimensional dynamical system for the
time-dependent speeds $p_i(t)$ and positions $q_i(t)$. We study the pulson
and peakon interactions analytically, and we determine their fate
numerically under adding viscosity.

\end{abstract}

\tableofcontents

\section{Introduction}\label{Intro}

\subsection{The b-family of fluid transport equations}

We shall analyze a one-dimensional version of active fluid transport
that is described by the following family of 1+1 evolutionary equations,
\begin{equation}\label{b-family}
m_t\
+\
\underbrace{\ \ um_x\ \
}
_{\hspace{-2mm}\hbox{convection}\hspace{-2mm}}\
+\
\underbrace{\ \ b\,u_xm\ \
}
_{\hspace{-2mm}\hbox{stretching}\hspace{-2mm}}\
=\
0
\,, \quad\hbox{with}\quad
u=g*m
\,,
\end{equation}
in independent variables time $t$ and one spatial coordinate $x$.

We shall seek solutions for the fluid velocity $u(x,t)$ that are defined
either on the real line and vanishing at spatial infinity, or on a
periodic one-dimensional domain.  Here $u=g*m$ denotes the convolution
(or filtering),
\begin{equation}\label{unused-1}
u(x)=\int_{-\infty}^\infty g(x-y)m(y)\,dy,
\end{equation}
which relates velocity $u$ to momentum density $m$ by
integration against kernel $g(x)$ over the real line.  We
shall choose $g(x)$ to be an even function, so that $u$ and $m$ have the
same parity.

The family of equations (\ref{b-family}) is characterized by the kernel
$g$ and the real dimensionless constant $b$, which is the ratio
of stretching to convective transport.  As we shall see, $b$ is also the
number of covariant dimensions associated with the momentum density $m$.
The function $g(x)$ will determine the traveling wave shape and length
scale for equation (\ref{b-family}), while the constant $b$ will provide a
balance or bifurcation parameter for the nonlinear solution behavior.
Special values of $b$ will include the first few positive and negative
integers.

The quadratic terms in equation (\ref{b-family}) represent the
competition, or balance, in fluid convection between nonlinear transport
and amplification due to $b-$dimensional stretching.
For example, if $m$ is fluid momentum (a one-form density in one
dimension) then $b=2$.
Equation (\ref{b-family}) with $b=2$ arises in the nonlinear dynamics
of shallow water waves, as shown in \cite{CH[1993]} and
\cite{DGH[2001]}.
Equation (\ref{b-family}) with $b=2$ and $b=3$ appears in the theory of
integrable partial differential equations \cite{CH[1993],DGH[2001],DHH[2002]}.
The three-dimensional analog of equation (\ref{b-family}) with $b=2$
was introduced in \cite{HMR[1998a],HMR[1998b]}. Applying the
proper viscosity to this three-dimensional analog with $b=2$ produces the
Navier-Stokes-alpha model of turbulence \cite{Chen-etal[1998]}. The 1D
version of this turbulence model is
\begin{equation}\label{viscous-b-eqn}
m_t\
+\
\underbrace{\ \ um_x\ \
}
_{\hspace{-2mm}\hbox{convection}\hspace{-2mm}}\
+\
\underbrace{\ \ b\,u_xm\ \
}
_{\hspace{-2mm}\hbox{stretching}\hspace{-2mm}}\
=\
\underbrace{\ \
\nu\,m_{xx}\
}
_{\hspace{-2mm}\hbox{viscosity}
}
\,, \quad\hbox{with}\quad
u=g*m
\,.
\end{equation}
We shall compare our analysis of equation (\ref{b-family}) with
numerical simulations of (\ref{viscous-b-eqn}) for small viscosity.

\subsection{Outline of the paper}

After summarizing previous investigations of particular cases in the
b-family (\ref{b-family}) of active transport equations, section
\ref{history&properties} discusses its symmetries and other general
properties such as parity and reversibility.  Section \ref{twaves}
discusses the traveling waves of equation (\ref{b-family}) and derives
their Pulson solutions, which may be  generalized functions for $b>1$.
Section \ref{PulsonInteract} analyzes the interaction dynamics of the
Pulson solutions for any positive $b$ and any $g$.  Section \ref{Peakons}
specializes the analysis of the Pulson solutions to the Peakons, for which
$g(x)=e^{-|x|/\alpha}$ is a peaked pulse of width $\alpha$, and $b$ is
taken to be arbitrary.
\rem{
In section \ref{IVP} we introduce numerics that illustrates
the different types of behavior that may arise in the initial value
problems for Peakon solutions with $b>0$, $b=0$ and $b<0$. These
illustrations of analytically verifiable cases also serve to validate
and ensure the accuracy of our numerical methods.
}
In section \ref{Viscosity} we add viscosity to the peakon equation,
and describe our numerical methods for illustrating the different types of
behavior that may arise in the initial value problems for Peakon solutions
with $b>0$, $b=0$ and $b<-1$.  Section \ref{PeakonsViscousFate} using these
numerical methods to determine how viscosity affects the fate of the
peakons. Section \ref{IVP} provides a synopsis of the figures. Section
\ref{Conclusions} summarizes the paper's main conclusions.

\section{History and general properties of the b-equation}
\label{history&properties}
Camassa and Holm \cite{CH[1993]} derived the following equation for
unidirectional motion of shallow water waves in a particular Galilean
frame,
\begin{equation}\label{CH-equation-withdisp}
\hspace{-3mm}
m_t\
+
\underbrace{\ \ um_x\ \
}
_{\hspace{-2mm}\hbox{convection}\hspace{-2mm}}\
+\
\underbrace{\ \ 2\,u_xm\ \
}
_{\hspace{-2mm}\hbox{stretching}\hspace{-2mm}}\
=\
\underbrace{
-\,c_0 u_x-\,\gamma\, u_{xxx}\
}
_{\hbox{dispersion}}
\quad
\&\quad
m=u-\alpha^2u_{xx}
\,.
\end{equation}
Here $m=u-\alpha^{\,2}u_{xx}$ is a momentum variable, partial
derivatives are denoted by subscripts, the constants
$\alpha^{\,2}$ and $\gamma/c_0$ are squares of length scales, and
$c_0=\sqrt{g'h}$ is the linear wave speed for undisturbed water of depth
$h$ at rest under gravity $g'$ at spatial infinity, where $u$ and $m$ are
taken to vanish. Any constant value $u=u_0$ is also a solution of
(\ref{CH-equation-withdisp}).

Equation (\ref{CH-equation-withdisp}) was derived using Hamiltonian
methods in \cite{CH[1993]} and was shown in \cite{DGH[2001]}
also to appear as a water wave equation at quadratic order in the 
standard
asymptotic expansion for shallow water waves in terms of 
their two small
parameters (aspect ratio and wave height). The famous 
Korteweg-de Vries
(KdV) equation appears at linear order in this 
asymptotic expansion and is
recovered from equation (\ref{CH-equation-withdisp}) when $\alpha^2\to0$.
Both KdV at linear order and its nonlocal, nonlinear generalization in
equation (\ref{CH-equation-withdisp}) at quadratic order in this expansion
have the remarkable property of being completely integrable by the
isospectral transform (IST) method. The IST properties of KdV solitons are
well known and these properties for equation (\ref{CH-equation-withdisp})
were studied in, e.g., \cite{CH[1993]} and \cite{BSS[2000]}.

When linear dispersion is absorbed by a Galilean transformation and a
velocity shift, equation (\ref{CH-equation-withdisp}) reduces to an active
transport equation that contains competing quadratically nonlinear terms
representing convection and stretching,
\begin{equation}\label{CH-equation-zerodisp}
m_t\
+\
\underbrace{\ \ um_x\ \
}
_{\hspace{-2mm}\hbox{convection}\hspace{-2mm}}\
+\
\underbrace{\ \ 2\,u_xm\ \
}
_{\hspace{-2mm}\hbox{stretching}\hspace{-2mm}}\
=\
0
\,, \quad\hbox{with}\quad
m=u-\alpha^2u_{xx}
\,.
\end{equation}
This is a special case of equation (\ref{b-family}) for which $b=2$ and
$g(x)=e^{-|x|/\alpha}$. The traveling wave solution of
(\ref{CH-equation-zerodisp}) is the ``peakon,''
$u(x,t)=ce^{-|x-ct|/\alpha}$ found in \cite{CH[1993]}, where
$e^{-|x|/\alpha}$ is the Green's function for the Helmholtz operator that
relates $m$ and $u$. The interactions among $N$ peakons are governed by the
$2N$ dimensional dynamical system for the speeds
$p_i(t)$ and positions $q_i(t)$ $i=1,\dots,N,$ appearing in the solution,
\begin{equation}\label{Superposed-peakons}
u(x,t)=\sum_{i=1}^N p_i(t)\,e^{-|x-q_i(t)|}
\,.
\end{equation}
As shown in Camassa and Holm \cite{CH[1993]}, a closed integrable
Hamiltonian system of ordinary differential equations for the speeds
$p_i(t)$ and positions $q_i(t)$ results upon substituting the
superposition of peakons (\ref{Superposed-peakons}) into
equation (\ref{CH-equation-zerodisp}). This integrable system governs the
dynamics of the peakon interactions.

A variant of equation (\ref{CH-equation-zerodisp}) with coefficient
$b=2\to{b}=3$,
\begin{equation}\label{DHH-equation}
m_t\
+\
\underbrace{\ \ um_x\ \
}
_{\hspace{-2mm}\hbox{convection}\hspace{-2mm}}\
+\
\underbrace{\ \ 3\,u_xm\ \
}
_{\hspace{-2mm}\hbox{stretching}\hspace{-2mm}}\
=\
0
\,, \quad\hbox{with}\quad
m=u-\alpha^2u_{xx}
\,,
\end{equation}
was first singled-out for further analysis by Degasperis and Procesi
\cite{Degasperis[1999]}. Degasperis, Holm and Hone \cite{DHH[2002]}
discovered that this
$2\to3$ variant of equation (\ref{CH-equation-zerodisp}) possesses
superposed peakon solutions (\ref{Superposed-peakons}) and also is
completely integrable by the isospectral transform method. Thus, the
$N-$dimensional pulson solution (\ref{Superposed-peakons}) is a
completely integrable dynamical system for
(\ref{DHH-equation}), as well, but with different
dynamics for the speeds $p_i(t)$ and positions $q_i(t)$ of the peakons.

Fringer and Holm \cite{FH[2001]} extended the zero-dispersion equation
(\ref{CH-equation-zerodisp}) for the peakons to the ``pulson'' equation,
\begin{equation}\label{Pulson-equation}
m_t\
+\
\underbrace{\ \ um_x\ \
}
_{\hspace{-2mm}\hbox{convection}\hspace{-2mm}}\
+\
\underbrace{\ \ 2\,u_xm\ \
}
_{\hspace{-2mm}\hbox{stretching}\hspace{-2mm}}\
=\
0
\,, \quad\hbox{with}\quad
u=g*m
\,.
\end{equation}
Here $u=g*m$ denotes the convolution (or filtering)
\begin{equation}\label{unused-2}
u(x)=\int_{-\infty}^\infty g(x-y)m(y)\,dy
\,,
\end{equation}
that relates velocity $u$ to momentum
density $m$ by integration against the kernel $g(x)$.  Fringer and Holm
\cite{FH[2001]} chose $g(x)$ to be an even function, so that $u$ and $m$
have the same parity. They studied the effects of the shape of the
traveling wave
$u(x,t)=cg(x-ct)$ on its interactions with other traveling waves in the
superposed solution,
\begin{equation}\label{unused-3}
u(x,t)=\sum_{i=1}^N p_i(t)\,g(x-q_i(t))
\,,\end{equation}
This superposed solution of traveling wave forms with time 
dependent
speeds $p_i(t)$ and positions $q_i(t)$, $i=1,\dots,N,$ 
revealed that the
nonlinear interactions among these pulsons  occur 
by elastic two-pulson
scattering, even though the Fringer-Holm pulson 
equation
(\ref{Pulson-equation}) is not integrable for an arbitrary 
choice of the
kernel $g$. When $g(x)=e^{-|x|/\alpha}$ is assumed, the 
pulson equation
for $b=2$ in  (\ref{Pulson-equation}) specializes to 
the peakon equation
for $b=2$ in 
(\ref{CH-equation-zerodisp}).

\subsection{Discrete symmetries: reversibility, parity and signature}

Equation (\ref{b-family}) for $m$ is reversible, or invariant under
$t\to-t$ and $u\to-u$. The latter implies $m\to-m$.  Hence, the
transformation $u(x,t)\to-u(x,-t)$ takes solutions into solutions,
and in particular, it reverses the direction and amplitude of the
traveling wave $u(x,t)=cg(x-ct)$.

We choose $g(x)$ to be an even function so that $m$ and $u=g*m$ both
have odd parity under mirror reflections.  Hence, equation
(\ref{b-family}) is invariant under the parity reflections
$u(x,t)\to-\,u(-x,t)$, and the solutions of even and odd parity form
invariant subspaces.

Equation (\ref{b-family}) implies a similar reversible, parity
invariant equation for the {\bf absolute value} $|m|$,
\begin{equation}\label{b-family|m|}
   |m|_t+\, u|m|_x+b\,u_x|m| = 0\,,
   \quad\hbox{with}\quad
   u=g*m
   \,.
\end{equation}
So, the positive and negative components
$m_\pm=\frac{1}{2}(m\pm|m|)$ satisfy equation (\ref{b-family})
separately.  Also, if $m$ is initially zero, it remains so.  This is
conservation of the signature of $m$.

\subsection{Lagrangian representation}

If $m^{1/b}$ is well-defined, equation (\ref{b-family}) may be written
as the conservation law
\begin{equation}\label{b-cons}
\partial_t\, m^{1/b}
+
\partial_x( m^{1/b} u )
=
0\,,
\end{equation}
and equation (\ref{b-family|m|}) for the absolute value implies
\begin{equation}\label{b-cons|m|}
\partial_t\, |m|^{1/b}
+
\partial_x( |m|^{1/b} u )
=
0\,.
\end{equation}
Adding and subtracting equations (\ref{b-cons}) and (\ref{b-cons|m|})
implies
\begin{equation}\label{b-family-mpm}
\partial_t\, (m^{1/b})_\pm
+
\partial_x\big( (m^{1/b})_\pm u \big)
=
0
\quad\hbox{with}\quad
(m^{1/b})_\pm=\frac{1}{2}(m^{1/b}\pm|m|^{1/b})
\,.
\end{equation}
Consequently, regions of positive and negative $m$ are transported by the same
velocity and their boundaries propagate so as to separately
preserve the two
integrals,
\begin{equation}
\int_{-\infty}^\infty (m^{1/b})_\pm
\,dx
\,.
\end{equation}
The common transport velocity allows a
transformation to Lagrangian
coordinates
$X_\pm$ defined
by
\begin{equation}
dX_\pm=(m^{1/b})_\pm(dx - udt)
\quad\hbox{so
that}\quad
\partial_tX_\pm + u \partial_x X_\pm
=0
\,.
\end{equation}
This formal transformation is not strictly
defined where $(m^{1/b})_\pm$
vanishes. However, by equation
(\ref{b-family-mpm}), regions where
$(m^{1/b})_\pm$ vanishes do not
propagate and do not
contribute to the integrated value
of
$X_\pm=\int_{-\infty}^x(m^{1/b})_\pm(y,0)\,dy$. Hence, these
regions
may be identified and excluded initially. The formal inverse
relation
holding in the remaining regions,
\begin{equation}
dx
=(m^{1/b})_\pm^{-1}\,dX _\pm + udt\,,
\end{equation}
implies
that
\begin{equation}
\frac{dx}{dt}\Big|_{X
_\pm}=u(x,t)
\,,
\end{equation}
so the Lagrangian trajectories
$x=x(X_\pm,t)$ of positive and negative
integrated initial values
of
$X_\pm=\int_{-\infty}^x(m^{1/b})_\pm(y,0)\,dy$ are transported by
the same
velocity $u=g*m$.

\subsection{Preservation of the norm
$\|m\|_{L^{1/b}}$ for $0\le{b}\le1$}

If $|m|^{1/b}$ is well-defined,
the continuity equation form
(\ref{b-cons|m|}) of equation
(\ref{b-family}) implies
conservation
of
\begin{equation}
\int_{-\infty}^\infty
|m|^{1/b}\,dx
=
\int_{-\infty}^\infty|m_0|^{1/b}\,dx
\,,
\quad\hbox{where}\quad
m_0(x)=m(x,0)
\,.
\label{|M|cons}
\end{equation}
This integral is conserved for all
$b$, but only defines a norm
(the $L^{1/b}$ norm $\|m\|_{L^{1/b}}$)
in the closed interval
$0\le{b}\le1$.  In the limit $b\to0$ this
becomes the $L_\infty$ norm
$|m|_{max}$.  Hence, when $b=0$ equation
(\ref{b-family}) has
both a maximum principle and a minimum principle
for $m$.  Such a
principle is meaningful only if $m^{1/b}$ is an
ordinary function,
e.g., if $m$ is not a generalized function, such
as the delta
functions that occur for the peakons we will discuss
below.

Thus, the $L^{1/b}$ norm $\|m\|_{L^{1/b}}$ is conserved by
equation
(\ref{b-family}) provided $|m|^{1/b}$ is well-defined for
the closed
interval $0\le{b}\le1$.  One may also define the
corresponding conserved
norm for $1/m$ in the closed interval
$-1\le{b}\le0$, provided
$|m|^{1/b}$ is well-defined on this
interval.

\subsection{Lagrangian representation for integer $b$}

Fluid  convection means transport of a quantity by the fluid motion.
Examples of transported fluid quantities are circulation (a
one-form) in Kelvin's theorem for the Euler equations and its
exterior derivative the vorticity (a two-form, by Stokes theorem) in the
Helmholtz equation.  For a Lagrangian trajectory $x(X,t)$ satisfying
$x(X,0)=X$ and
\begin{equation}
dx
=(m^{1/b})^{-1}dX + udt\,,
\end{equation}
we have seen that the conservation law (\ref{b-cons}) implies
\begin{equation}\label{m-eqn}
m^{1/b}(x,t)dx
=
m^{1/b}(X,0)dX
\,,
\end{equation}
provided that $m^{1/b}$ is a well defined function. The last issue may be
circumscribed as follows when $b$ is an integer. In 1D, higher order
differential forms may be created by using the direct, or tensor, product,
e.g., $dx\otimes dx = dx^{\otimes2}$. Consequently, the tensor product of
each side of equation (\ref{m-eqn}) $b$ times gives%
\footnote{Cases with integer values of $b$ will allow $m$ to be a
generalized function. Cases with non-integer values of $b$ will revert to
equation (\ref{m-eqn}) for which $m$
is required to be a classical function. }
\begin{equation}\label{otimes-b}
m(x,t)dx^{\otimes{b}}
=
m(X,0)dX^{\otimes{b}}
\,.
\end{equation}
Taking the partial time derivative of this equation at constant Lagrangian
coordinate $X$ and using $dx/dt|_X=u$ yields equation (\ref{b-family}) in
the form
\begin{equation}
\frac{d}{dt}\Big|_X\Big(m(x,t)dx^{\otimes{b}}\Big)
=
(m_t + um_x + b\,u_xm)dx^{\otimes{b}}
=
0
\,.
\end{equation}
Thus, when the parameter $b$ in equation (\ref{b-family}) is
an integer, it may be regarded geometrically as the number of dimensions
that are brought into play by coordinate transformations of the quantity
$mdx^{\otimes{b}}$ associated with $m$. Cases of equation (\ref{b-family})
with negative integer $b<0$ may be interpreted  as
\begin{equation}\label{unused-4}
\frac{d}{dt}\Big|_X
\Big(m\big(\partial_x\big)^{\otimes(-b)}\,\Big)=0
\,.
\end{equation}
For example, the case $b=-1$ may be written as
\begin{equation}
\frac{d}{dt}\Big|_X(m\partial_x)
=(m_t + um_x-u_xm)\big(\partial_x\big)
=0
\,,
\end{equation}
in which the difference of terms $(um_x-u_xm)\partial_x$ is the
commutator of the vector fields $u\partial_x$ and $m\partial_x$ on the real
line. The rest of the paper will remain in the Eulerian (spatial)
representation.

\subsection{Reversibility and Galilean covariance}

Equation (\ref{b-family}) is reversible, i.e., it is invariant under the
discrete transformation $u(x,t)\to-u(x,-t)$.  Equation (\ref{b-family}) is
also Galilean-covariant for all $b$. In fact, equation (\ref{b-family})
keeps its form under transformations to an {\it arbitrarily} moving
reference frame for all $b$. This includes covariance under transforming to
a {\it uniformly} moving Galilean frame. However, only in the case
$b=0$ is equation (\ref{b-family}) Galilean {\it invariant}, assuming that
$m$ Galileo-transforms in the same way as $u$. In this case, equation
(\ref{b-family}) transforms under
\begin{equation}\label{xtrans}
\hspace{-3mm}
t\to t+t_0\,, \quad
x\to x+x_0+ct\,, \quad
u\to u+c+u_0\,, \quad
m\to m+c+u_0\,,
\end{equation}
to the form
\begin{equation}\label{b-family-xtrans}
m_t+\, um_x+bu_x\,m
+u_0m_x
+bu_x(c+u_0)
=0
\,, \quad\hbox{with}\quad
u=g*m
\,.
\end{equation}
Thus, equation 
(\ref{b-family}) is invariant under space and time 
translations
(constants $x_0$ and $t_0$), covariant under 
Galilean
transforms (constant $c$),
and acquires linear dispersion 
terms under velocity shifts (constant $u_0$).
Equation 
(\ref{b-family}) regains Galilean invariance if $m$ is
Galilean 
invariant.
However, the dispersive term $u_0m_x$ introduced
by the constant
velocity shift $u_0\ne0$
breaks the reversibility of equation (\ref{b-family}) even if $m$ is
invariant under this shift.

\rem{
\subsection{Dual cases under $L^2$}

Cases $b$ and $b\,'$ with $b+b\,'=1$ are natural duals under the
$L^2$ pairing
\begin{equation}\label{l2-pairing}
      <m,u>=\int mu\,dx\,.
\end{equation}
The case $b=0$ is $(\partial_t+\pounds_u)m=m_t+um_x=0$, describing 1D
convection of the (active) scalar $m$ by the velocity $u$.  Its dual
under $L^2$ pairing is the case $b=1$ for convection of the (active)
density $mdx$ by the velocity $u$.  The case $b=2$ may be written as
$m_t+{\rm ad}^*_um=0$ where ad$^*_um=(um)_x+u_xm$ is (minus) the dual
of the ad-operation under the $L^2$ pairing (\ref{l2-pairing}).  In this
case, the quantity $m(dx)^2$ is interpreted as a 1D momentum (a one-form
density, dual to velocity vector fields) and the pairing
(\ref{l2-pairing}) is (twice) the conserved kinetic energy.  The dual
to this case is the case $b\,'=-1$, for which $m$ and $u$ are both 1D
vector fields.  Equation (\ref{b-family}) for $b=2$ also has an
Euler-Poincar\'e variational principle and an associated Lie-Poisson
Hamiltonian structure.  See Holm, Marsden and Ratiu [1998a,1998b]
\cite{HMR[1998a],HMR[1998b]} for more discussion.
}

\subsection{Integral momentum conservation}

Equation (\ref{b-cons}) implies that $M=\int_{-\infty}^{\infty}
m\,dx$ is conserved for any
$g$ when $b=1$.  However, when $g(x)$ is even, the family of equations
(\ref{b-family}) also conserves the total momentum integral $M$ for
any $b$.  This is shown by directly calculating from (\ref{b-family})
that
\begin{equation}
\frac{d}{dt}\int_{-\infty}^\infty m(x) \,dx
=
(1-b)\int_{-\infty}^\infty\int_{-\infty}^\infty
m(x)g\,'(x-y)m(y)\,dx\,dy
=0
\,,
\end{equation}
in which the double 
integral vanishes as the product of an even
function and an odd 
function under interchange of $x$ and $y$, when
$g\,'(-x)=-g\,'(x)$. Hence, for even $g(x)$,
$M=\int_{-\infty}^{\infty} m\,dx$ is conserved for either periodic
or vanishing boundary conditions and for any $b$.
We shall assume henceforth that $g(x)$ is even and,
moreover, that
the integral $\int
m\,g*m\,dx$ is sign-definite, so that it defines a
norm,
\begin{equation}\label{KE-norm}
\|m\|_g^2
=
\int_{-\infty}^\infty
m\,g*m\,dx
=
\int_{-\infty}^\infty
\int_{-\infty}^\infty
m(x)\,g(x-y)\,m(y)\,dx\,dy
\ge0
\,.
\end{equation}
This norm is conserved by equation (\ref{b-family}) when $b=2$.

\section{Traveling waves and generalized functions}\label{twaves}

Its invariance under space and time translations ensures that equation
(\ref{b-family}) admits traveling wave solutions for any
$b$.  Let us write the traveling wave solutions
as
\begin{equation}\label{unused-6}
u=u(z)
\quad\hbox{and}\quad
m=m(z)
\,,
\quad\hbox{where}\quad
z=x-ct
\,,
\end{equation}
and
let prime $\,'$ denote $d/dz$.

\subsection{Case $b=0$}

\subsubsection{Pulsons for $b=0$}
For $b=0$, equation 
(\ref{b-family}) is Galilean invariant and
its traveling wave 
solutions satisfy
\begin{equation}\label{bzero-eqn}
(u(z)-c)m'(z)=0\,,
     \quad
     z=x-ct\,,
\end{equation}
where 
prime $\,'$ denotes $d/dz$.
Equation (\ref{bzero-eqn}) admits 
generalized functions
$m\,'(z)\simeq\delta(z)$ matched by $u-c=0$ at 
$z=0$.  The velocity $u$
is given by the integral of the Green's 
function that relates $m$ and
$u=g*m$,
\begin{equation}
     u-c 
\simeq c 
\Big[\int\!\!g(y)\,dy\Big]^z_0\,.
\end{equation}

\subsubsection{Peakons, 
ramps, and cliffs for $b=0$}

When $g(x)=e^{-|x|/\alpha}$ (the 
Green's function for the 1D Helmholtz
operator), we have 
$m=u-\alpha^2u_{xx}$.  Consequently, the 
equation
$m'=u'-\alpha^2u'''=\pm2\delta(z)$ with  $u-c=0$ at $z=0$ is 
satisfied
by
\begin{equation}
     u-c = \pm 
c\Big[\int\!\!e^{-|y\,|/\alpha}\,dy\Big]^z_0
         = \pm c\,\,{\rm 
sgn}(z)\Big(1-e^{-|z|/\alpha}\Big)\,.
\end{equation}
This represents 
a rightward moving traveling wave that connects the left
states 
$u-c=\pm c$ to the same two right 
states.

\begin{definition}[Peakons]
The symmetric connections $u=\pm 
ce^{-|z|/\alpha}$, with a jump in derivative
at $z=0$, are the 
peakons, for which $m=u-\alpha^2u_{xx}$ 
and
$g(x)=e^{-|x|/\alpha}$.
\end{definition}

\begin{definition}[Cliffs]
The 
antisymmetric connections $u=\pm c\,{\rm
sgn}(z)(1-e^{-|z|/\alpha})$ 
(with $u-c=\pm c$ connecting to
$u-c=\mp c$), with no jump in 
derivative at $z=0$, are the regularized
shocks (cliffs).  These 
propagate rightward but may face either leftward
or rightward, 
because equation (\ref{b-family}) in the absence of
viscosity has no 
entropy condition that would distinguish between leftward
and 
rightward facing 
solutions.
\end{definition}

\begin{definition}[Ramps]
Equation 
(\ref{b-family}) also has ramp-like similarity 
solutions
$u\simeq{x/t}$ when $g(x)=e^{-|x|/\alpha}$ for any $b$. 
These may emerge
in the initial value problem for the peakon case of 
equation
(\ref{b-family}) and interact with the peakons and 
cliffs.
\end{definition}

\begin{remark}[First integral for $b=0$ 
traveling waves]
For $b=0$, the traveling wave equation 
(\ref{bzero-eqn}) apparently has
only the first integral for 
$m=u-\alpha^2u_{xx}$,
\begin{equation}\label{unused-7}
(u-c)(u-\alpha^2u'')-\frac{u^2}{2} + \frac{\alpha^2}{2}{u'\,}^2
=
        K
        \,.
     \end{equation}
     Thus, perhaps 
surprisingly, we have been unable to find a second
     integral for 
the traveling wave equation for peakons when 
$b=0$.
\end{remark}

\begin{remark}[Reversibility]
Reversibility 
means that equation (\ref{b-family}) is invariant under the
transformation
$u(x,t)\to-u(x,-t)$. Consequently, the rightward traveling waves have
leftward moving counterparts under the symmetry $c\to-c$.  The case of
constant velocity $u=\pm{c}$ is also a solution.
\end{remark}\bigskip

\remfigure{
\begin{figure}
\begin{center}
     \leavevmode {
        \hbox{\epsfig{
           figure=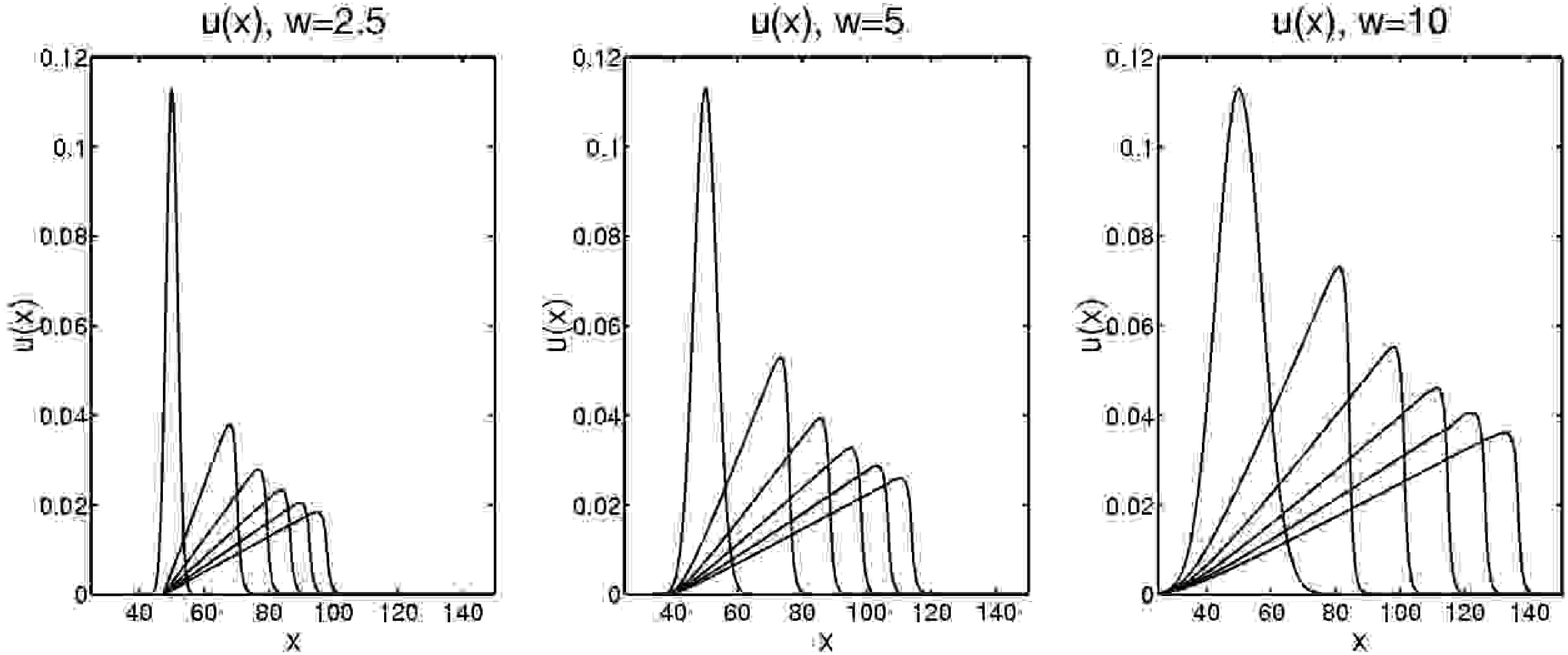, scale=0.3
        }}
     }
\caption{\label{Gauss-ic-pkn/a1,b0/w2.5-10}
   {\bf Ramps and cliffs for $b=0$.}
   Inviscid $b$-family,
   $b=0$,
   $\alpha=1$,
   initial width $w=2.5,5,10$.
\rem{\sf
For $b=0$, evolving velocity profile under the peakon equation for a set
of Gaussian initial conditions of increasing width $w=2.5,5,10$, for
$\alpha=1$.  Times are T=0, 500, 1000, 1500, 2000, 2500.
}
}
\end{center}
\end{figure}
}

Figure \ref{Gauss-ic-pkn/a1,b0/w2.5-10} shows
that the ramp and cliff
pattern develops in the
velocity profile under the peakon equation
(\ref{b-family}) with
$g(x)=e^{-|x|/\alpha}$ for a set of Gaussian
initial conditions
$(5\sqrt\pi)^{-1}\exp(-(x-50)^2/w)$
of increasing
widths $w=2.5,5,10$,
for $\alpha=1$ and $b=0$.  Apparently, the ramp
solution is
numerically stable, but the coexisting peakon solution is
not stable in
this case. A complete stability analysis of these
various solutions is
outside the scope of the present paper. Instead
we shall investigate the
solutions of equation (\ref{b-family}) by
numerically integrating
selected
examples.

\subsection{Case
$b\ne0$}

For
$b\ne0$, the conservation law (\ref{b-cons}) for traveling
waves
becomes,
\begin{equation}\label{unused-8}
     \Big((u-c)m^{1/b}\Big)\,'=0\,,
\end{equation}
which yields
after one integration
\begin{equation}\label{bnonzero-eqn}
(u-c)^bm=K\,,
\end{equation}
where $K$ is the first integral.  For
$g(x)=e^{-|x|/\alpha}$, so that
$m=u-\alpha^2u_{xx}$, this
becomes
\begin{equation}\label{1stConstant}
(u-c)^b(u-\alpha^2u\,'')=K\,.
\end{equation}
For $u-c\ne0$ we rewrite
this as
\begin{equation}\label{unused-9}
     \alpha^2u\,''=u-K(u-c)^{-b}\,
\end{equation}
and integrate again
to give the second integral in two separate cases,
\begin{eqnarray*}\label{unused-10}
\alpha^2{u\,'\,}^2
     =
     \left\{
     \begin{array}{l}
     u^2 -
\frac{2K}{1-b}(u-c)^{1-b} + 2H
     \quad\hbox{for}\quad b\ne1\,,

\\[4pt]
     u^2 - 2K\log(u-c) + 2H
     \quad\hbox{for}\quad b=1\,.

\end{array}
     \right.
\end{eqnarray*}
We shall rearrange this into
quadratures:
\begin{equation}\label{tw-soln/bne1}
     \pm
\,\frac{dz}{\alpha} =
     \frac{du}{\Big[u^2 -
\frac{2K}{1-b}(u-c)^{1-b} + 2H\Big]^{1/2}}
     \quad\hbox{for}\quad
b\ne1
     \,,
\end{equation}
and
\begin{equation}\label{tw-soln/b=1}
\pm \,\frac{dz}{\alpha}
     =
     \frac{du}{\Big[u^2 - 2K\log(u-c) +
2H\Big]^{1/2}}
     \quad\hbox{for}\quad b=1
\,.
\end{equation}
For $b=1$ and $K\ne0$, the integral in equation
(\ref{tw-soln/b=1})
     is transcendental.

\subsubsection{Special
cases of traveling waves for $b\ne0$}

\begin{itemize}

\item
For
$K=0$ the two quadratures (\ref{tw-soln/bne1}) and
(\ref{tw-soln/b=1})
are equal, independent of
$b$, and elementary,
thereby yielding the traveling wave
solutions
\begin{equation}\label{DeformedPkn}
     e^{- |z|/\alpha}
=
     \frac{u+\sqrt{u^2+2H}}{c+\sqrt{c^2+2H}}\,,
\end{equation}
with
$u-c=0$ at $z=0$.

\item
For $H=0$ equation (\ref{DeformedPkn})
recovers the peakon traveling
wave.

\item
For $H>0$ equation 
(\ref{DeformedPkn}) gives a rightward 
moving
traveling wave that is a 
continuous deformation of the 
peakon.

\item
For $H>0$ and $c=0$ 
equation (\ref{DeformedPkn}) 
gives stationary
solutions of the 

form
\begin{equation}\label{stat-pkn-like}
    u+\sqrt{u^2+2H}\simeq 
e^{- 
|z|/\alpha}\,.
\end{equation}

\end{itemize}

\subsection{Case 
$b>0$}

\subsubsection{Pulsons 
for $b>0$}\label{pulson-intro}

Equation (\ref{b-family}) for $b>0$ 
has nontrivial solutions vanishing
as $|z|\to\infty$ that allow $K=0$ 
in equation (\ref{bnonzero-eqn}), 
so
that
\begin{equation}\label{pulson-cond} 
(u-c)^bm=0\,.
\end{equation}
This admits the generalized function 
solutions
\begin{equation}\label{pulson-tw-soln}
     m=c\delta(z)
 \quad\hbox{and}\quad
     u=g*m=cg(z)
     \,,
\end{equation}
matched 
by $u-c=0$ at $z=0$.
This is the {\bf  pulson traveling wave}, whose 
shape in $u$ is given by the
kernel $g$.  The constant velocity case 
$u=c$ is a trivial
traveling wave.

\begin{remark}[Pulson and peakon 
traveling waves]
The pulson solution (\ref{pulson-tw-soln}) requires 
$g(0)=1$ and $g\,'(0)=0$.
We shall assume for definiteness
that the 
even function $g(z)$ achieves its maximum at $g(0)=1$, so that
the 
symmetric
pulson traveling wave $u(x,t)=cg(x-ct)$ moves at the speed 
of its
maximum, which occurs at its center of symmetry. For example, 
the
peakon $u(x,t)=ce^{-|x-ct|}$ moves at the speed of its 
peak.
\end{remark}

\subsubsection{Peakons for 
$b>1$}\label{peakon-intro}

Equation (\ref{1stConstant}), for which 
$g(x)=e^{-|x|/\alpha}$, yields
the peakon traveling 
wave
\begin{equation}\label{peakon-soln}
     u(z)=ce^{-|z|/\alpha}
\quad\hbox{and}\quad
m(z)=u-\alpha^2u\,''=2c\delta(z/\alpha)\,
\end{equation}
when 
$K=0$.

\remfigure{
\begin{figure}
\begin{center}
\leavevmode 
{
\hbox{\epsfig{figure=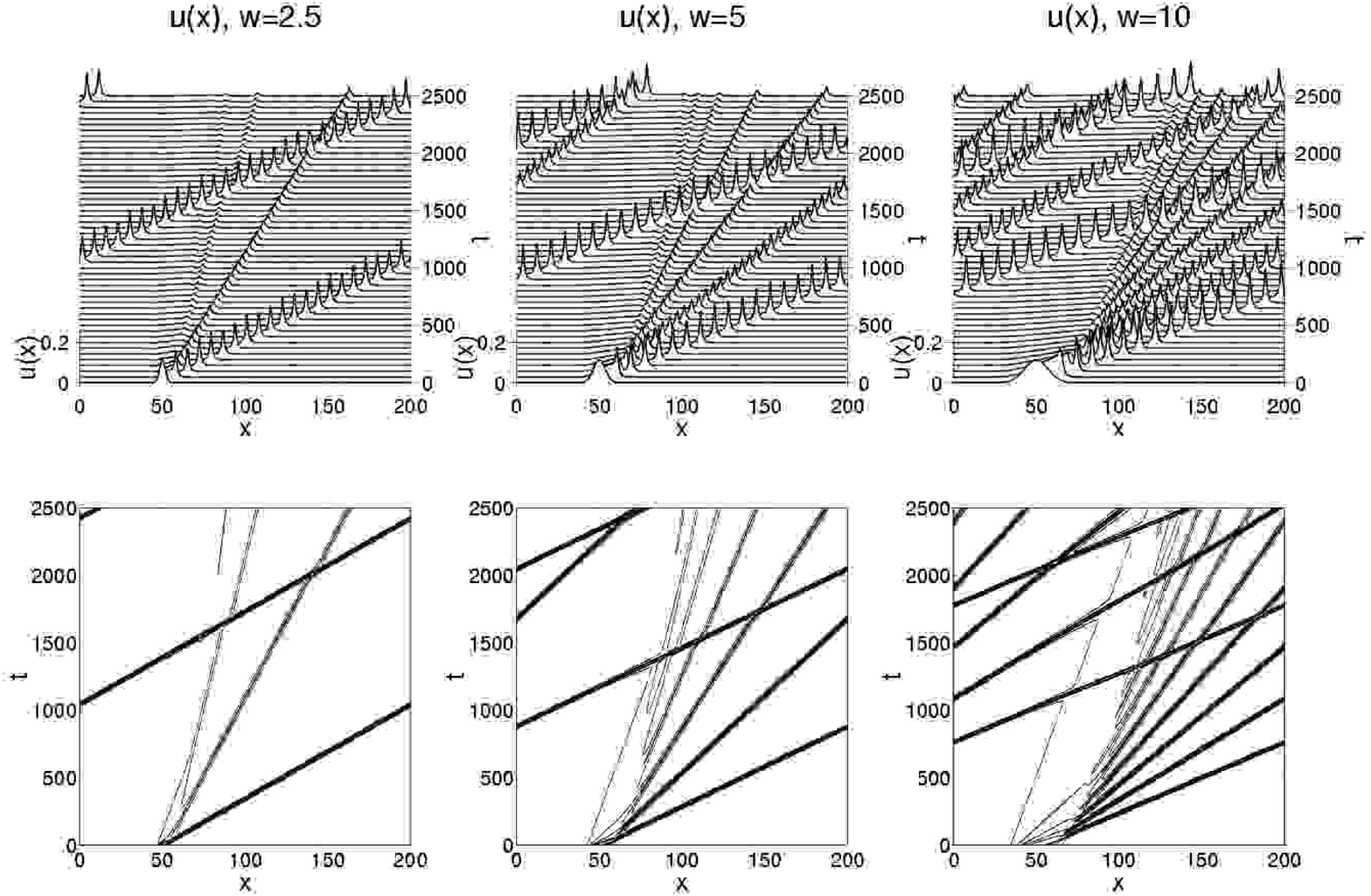, scale=0.3}}
}
\caption{\label{Gauss-ic-pkn/a1,b2/w2.5-10}
   {\bf Peakons for $b=2$.}
   Inviscid $b$-family,
   $b=2$,
   $\alpha=1$,
   initial width $w=2.5,5,10$.
\rem{\sf
For $b=2$, evolving velocity profile under the peakon equation for a set of 
Gaussian initial conditions of increasing width $w=2.5,5,10$ for $\alpha=1$.
}
}
\end{center}
\end{figure}
}

\remfigure{
\begin{figure}
\begin{center}
\leavevmode {
\hbox{\epsfig{figure=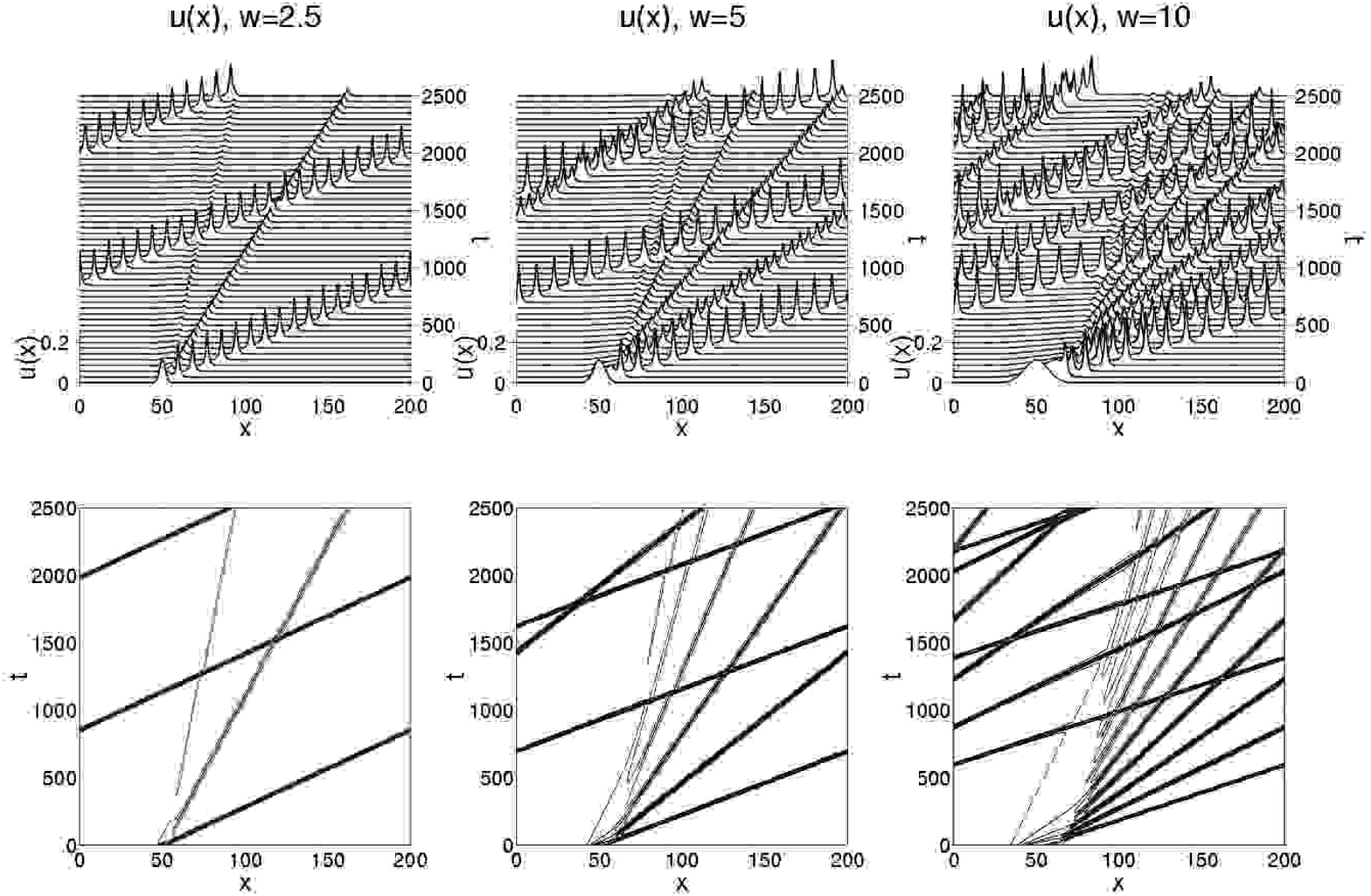, scale=0.3}}
    }
\caption{\label{Gauss-ic-pkn/a1,b3/w2.5-10}
   {\bf Peakons for $b=3$.}
   Inviscid $b$-family,
   $b=3$,
   $\alpha=1$,
   initial width $w=2.5,5,10$.
\rem{\sf
For $b=3$, evolving velocity profile under the peakon equation for a set
of Gaussian initial conditions of increasing width $w=2.5,5,10$ for
$\alpha=1$.
}
}
\end{center}
\end{figure}
}

Figures \ref{Gauss-ic-pkn/a1,b2/w2.5-10} and
\ref{Gauss-ic-pkn/a1,b3/w2.5-10} show
the effects of varying the width $w$ of a Gaussian initial condition
$(5\sqrt\pi)^{-1}\exp(-(x-50)^2/w)$ for the peakon equation
in a periodic domain, when $\alpha=1$
and $b=2,3$.  As the width of the initial Gaussian increases, the
figures show that more peakons of width $\alpha=1$ are emitted.  (This is
consistent with conservation of momentum.) The peakons are observed to be
stable for $b>1$, they propagate as solitary traveling waves, and they
interact elastically. We shall discuss the peakon interactions in
more detail in sections \ref{Peakons} through \ref{IVP}.

\subsection{Case $b<0$}

We shall examine the cases $b=-0.5,-1,-2,-3,-4$.  Numerical results for
$b=-2$ and $b=-3$ are described in section \ref{numerical-b-2-3}.  For
other values of $b<0$ the analysis is similar, but it involves less elementary
considerations such as transcendental or hyperelliptic functions.  The numerics
shown later will demonstrate that the elementary solutions discussed here, many
of them stationary, do tend to emerge in numerical integrations of the initial
value problem for equation (\ref{b-family}) with $b\le-1$.

\subsubsection{Case $b=-1/2$}

Figure \ref{Gauss-ic-pkn/a1,b-0.5/w10-20} shows that a ramp and cliff pattern
develops in the velocity profile under the peakon equation (\ref{b-family})
with $g(x)=e^{-|x|/\alpha}$ for a set of Gaussian initial conditions
$(5\sqrt\pi)^{-1}\exp(-(x-100)^2/w)$ of increasing width $w=10,15,20$, for
$\alpha=1$ and $b=-1/2$.  Apparently, the ramp solution is numerically stable
for $b=-1/2$.

\remfigure{
\begin{figure}
\begin{center}
    \leavevmode {
\hbox{\epsfig{
figure=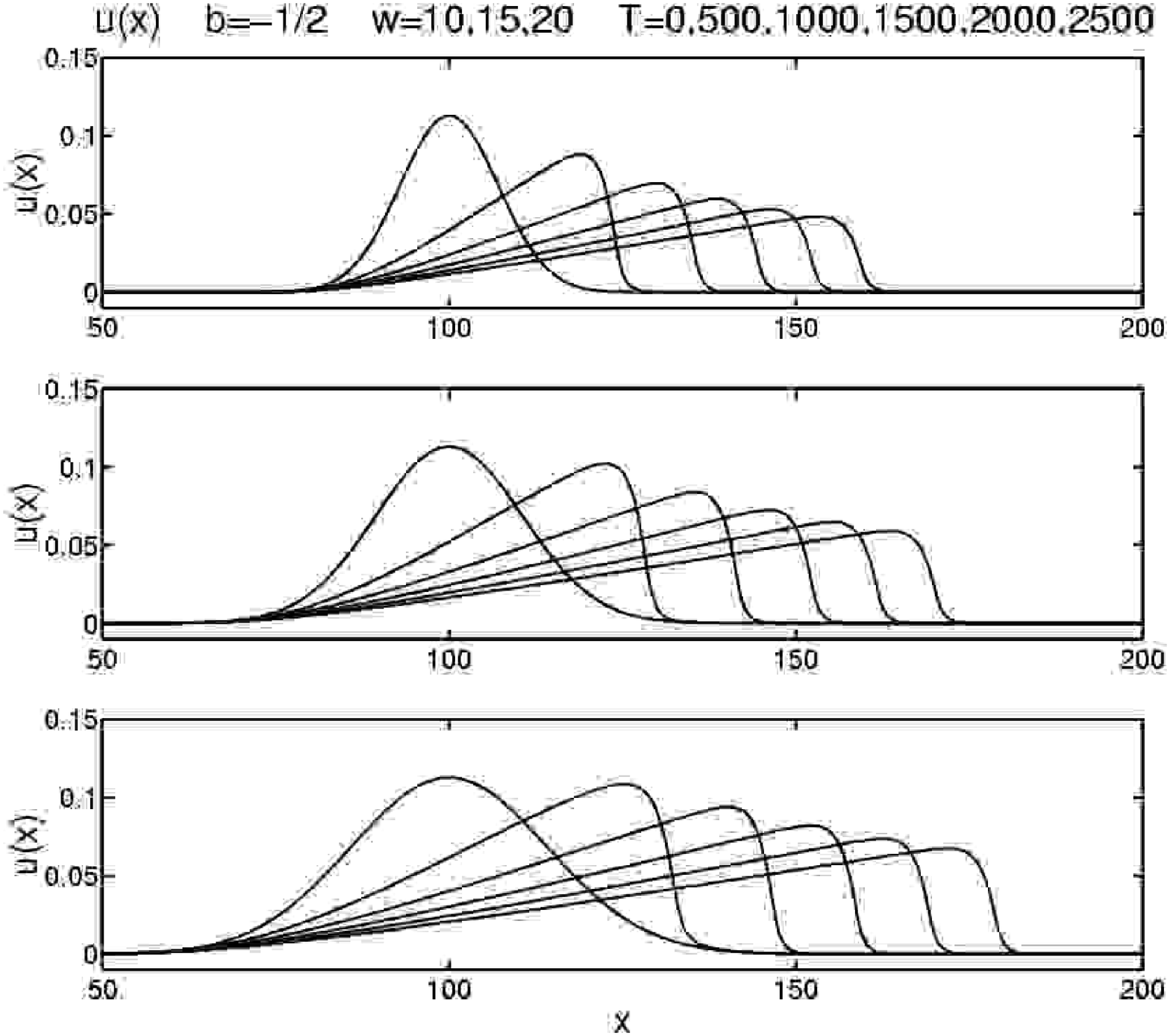, scale=0.4
}}
    }
\caption{\label{Gauss-ic-pkn/a1,b-0.5/w10-20}
   {\bf Ramps and cliffs for $b=-1/2$.}
   Inviscid $b$-family,
   $b=-1/2$,
   $\alpha=1$,
   initial width $w=10,15,20$.
\rem{\sf
For $b=-1/2$, evolving velocity profile under the peakon equation for a
set of Gaussian initial conditions of increasing width $w=10,15,20$
for $\alpha=1$.  Times are T=0, 500, 1000, 1500, 2000, 2500.
}
}
\end{center}
\end{figure}
}

\subsubsection{Case $b=-1$}

For $b=-1$, equation (\ref{tw-soln/bne1}) becomes
\begin{equation}\label{unused-a}
    \pm
\,\frac{dz}{\alpha}
     =
     \frac{du}{\Big[u^2 - K(u-c)^{2} +
2H\Big]^{1/2}}
     \,,
\end{equation}
which integrates to
\begin{equation}\label{unused-11}
e^{- |z|/\alpha}
     =
     \frac{u+\sqrt{u^2 - K(u-c)^{2} + 2H} +
Kc}{c+\sqrt{c^2+2H}+ Kc}
     \,,
\end{equation}
with $u-c=0$ at
$z=0$.  ($K=0$ and $H=0$ recovers the peakon
traveling
wave.)

\rem{
\paragraph{Stationary solutions.} Stationary
solutions of equation
(\ref{tw-soln/bne1}) with $c=0$, $K\ne0$ and
$H\ne0$ also exist. Their
precise forms depend on the sign of
$(1-K)/(2H)$ and will not be pursued
farther
here.
}

\begin{remark}[Stationary
plane wave solutions for $b=-1$]
Equation (\ref{b-family}) for $b=-1$
is satisfied for any
wavenumber $k$ by,
\begin{equation}\label{unused-13}
m=\cos(k(x-ct)+\phi_0)
     \quad\hbox{and}\quad
u=\hat{g}(k)\cos(k(x-ct)+\phi_0)
     \,,
\end{equation}
where $\hat{g}(k)$ is the
Fourier transform of the kernel $g(x)$ and
$\phi_0$ is a constant 
phase shift. In the absence of linear 
dispersion,
these solutions are 
stationary, $c=0$. When linear 
dispersion is
added to equation 
(\ref{b-family}), these solutions 
are the 1D analogs of
Rossby waves
in the 2D quasigeostrophic
equations.
\end{remark}

\remfigure{
\begin{figure}
\begin{center}
\leavevmode {
\hbox{\epsfig{figure=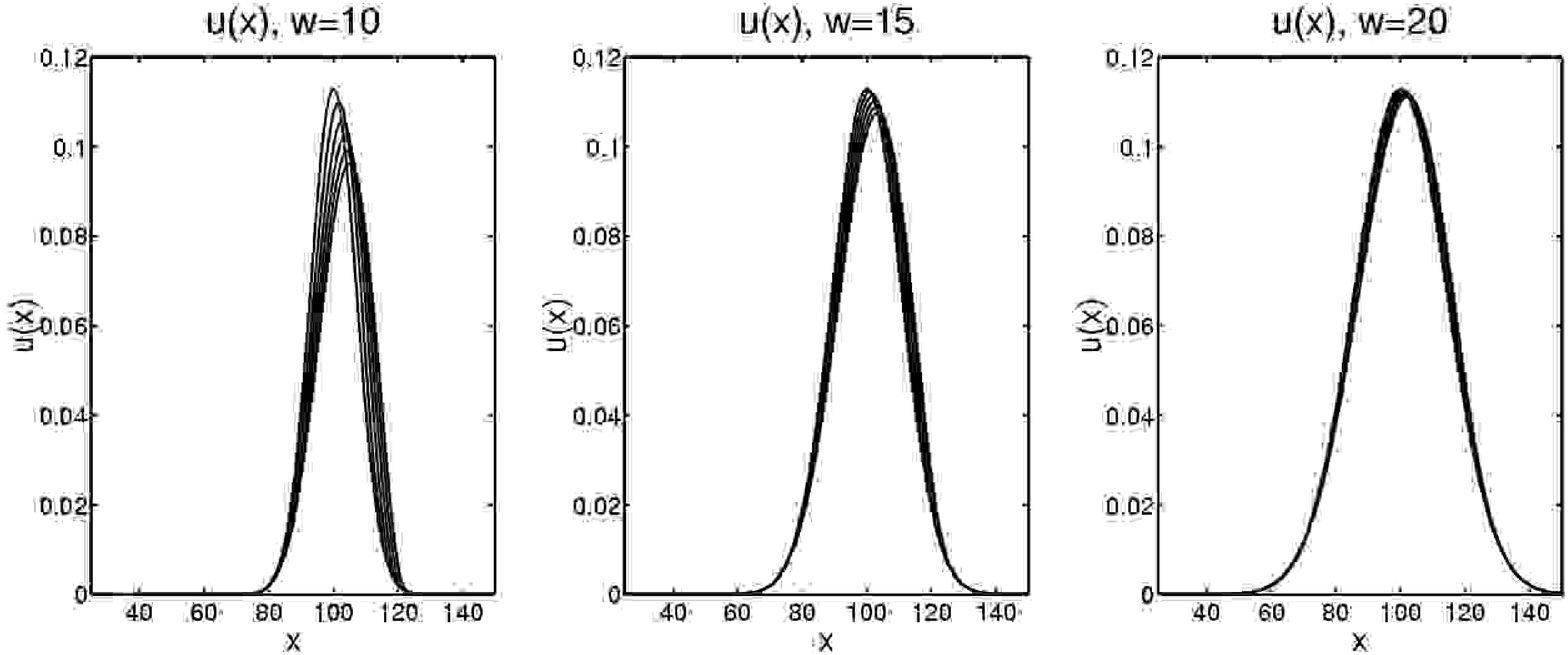, scale=0.3}}
    }
\caption{\label{Gauss-ic-pkn/a1,b-1/w10-20}
   {\bf Stationary solutions for $b=-1$.}
   Inviscid $b$-family,
   $b=-1$,
   $\alpha=1$,
   initial width $w=10,15,20$.
\rem{\sf
For
$b=-1$, evolving velocity profile under the peakon equation for
a set
of Gaussian initial conditions of increasing width $w=10,15,20$
for
$\alpha=1$.  Times are T=0, 500, 1000, 1500, 2000,
2500.
}
}
\end{center}
\end{figure}
}

Figure
\ref{Gauss-ic-pkn/a1,b-1/w10-20} shows the velocity profiles
under
evolution by the peakon equation, (\ref{b-family})
with
$g(x)=e^{-|x|/\alpha}$, for a set of Gaussian initial
conditions
of increasing width $w=10,15,20$ for $\alpha=1$ and
$b=-1$.
Evidently, the coexisting peakon solution for $b=-1$ does not
emerge
because $K\ne0$ and $H\ne0$ for this initial condition.
Instead, the
stable solution is essentially stationary with a slight 
rightward 
drift and
leaning slightly to the right. The reason for 
this 
lethargic propagation
becomes clear upon writing the b-equation 
solely in terms of the velocity
$u(x,t)$ 
as
\rem{
\begin{equation}\label{b-family-u}
u_t + 
(b+1)uu_x
=
\alpha^2(u_{xxt}+uu_{xxx}+bu_x u_{xx} 
)
\,.
\end{equation}
}
\begin{eqnarray}\label{b-family-u}
u_t + 
(b+1)uu_x
&=&
\alpha^2(u_{xxt}+uu_{xxx}+bu_x 
u_{xx})
\\
&=&
\alpha^2\partial_x
\Big(u_{xt} + u u_{xx} + 
\frac{b-1}{2} u_x^2 
\Big)
\nonumber\\
&=&
\alpha^2\partial_x^2
\Big(u_{t} + u u_{x} + 
\frac{b-3}{2} u_x^2 
\Big)
\,.
\nonumber
\end{eqnarray}

\begin{remark}[$b=-1$ is a 
turning point]
When $b=-1$ the nonlinear steepening term $(b+1)uu_x$ 
vanishes in
(\ref{b-family-u}) and the residual propagation is due to
its nonlinear ``curvature terms'' with higher order derivatives.
In the parameter regime $b>-1$ (resp. $b<-1$) the solutions of
equation (\ref{b-family}) or (\ref{b-family-u})  move rightward
(resp. leftward), provided the curvature terms on the right hand side
of equation (\ref{b-family-u}) are either negative, or sufficiently small.
\end{remark}

\begin{remark}[Short wave limit equation]
The high wavenumber, or short wave, limit of equation (\ref{b-family-u})
is
\begin{equation}
\partial_x^2
\Big(u_{t} + u u_{x} + \frac{b-3}{2} u_x^2 \Big) = 0
\,.
\end{equation}
For $b=2$ and $\lim_{x\to-\infty}u_x=0$, this integrates to become the
Hunter-Saxton equation \cite{HunterSaxton[1991]}. For $b=3$, it is the second
derivative of the Burgers equation.
\end{remark}

\subsubsection{Case $b=-2$ stationary solutions}

For $b=-2$, the traveling wave quadrature (\ref{tw-soln/bne1}) becomes an
elliptic integral
\begin{equation}\label{unused-14}
     \pm \,\frac{dz}{\alpha}
=
\frac{d(u-c)}{\Big[u^2 - \frac{2K}{3}(u-c)^{3} + 2H\Big]^{1/2}}
\,.
\end{equation}
The hyperbolic limit of this equation for $H=0$
vanishes at infinity for the stationary solution
($c=0$) to give
\begin{equation}\label{hyper-limit-b-2}
u(z)= \frac{3}{2K}{\rm sech}^2 \frac{z}{2\alpha}
\,.
\end{equation}

\rem{
\begin{eqnarray}
\begin{array}{l}
u(x,t)=-\kappa\Big(k^2(1-\tanh^2k(x-ct))+\frac{2}{3}(k^2-1/4)^2\Big),
\\[4pt]
\hbox{where}\quad c=-\frac{2k}{3}(k^4-1/16).
\end{array}
\end{eqnarray}
---------------------------------------------------------------------
\comment{

It will turn out that the transient approach to these stationary

solutions will \fbox{dominate} the numerics for $b\le\approx-1$.
}
---------------------------------------------------------------------
}

\remfigure{
\begin{figure}
\begin{center}
\leavevmode {
\hbox{\epsfig{figure=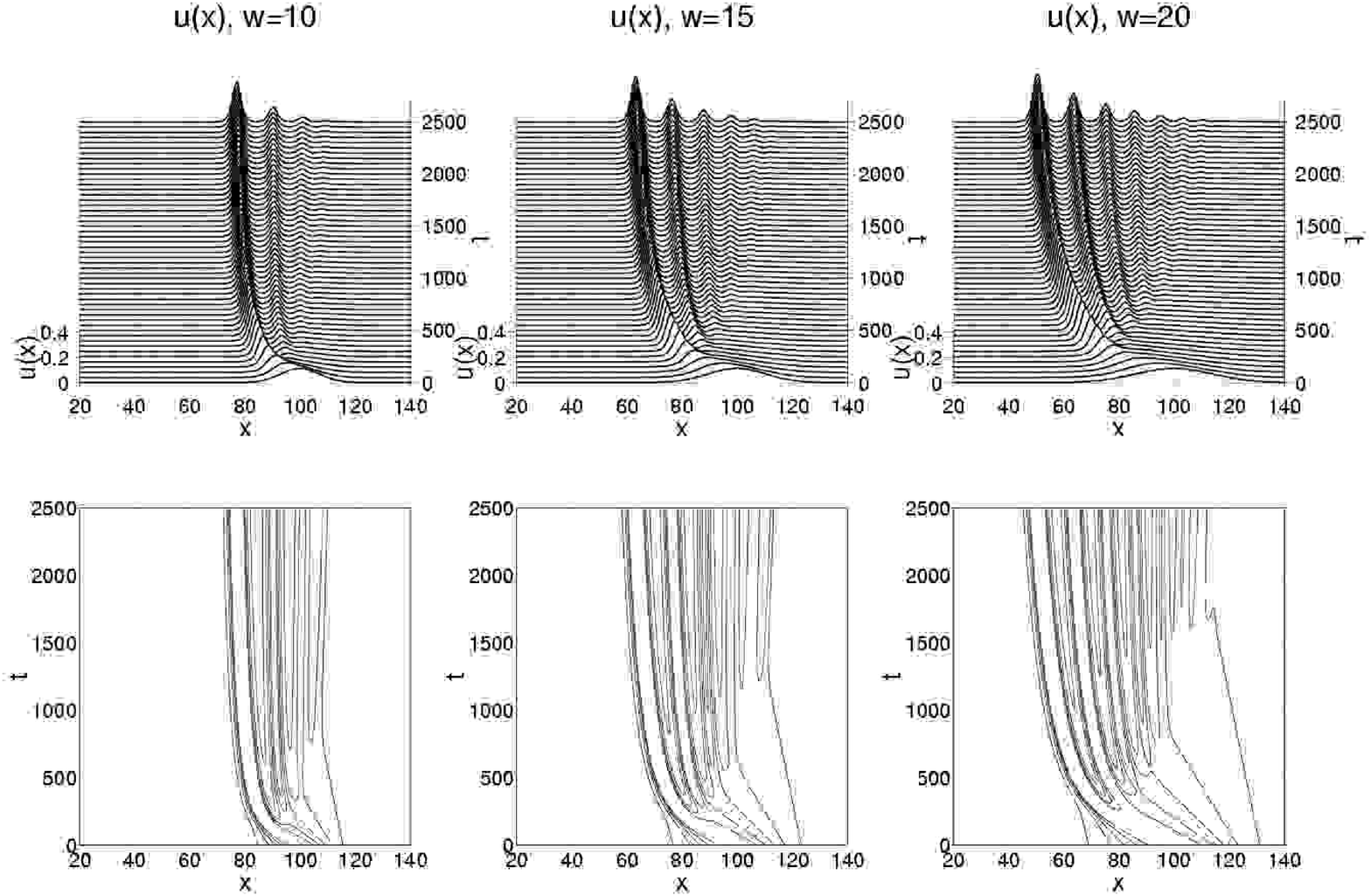, scale=0.3}}
    }
\caption{\label{Gauss-ic-pkn/a1,b-2/w10-20}
   {\bf Stationary solutions for $b=-2$.}
   Inviscid $b$-family,
   $b=-2$,
   $\alpha=1$,
   initial width $w=10,15,20$.
\rem{\sf
For
$b=-2$, evolving velocity profile under the peakon equation for
a set
of Gaussian initial conditions of increasing width $w=10,15,20$
for
$\alpha=1$.
}
}
\end{center}
\end{figure}
}

\subsubsection{Case $b=-3$ stationary solutions}

For $b=-3$, the hyperbolic limit $H=0$ of equation
(\ref{tw-soln/bne1})
is
\begin{equation}\label{tw-soln/bne1/H=0}
\pm \,\frac{dz}{\alpha}
=
\frac{du}{\Big[u^2 -
\frac{K}{2}(u-c)^{4}\Big]^{1/2}}
\,,
\end{equation}
which for $c=0$
is
\begin{equation}\label{tw-soln/b=-3/H=0/c=0}
    \pm
\,\frac{dz}{\alpha}
=
\frac{du}{u\Big[1 -
\frac{K}{2}u^2\Big]^{1/2}}
\,,
\end{equation}
and may be integrated
in closed form to obtain a continuous deformation
of the
peakon,
\begin{equation}\label{unused-15}
\frac{e^{- |z|/\alpha}}{1+\sqrt{1 -
K/2}}
=
\frac{u}{1+\sqrt{1 - Ku^{2}/2}}
\,,\
\hbox{for}\
b=-3\,, \ c=0 \hbox{ and }
H=0.
\end{equation}
Rearranging equation (\ref{unused-15}) and scaling $u$ by $u_0$ gives,
\begin{equation}
u(z) = \frac{u_0}{
\frac{A}{2}e^{|z|/\alpha} + (1-\frac{A}{2})e^{-|z|/\alpha}}\,,
\end{equation}
with $A=1+\sqrt{1-K/2}$, so that $A\in[1,2]$ for $K\in[0,2]$.  For $A=1$,
we have $u(z) = u_0\,{\rm sech}\,(z/\alpha)$.  And for $A=2$, we recover
the stationary peakon, $u(z) = u_0\,e^{-|z|/\alpha}$.

\remfigure{
\begin{figure}
\begin{center}
     \leavevmode {
        \hbox{\epsfig{figure=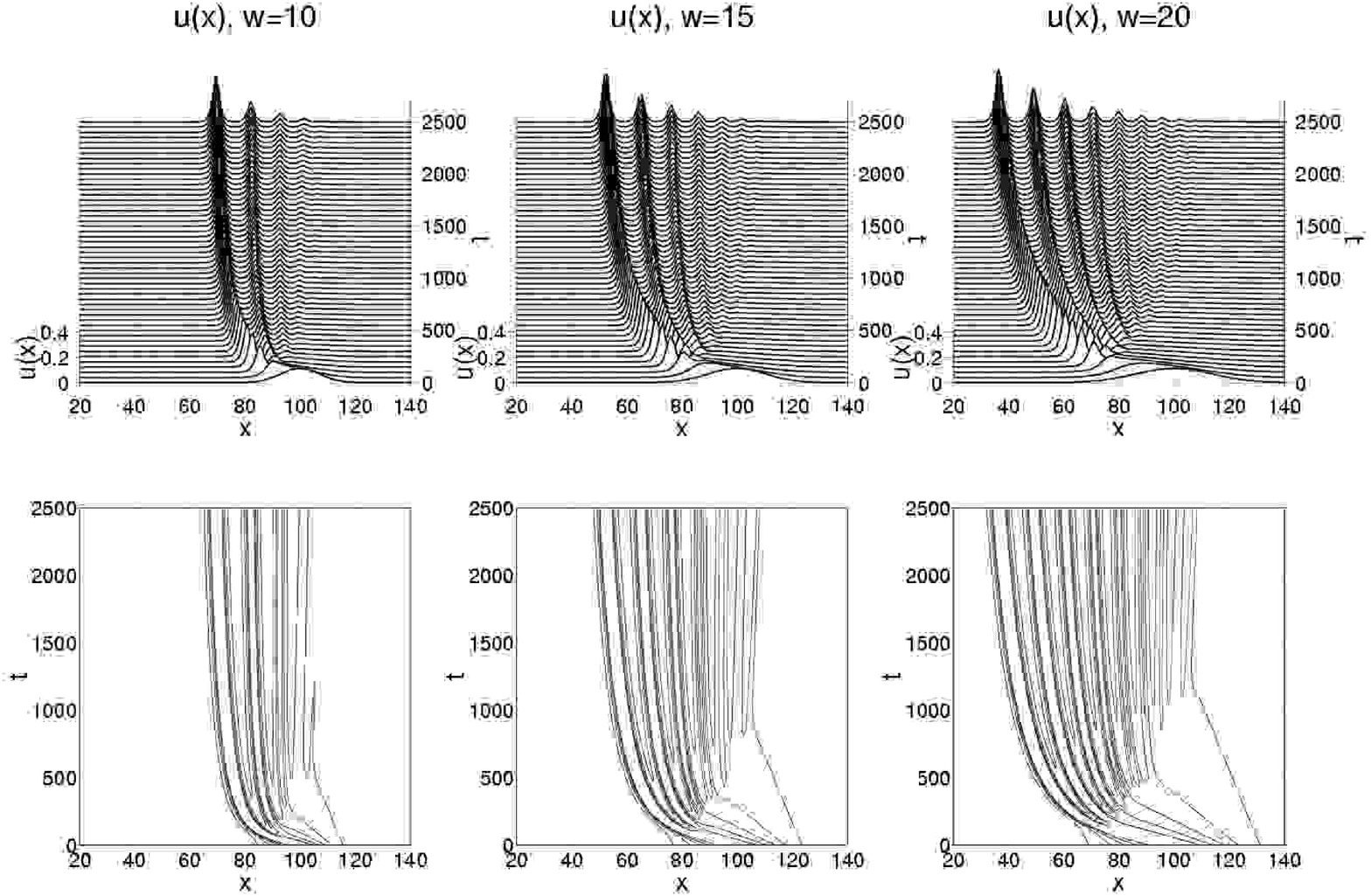, scale=0.3}}
     }
\caption{\label{Gauss-ic-pkn/a1,b-3/w10-20}
   {\bf Stationary solutions for $b=-3$.}
   Inviscid $b$-family,
   $b=-3$,
   $\alpha=1$,
   initial width $w=10,15,20$.
\rem{\sf
For $b=-3$, evolving velocity profile under the peakon equation for
a set of Gaussian initial conditions of increasing width $w=10,15,20$
for $\alpha=1$.
}
}
\end{center}
\end{figure}
}

\subsubsection{Case $b=-4$ stationary solutions}

For $b\le-4$ the analytical expressions for the cnoidal waves become
less elementary, because the integral in equation (\ref{tw-soln/bne1})
is then hyperelliptic.
However, our numerics show that
the dynamical behavior for $b=-4$ is similar to that of the cases $b=-2$ and
$b=-3$ shown in Figures
\ref{Gauss-ic-pkn/a1,b-2/w10-20}-\ref{Gauss-ic-pkn/a1,b-3/w10-20}. Namely,
a series of transient leftward propagating pulses, or {\it leftons},
of width alpha emerge and
tend to a nearly steady state.  Consistent with momentum (area)
conservation and
the tendency toward pulses of width alpha, the number of emerging leftons
increases with the width of the initial Gaussian.  At a longer time
scale, this train of pulses appears to tend toward stationary ($c\to0$).

\subsubsection{Numerical Results for $b=-2$ and $b=-3$}\label{numerical-b-2-3}

Figures \ref{Gauss-ic-pkn/a1,b-2/w10-20} and \ref{Gauss-ic-pkn/a1,b-3/w10-20}
show that a series of leftons in the velocity profile emerges
under the peakon equation for a set of Gaussian initial conditions of 
increasing
width $w=10,15,20$, for $\alpha=1$ and $b=-2,-3$.  Apparently these are not
peakons, because the velocity at which they move is not equal to their height.
The leftons emerge from the initial Gaussian in order of height
and then tend toward a nearly stationary state.  The number of emerging pulses
increases with the width of the initial Gaussian, as expected from momentum
(area) conservation and the tendency toward pulses of width alpha, and the
leftward speed of the emerging pulses increases with the magnitude of $b$.  The
latter is consistent with the coefficient $(b+1)$ of the nonlinearity in
equation (\ref{b-family-u}) as $b$ becomes more negative.

Figure \ref{Gauss-ic-pkn/a1,b-2/w10} shows the leftons at time $T=2500$,
versus $u(x) \simeq {\rm sech}^2(x/(2\alpha))$ for $b=-2$,
and versus $u(x) \simeq {\rm sech}(x/\alpha)$ for $b=-3$.  By this time,
the leftons have become stationary solutions with $c=0$ for both $b=-2$
and $b=-3$.

\remfigure{
\begin{figure}
\begin{center}
     \leavevmode {
        \hbox{\epsfig{figure=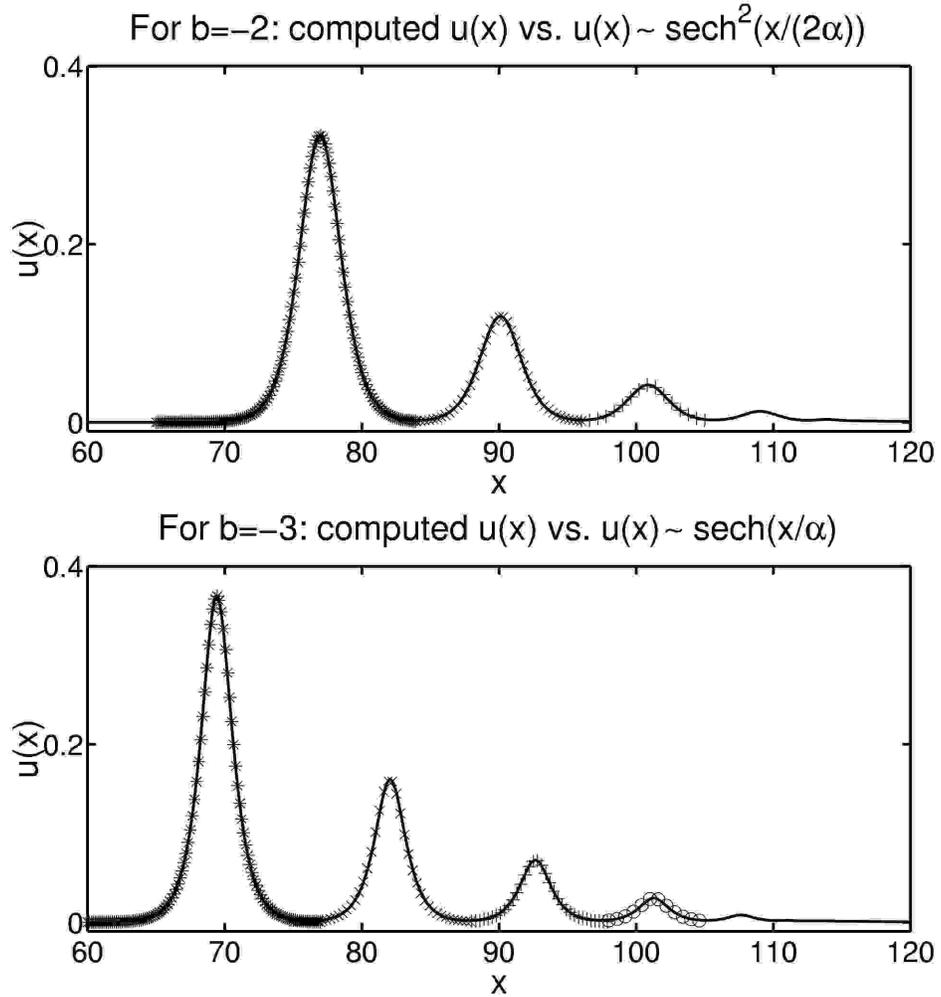, scale=0.5}}
     }
\caption{\label{Gauss-ic-pkn/a1,b-2/w10}
   {\bf Stationary solutions for $b=-2$
      versus $u(x)\simeq{\rm sech}^2(x/(2\alpha))$,
   and for $b=-3$
      versus $u(x)\simeq{\rm sech}(x/\alpha)$.}
   Inviscid $b$-family,
   $b=-2,-3$,
   $\alpha=1$,
   initial width $w=10$.
}
\end{center}
\end{figure}
}

\section{Pulson
interactions for
$b>0$}\label{PulsonInteract}

As
we have seen in section \ref{pulson-intro}, the b-family of
equations
(\ref{b-family}) admits the pulson traveling wave
solution
(\ref{pulson-tw-soln}) for
$b>0$.  The interaction dynamics
among $N$ of these pulsons is obtained by
superposing the traveling
wave solutions
$u(x,t)=cg(x-ct)$
as
\begin{equation}\label{pulson-soln}
u(x,t)=\sum_{i=1}^Np_i(t)g(x-q_i(t))
\quad\hbox{and}\quad
m(x,t)=\sum_{i=1}^Np_i(t)\delta(x-q_i(t))\,,
\end{equation}
for 
any $b>0$ and $u=g*m$, where the function $g$
is 
even so that 
$g\,'(0)=0$ and bounded and we may
set $g(0)=1$.  For 
these 
superpositions of pulsons to be exact solutions,
the time 
dependent
parameters $p_i(t)$ and $q_i(t)$ must satisfy the
following
$N-$dimensional particle dynamics equations obtained by
substituting
(\ref{pulson-soln}) into equation (\ref{b-family}),
\begin{eqnarray}
\label{ODEeqn-p}
\dot{p}_i &=& (1-b)\,p_i\sum_{j=1}^{N} p_j\,
g\,'(q_i-q_j)
=
(1-b)\,\frac{\partial G_N}{\partial
q_i}\,,
\\
\dot{q}_i &=& \sum_{j=1}^{N} p_j\,
g(q_i-q_j)
=
\frac{\partial G_N}{\partial
p_i}
\,.\label{ODEeqn-q}
\end{eqnarray}
Here the generating function $G_N$ is obtained by restricting the
norm $\|m\|_g^2$ in (\ref{KE-norm}) to the class of
superposed traveling
wave solutions (\ref{pulson-soln}), as
\begin{equation} \label{G_N-def}
G_N = \frac{1}{2}\sum_{i,j=1}^{N} p_i p_j\, g(q_i-q_j)
\,.
\end{equation}
Thus, the symmetric kernel $g(x)$ determines the shape of the
traveling wave solutions (\ref{pulson-soln}),
and these traveling waves interact nonlinearly via the
pulson dynamics of $p_i(t)$ and $q_i(t)$ with $i=1,\dots,N$ in equations
(\ref{ODEeqn-p}) and (\ref{ODEeqn-q}) for $b>0$. We shall see
that the character of these interactions depends vitally on the value of $b$.

\subsection{Pulson interactions for $b=2$}

When $b=2$, equations
(\ref{ODEeqn-p}) and (\ref{ODEeqn-q}) describe the canonical dynamics
of a Hamiltonian system with $N$ degrees of freedom.  These are the
geodesic pulson equations studied in Fringer and Holm \cite{FH[2001]}, in
which the following results are obtained,
\begin{description}
\item$\bullet$
Equation (\ref{b-family}) conserves the kinetic energy
$\frac{1}{2}\|m\|_g^2 =
\frac{1}{2}\int_{-\infty}^\infty m\,g*m\,dx$.
\item$\bullet$
Equations (\ref{ODEeqn-p})
and (\ref{ODEeqn-q}) describe canonical geodesic motion in an
$N-$dimensional configuration space whose co-metric is
$g^{ij}(q)=g(q_i-q_j)$.
\item$\bullet$
The generating function $G_N$ is the kinetic energy
Hamiltonian for the canonical geodesic motion.
\item$\bullet$
The solutions in (\ref{pulson-soln}) behave as particle-like pulses
whose pairwise interactions as determined by equations (\ref{ODEeqn-p})
and (\ref{ODEeqn-q}) comprise nonlinear, but elastic,
scattering events.
\item$\bullet$
The pairwise interactions for the pulsons can be solved analytically
for any symmetric function $g(x)$.
\end{description}

\begin{remark}
As we shall show, the last two statements also hold for any $b>1$.
\end{remark}

\remfigure{
\begin{figure}
\begin{center}
     \leavevmode {
\hbox{\epsfig{figure=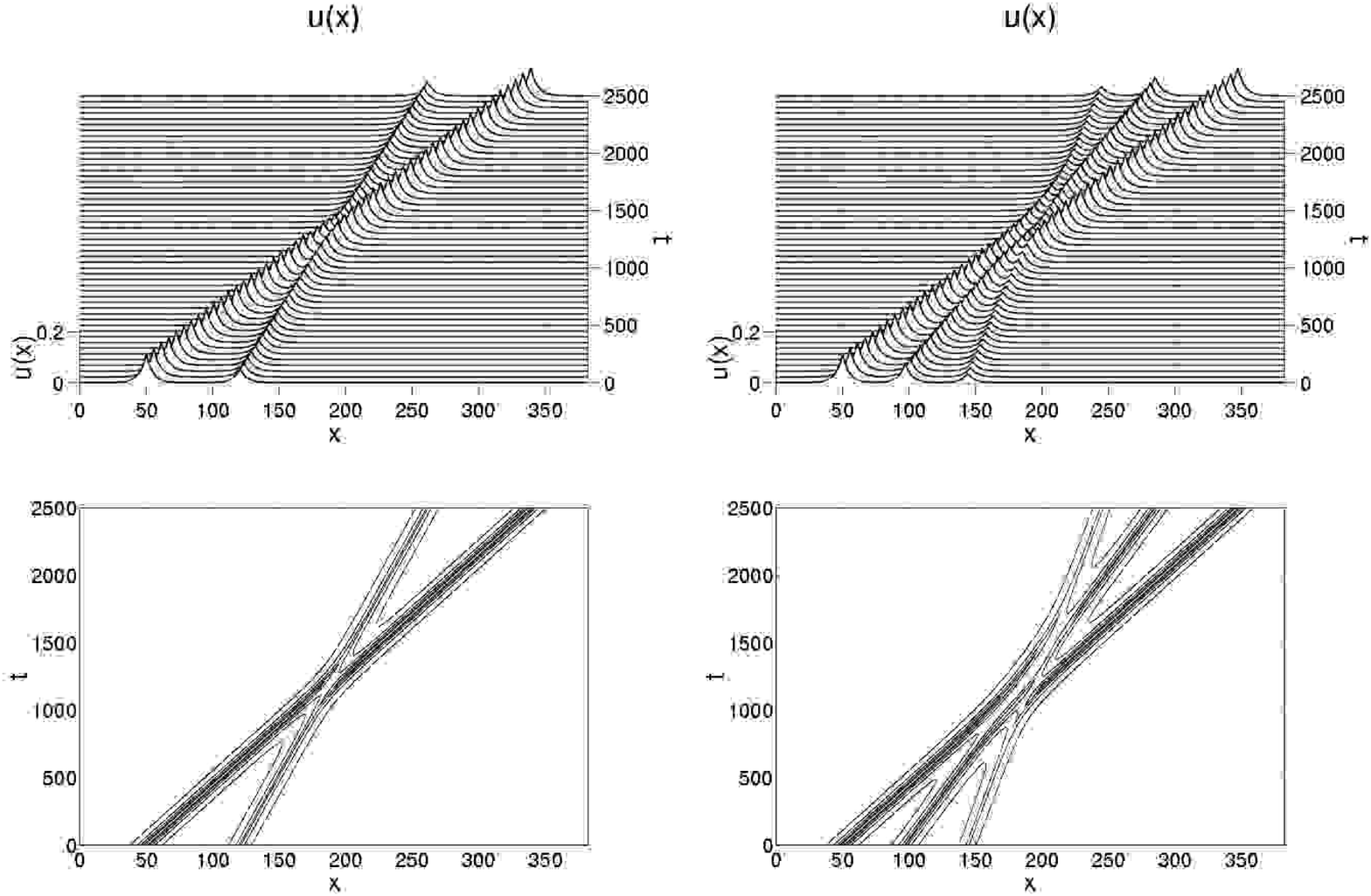, scale=0.3}}
     }
\caption{\label{2-3-pkn-ic/a1,b2}
   {\bf Peakons of width $\alpha$ for $b=2$: collisions.}
   Inviscid $b$-family,
   $b=2$,
   $\alpha=5$,
   initial width $w=5$.
\rem{\sf
2-peakon and 3-peakon interactions for $b=2$.  $\alpha=5$, and the peakons
have width $w=5$. The 3-peakon interaction decomposes into a series of 2-peakon
interactions.
}
}
\end{center}
\end{figure}
}

\rem{
\subsection{Pulson interactions for $b=3$}

\marginpar{\footnotesize\bf $\P\ $   \fbox{For D2H}}
\comment{
This section wasn't in your original paper, but it seems natural
to mention the $b=3$ case next to the $b=2$ case, and to say what
we can about it.
}

\marginpar{\footnotesize\bf $\P\ $   \fbox{For D2H}}
\comment{WRITE STUFF HERE about the Pulson interactions for $b=3$.}
}

\remfigure{
\begin{figure}
\begin{center}
     \leavevmode {
\hbox{\epsfig{figure=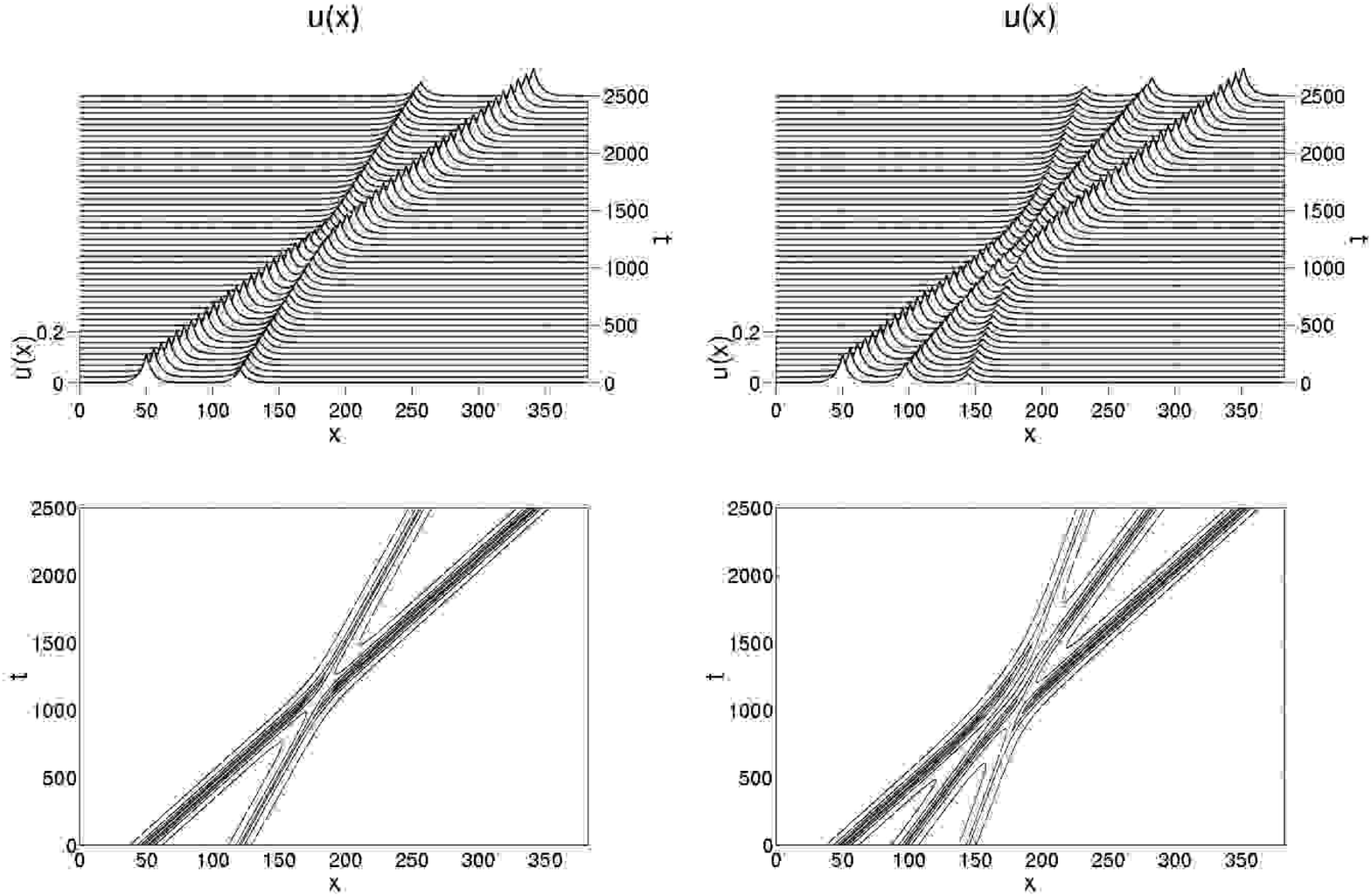, scale=0.3}}
     }
\caption{\label{2-3-pkn-ic/a1,b3}
   {\bf Peakons of width $\alpha$ for $b=3$: collisions.}
   Inviscid $b$-family,
   $b=3$,
   $\alpha=5$,
   initial width $w=5$.
\rem{\sf
2-peakon and 3-peakon interactions for $b=3$.  $\alpha=5$, and the peakons
have width $w=5$. Again, the 3-peakon interaction decomposes into a series of
2-peakon interactions.
}
}
\end{center}
\end{figure}
}

\subsection{Peakon interactions for $b=2$ and $b=3$: numerical results}
\begin{itemize}
\item
Figure \ref{2-3-pkn-ic/a1,b2} shows the evolution of the velocity profiles
in the 2-peakon and 3-peakon interactions for $b=2$, with
$g(x)=e^{-|x|/\alpha}$ and a periodic domain.  The 3-peakon interaction
decomposes into a series of 2-peakon interactions. These simulations verify the
analytical results for the 2-peakon interaction to three significant
figures over
propagation distances of about sixty peakon widths.
\item
Figure \ref{2-3-pkn-ic/a1,b3} shows the evolution of the 
velocity profiles
in the 2-peakon and 3-peakon interactions for 
$b=3$, with
$g(x)=e^{-|x|/\alpha}$ and a periodic 
domain.

\remfigure{
\begin{figure}
\begin{center}
\leavevmode {
       \hbox{\epsfig{figure=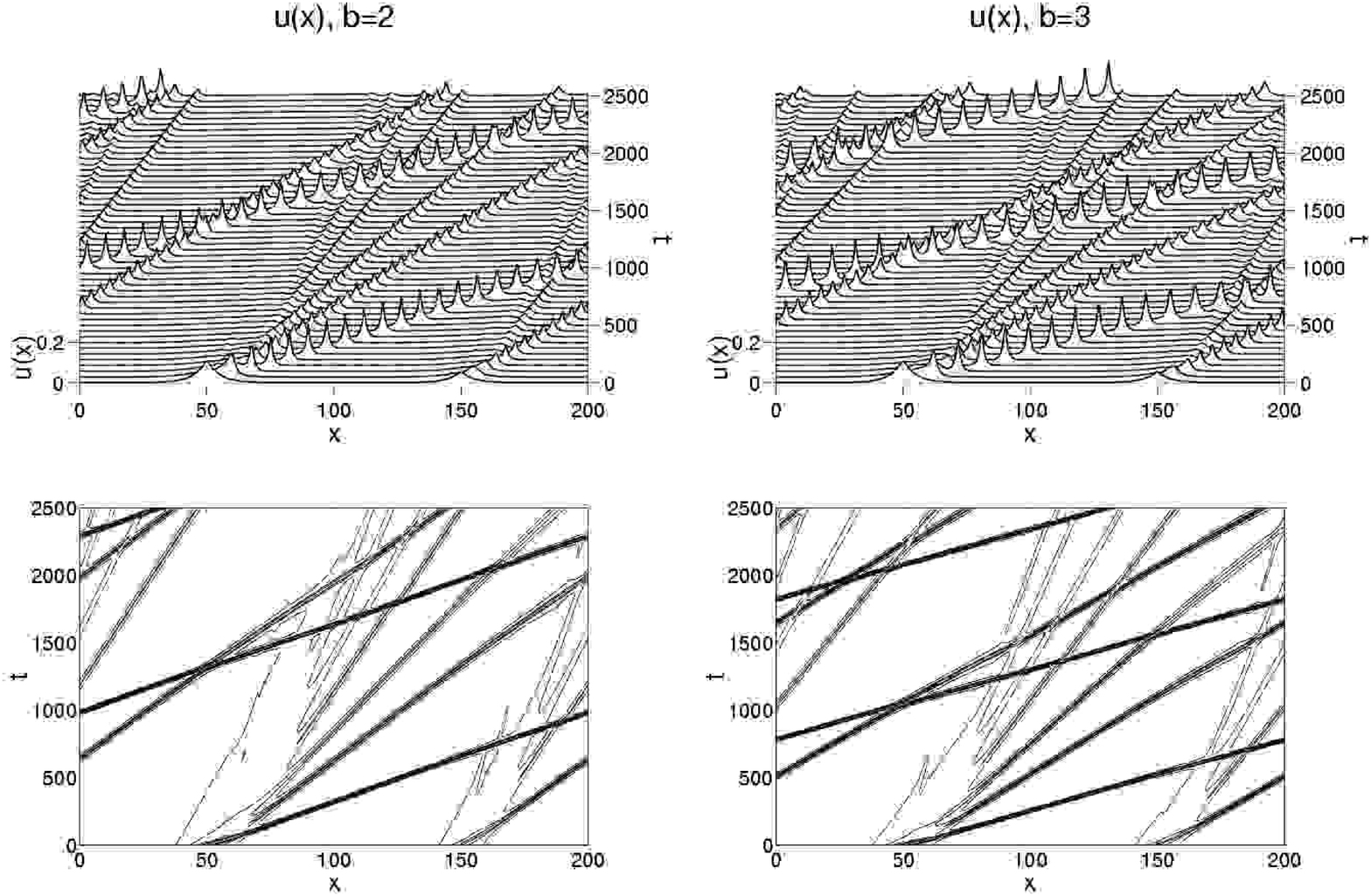, scale=0.3}}
    }
\caption{\label{2-pkn/a1,b2-3/w5}
   {\bf Peakons of width $>\alpha$ for $b=2,3$:
      emergence of width-$\alpha$ peakons.}
   Inviscid $b$-family,
   $b=2,3$,
   $\alpha=1$,
   initial width $w=5$.
\rem{\sf
Evolving 
patterns of the velocity profile under the peakon equation
for an 
initial pair of peakons of width $w=5$ for
$\alpha=1$ and 
$b=2,3$.
}
}
\end{center}
\end{figure}
}
\item
Figure 
\ref{2-pkn/a1,b2-3/w5} shows that peakons of initial width
greater 
than $\alpha$ break up into peakons of width $\alpha$ under
the 
evolution of the peakon equation in a periodic domain at fixed
values 
of $\alpha=1$ and $b=2,3$.  The emitted peakons are stable,
propagate 
as solitary traveling waves, and interact elastically.
Conversely, a 
peakon or other initial condition that is narrower than $\alpha$
will 
decompose into two oppositely moving trains of peakons 
and
antipeakons, each of width $\alpha$.
\end{itemize}

\subsection{Pulson-Pulson interactions for $b>0$ and symmetric $g$}

For $N=2$, the pulson dynamics in equations (\ref{ODEeqn-p}) and
(\ref{ODEeqn-q}) for $b>0$ reduces to
\begin{equation}\label{unused-c}
\frac{dp_1}{dt}=(1-b)\frac{\partial G}{\partial q_1}
\,, \quad
\frac{dp_2}{dt}=(1-b)\frac{\partial G}{\partial q_2},
\end{equation}
\begin{equation}
\frac{dq_1}{dt}=\frac{\partial G}{\partial p_1}, \quad
\frac{dq_2}{dt}=\frac{\partial G}{\partial p_2},
\end{equation}
and the generating function from (\ref{G_N-def}) is given by
\begin{equation}\label{gen-fct}
G=\frac{1}{2}(p_1^2+p_2^2)+p_1p_2\,g(q_1-q_2)
\,.
\end{equation}
(This is the Hamiltonian and the equations are canonical only for $b=2$,
which includes the Camassa-Holm case for which $g(q_1-q_2)=e^{-|q_1-q_2|}$
gives the peakon solutions.)

\subsubsection*{Conservation laws and reduction to quadrature}
Besides the total momentum
\begin{equation}\label{unused-e}
P=p_1+p_2
\,,
\end{equation}
the two-pulson system for $b>0$ and symmetric $g$ also conserves a
second quantity that is quadratic in  $p_1$ and $p_2$, namely
\begin{equation} \label{2ndconst}
H=p_1p_2\big(1-g(q_1-q_2)\big)^{b-1}
\,.
\end{equation}

For a Hamiltonian system with two degrees of freedom this second
conservation law would
be enough to ensure integrability, by Liouville's theorem. Even
in the present case of $b>0$ without a Hamiltonian structure, this
will be sufficient to reduce the 2-pulson system to 
quadratures.%
\footnote{When $b=1$, the momenta $p_1$ and $p_2$ 
are
separately conserved and the problem immediately reduces to 
quadratures
in $q=q_1-q_2$ and $Q=q_1+q_2$.}

Following the 
analysis for the case $b=2$ and arbitrary $g$ in
Fringer and Holm 
\cite{FH[2001]}, we introduce sum and difference
variables 
as
\begin{equation}\label{unused-f}
P=p_1+p_2
\,, \quad
Q=q_1+q_2
\,, 
\quad
p=p_1-p_2
\,, \quad
q=q_1-q_2
\,.
\end{equation}
In these 
variables, the generating function (\ref{gen-fct}) 
becomes
\begin{equation}\label{unused-g}
G=\frac{1}{4}P^2(1+g(q))+\frac{1}{4}p^2(1-g(q))
\,,
\end{equation}
and 
the second constant of motion (\ref{2ndconst}) 
becomes
\begin{equation} 
\label{2ndConstant}
H=\frac{1}{4}(P^2-p^2)\big(1-g(q)\big)^{b-1}
\,.
\end{equation}
Likewise, 
the 2-pulson equations of motion transform to
sum and difference 
variables as
\begin{eqnarray*}
\begin{array}{l}\label{unused-aaa}
 
\frac{dP}{dt} = 2(1-b)\frac{\partial G}{\partial Q}=0\,,\cr
 
\frac{dQ}{dt} =  2\frac{\partial G}{\partial P}=P(1+g(q))\,,\cr
 
\frac{dp}{dt} = 2(1-b)\frac{\partial G}{\partial q}
 
= \frac{1}{2}(1-b)(P^2-p^2)g\,'(q)\,,\cr
     \frac{dq}{dt} = 
2\frac{\partial G}{\partial p} = 
p(1-g(q)).
\end{array}
\end{eqnarray*}

Eliminating $p^2$ between the 
formula for $H$ and the equation of motion for
$q$ 
yields
\begin{equation}\label{unused-bbb}
\left(\frac{dq}{dt}\right)^2=P^2\big(1-g(q)\big)^2-4H\big(1-g(q)\big)^{3-b}
\,.
\end{equation}
We 
rearrange this into the following quadrature,
\begin{equation} 
\label{pulson-quad}
dt=\frac{dg(q)}{g\,'(q)\sqrt{Z}}, 
\qquad
Z=P^2\big(1-g(q)\big)^2-4H\big(1-g(q)\big)^{3-b}
\,.
\end{equation}
This 
simplifies to the quadratic $Z=P^2\big(1-g(q)\big)^2-4H$ when
$b=3$. 
For the peakon case, we have  $g(q)=e^q$ so that
$g\,'(q)=g(q)$ and 
the quadrature (\ref{pulson-quad}) simplifies to an
elementary 
integral for $b=-1,0,1,2,3$.  Having obtained $q(t)$ from
the 
quadrature, the momentum difference $p(t)$  is found 
from
(\ref{2ndConstant}) via the algebraic 
expression
\begin{equation}
p^2 = P^2 - 
\frac{4H}{\big(1-g(q)\big)^{b-1}}
\,,
\end{equation}
in terms of $q$ 
and the constants of motion $P$ and $H$.
Finally, the sum $Q(t)$ is 
found by  a further quadrature.
The remainder of the solution for 
arbitrary $b$ and $g$ closely
follows Fringer and
Holm 
\cite{FH[2001]} for the case $b=2$.

Upon writing the quantities $H$, 
$P$ and $G$ as
\begin{equation}\label{unused-ccc}
H=c_1c_2, \quad 
P=c_1+c_2,
\quad
G=\frac{1}{2}c_1^2+\frac{1}{2}c_2^2=\frac{1}{2} P^2 
- H
\,,
\end{equation}
in terms of the asymptotic speeds of the 
pulsons, $c_1$ and $c_2$, we find
the relative momentum 
relation,
\begin{equation}\label{p-square-eq}
p^2 = (c_1+c_2)^2 - 
\frac{4c_1c_2}{\big(1-g(q)\big)^{b-1}}
\,.
\end{equation}
This 
equation has several implications for the qualitative properties 
of
the 2-pulson collisions.

\begin{definition}
Overtaking, or 
rear-end, pulson collisions satisfy $c_1c_2>0$,
while head-on pulson 
collisions satisfy $c_1c_2<0$.
\end{definition}

The pulson order 
$q_1<q_2$ is preserved in an overtaking, or rear-end,
collision when 
$b>1$. This follows, as

\begin{proposition}[Preservation of pulson 
order]
For overtaking, or rear-end, collisions when $b>1$, the 
2-pulson dynamics
preserves the sign condition 
$q=q_1-q_2<0$.
\end{proposition}

\paragraph{Proof.} Suppose the 
peaks were to overlap in a
collision for $b>1$, thereby producing 
$q=0$ during a collision. The
condition $g(0)=1$ implies the second 
term in (\ref{p-square-eq}) diverges
for $b>1$ when the overlap 
occurs. However, this divergence would
contradict $p^2\ge0$.
\hfill 
$\Box$\bigskip

Consequently, seen as a collision between two 
initially well-separated
``particles'' with initial speeds $c_1$ and 
$c_2$, the separation $q(t)$
reaches a  nonzero distance of closest 
approach $q_{min}$ in an
overtaking, or rear-end, collision that may 
be expressed in terms of the
pulse shape 
as,

\begin{corollary}[Minimum separation 
distance]\label{cor-q-min}
The minimum separation distance reachable 
in two-pulson collisions with
$c_1c_2 >0$ is given 
by,
\begin{equation}\label{q-min}
\big(1-g(q_{min})\big)^{b-1}
=
\frac{4c_1c_2}{(c_1+c_2)^2}
\,.
\end{equation}
\end{corollary}

\paragraph{Proof.} 
Set $p^2=0$ in equation (\ref{p-square-eq}).
\hfill 
$\Box$

\begin{remark}
We shall use result (\ref{q-min}) later in 
checking the accuracy of our
numerical simulations of these 
two-pulson interactions.
\end{remark}

\begin{proposition}[Head-on 
collisions admit $q\to0$]
The 2-pulson dynamics allows the overlap 
$q\to0$ when $b>1$ in 
head-on
collisions.
\end{proposition}

\paragraph{Proof.} Because 
$p^2\ge0$, the overlap $q\to0$ implying $g\to1$
is only possible in 
equation (\ref{p-square-eq}) with $b>1$ for
$c_1c_2<0$. That is, for 
the head-on collisions.
\hfill $\Box$

\paragraph{Remarks about 
head-on collisions.}
For $b>1$, equation (\ref{p-square-eq}) implies 
that $p^2\to\infty$ 
diverges when
$q\to0$ in head-on collisions. The 
case $b=1$ is regular  and equation
(\ref{p-square-eq}) reduces to 
the constant relation $p^2=(c_1-c_2)^2$. For
$0<b<1$, the quantity 
$p^2$ no longer diverges  when $q\to0$
and the solution for the 
relative momentum in head-on collisions is 
again
regular.

\subsection{Pulson-antiPulson 
interactions for $b>1$ and symmetric 
$g$}

\paragraph{Head-on 
Pulson-antiPulson collision.}
We consider the special case of {\it 
completely antisymmetric}
pulson-antipulson collisions, for 
which
$p_1 = -p_2 = p/2$ and $q_1 = -q_2 = q/2$ (so that $P=0$ and 
$Q=0$).
In this case, the quadrature formula (\ref{pulson-quad}) 
reduces 
to%
\footnote{
For 
$b=3$, the quadrature formula (\ref{quadrat-2}) for the 
separation
distance in the pulson-antipulson collision reduces to 
straight line 
motion,
$q(t)-q(t_0)=\pm2c(t-t_0)$.}
\begin{equation} 
\label{quadrat-2}
\pm (t-t_0) = \frac{1}{\sqrt{-4 
H}}
\int_{q(t_0)}^{q(t)} 
\frac{dq\,'}{\big(1-g(q\,')\big)^{(3-b)/2}}\,,
\end{equation}
and the 
second constant of motion in (\ref{2ndConstant}) 
satisfies
\begin{equation} 
\label{2ndConstant-ppbar}
-4H=p^2\big(1-g(q)\big)^{b-1}
\,.
\end{equation}
After 
the collision, the pulson and antipulson separate
and travel 
oppositely apart; so that asymptotically in time
$g(q)\to0$, 
$p\to2c$, and
$H\to-{c}^2$,  where $c$ (or $-c$) is the asymptotic 
speed (and
amplitude) of the pulson
(or antipulson).  Setting 
$H=-{c}^2$ in equation
(\ref{2ndConstant-ppbar}) gives a
relation for 
the pulson-antipulson $(p,q)$ phase trajectories for any 
kernel,
\begin{equation} \label{p_vs_q_eqn}
p = \pm\,\frac{2 
c}{\big(1-g(q)\big)^{(b-1)/2}}\,.
\end{equation}
Notice that $p$ 
diverges for $b>1$ (and switches branches of the square
root) when 
$q\to 0^+$, because $g(0) = 1$.  In contrast, $p$ remains
constant 
for $b=1$ and it vanishes for $b<1$ (and again switches branches 
of
the square root) when $q\to 0^+$. Note that our convention for 
switching
branches of the square root allows us to keep $q>0$ 
throughout, so the
particles retain their order.

\paragraph{Remark 
about preservation of particle identity in
collisions.} The
relative 
separation distance $q(t)$ in pulson-antipulson collisions
is 
determined
by following a phase point along a level surface of the 
second
constant of motion $H$ in
the phase space with coordinates 
$(q,p)$. Because $H$ is quadratic, the
relative momentum $p$ has two 
branches on such a level surface, as
indicated by the
$\pm$ sign in 
equation (\ref{p_vs_q_eqn}). At the pulson-antipulson
collision 
point,
both $q\to0^+$ and either $1/p\to0^+$ or $p\to0^+$, so 
following a
phase point through
a collision requires that one choose 
a convention for which branch of
the level surface is
taken after the 
collision. Taking the convention that $p$ changes sign
(corresponding 
to a ``bounce''), but $q$ does not change sign (so the
``particles'' 
keep their identity) is convenient, because it 
allows
the phase 
points to be
followed more easily through multiple 
collisions. This 
choice is also
consistent with the pulson-pulson and 
antipulson-antipulson collisions. In
these other ``rear end'' 
collisions, as implied by equation
(\ref{p-square-eq}), the 
separation distance always
remains positive and again the particles 
retain their identity.

\begin{theorem}
[Pulson-antiPulson exact 
solution]
The exact analytical solution for the pulson-antipulson 
collision for any
$b$ and any symmetric $g$ may be written as a 
function of position $x$ and
the separation between the pulses $q$ 
for any pulse shape or kernel $g(x)$
as
\begin{equation}\label{vel_collide_eqn}
u(x,q) =
\frac{c}{\big(1-g(q)\big)^{(b-1)/2}}
\Big[g(x+q/2) -
g(x-q/2)\Big],
\end{equation}
where $c$ is the pulson speed at sufficiently large separation and the
dynamics of the separation $q(t)$ is given by the quadrature
(\ref{quadrat-2}) with $\sqrt{-4H}=2c$.
\end{theorem}
\paragraph{Proof.}
The solution (\ref{pulson-soln}) for the velocity $u(x,t)$ in the head-on
pulson-antipulson collision may be expressed in this notation as
\begin{equation}
u(x,t) = \frac{p}{2}g(x+q/2) - \frac{p}{2}g(x-q/2)\,.
\end{equation}
In using equation (\ref{p_vs_q_eqn}) to eliminate $p$ this
solution becomes equation (\ref{vel_collide_eqn}).
\hfill $\Box$\bigskip

\remfigure{
\begin{figure}
\begin{center}
     \leavevmode {
        \hbox{\epsfig{figure=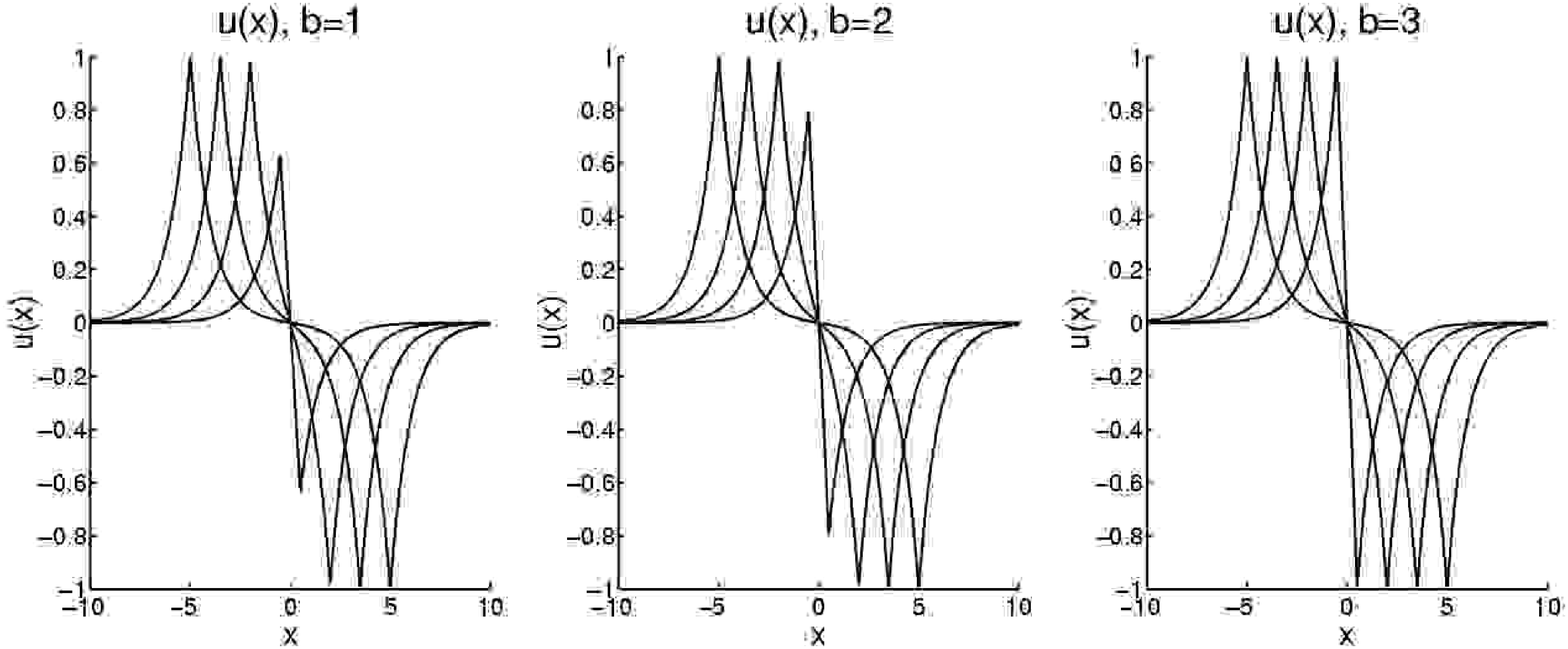, scale=0.3}}
     }
\caption{\label{pkn-anti-pkn}
   {\bf Peakon-antipeakon collisions for $b=1,2,3$.}
   Inviscid $b$-family,
   $b=1,2,3$,
   $\alpha=1$,
   initial width $w=1$.
\rem{\sf
Exact solutions for the Peakon-antiPeakon collision for $b=1$, $b=2$,
and $b=3$ at four successive times.
}
}
\end{center}
\end{figure}
}

Figure \ref{pkn-anti-pkn} shows the exact solutions for the
Peakon-antiPeakon collision in the cases $b=1$, $b=2$, and $b=3$.
The positive and negative peaks approach each other until the solution
develops a negative vertical slope in finite time. As the separation
$q\to0$, the positive and negative peaks ``bounce,'' thereby reversing
polarity, after which they separate in opposite directions.

\subsection{Specializing Pulsons to Peakons for $b=2$ and $b=3$}

We now restrict to $g(x)=e^{-|x|}$, the
Green's function for the 1D Helmholtz operator satisfying
\begin{equation}\label{Helm-op}
\Big(1-\frac{d^2}{dx^2}\Big)e^{-|x|}=2 
\delta(x)
\,.
\end{equation}
In this case, $m=u-\alpha^2u_{xx}$, the 
pulson traveling wave solution
is given by 
$u(x,t)=cg(x-ct)=ce^{-|x-ct|}$, has a discontinuity in
derivative at 
its peak, and is called the peakon.  For $b=2$ and $b=3$
in the 
peakon case the main results are,
\begin{itemize}
\item
For $b=2$ and 
$g(x)=e^{-|x|}$, equation (\ref{b-family}) becomes 
the
zero-dispersion limit of the integrable Camassa-Holm equation 
for
shallow water waves discovered in Camassa and Holm 
\cite{CH[1993]}. Upon
restoring its linear dispersion, this equation 
was recently proved to be
a higher-order accurate asymptotic 
description of shallow water waves in
Dullin et al. 
\cite{DGH[2001]}.
\item
For $b=3$ and  $g(x)=e^{-|x|}$, equation 
(\ref{b-family}) becomes the
integrable partial differential equation 
studied in Degasperis,
Holm and Hone \cite{DHH[2002]}.
\item
When $g(x)=e^{-|x|}$ the $N-$peakon dynamics for both $b=2$ and $b=3$ turns
out to be integrable -- as is the dynamics of the original PDE
(\ref{b-family}). The solutions of the initial value problem for
(\ref{b-family}) for both $b=2$ and $b=3$ may be found analytically by
using the Isospectral Scattering Transform (IST) method.
\item
The two cases $b=2$ and $b=3$ have quite different isospectral eigenvalue
problems. These are discussed in Camassa and Holm \cite{CH[1993]} and in
Dullin et al. \cite{DGH[2001]} for the case $b=2$, and
in Degasperis,  Holm and Hone \cite{DHH[2002]} for the case
$b=3$.  See also Beals, Sattinger and Smigialski \cite{BSS[2000]}
for a discussion of solving the inverse isospectral problem using classical
methods for the case $b=2$.
\end{itemize}

\section{Peakons of width $\alpha$ for arbitrary $b$}\label{Peakons}

When $g=e^{-|x|/\alpha}$, we may invert the velocity-momentum
relation $u=g*m$ by using the Green's function expression
(\ref{Helm-op}) with the Helmholtz operator to find  $m=u-\alpha^2
u_{xx}$.  Hence,  equation (\ref{b-family}) may be rearranged into the
local {\bf momentum conservation law},
\begin{equation}\label{b-family-peakon-momentum}
m_t
=-\, \frac{\partial}{\partial x}
\Big(
mu
+ \frac{b-1}{2} u^2
- \frac{b-1}{2} \alpha^2 u_x^2\Big)
\,.
\end{equation}
This conservation law for peakons may also be rewritten in {\bf convection
form}:
\begin{equation} \label{convect-b-eqn}
u_t + u u_x
=
- \,\tau_x
\quad\hbox{with}\quad
(1-\alpha^2\partial_x^2)\tau
=
\frac{b}{2}u^2+\frac{3-b}{2}\alpha^2u_x^2
\,.
\end{equation}
The two forms (\ref{b-family-peakon-momentum}) 
and
(\ref{convect-b-eqn}) of the b-family of equations 
(\ref{b-family})
suggest that values $b=0,1,3$ are special. These 
values of $b$ are
natural candidates for boundaries, or bifurcation 
points for changes
in solution
behavior.

Equation 
(\ref{convect-b-eqn}) describes peakons of
shape 
$g(x)=e^{-|x|/\alpha}$. This peakon equation
will form the basis of 
the rest of our 
study.

\subsection{Slope 
dynamics for Peakons: inflection points and the 
steepening lemma when 
$1<b\le3$}

We 
shall consider solution dynamics of equation (\ref{convect-b-eqn})
in 
the peakon case satisfying (\ref{peakon-soln}), or 
equivalently,
equation (\ref{b-family}) with $g(x)=e^{-|x|/\alpha}$, 
which 
satisfies
\begin{equation}\label{pkn-g}
(1-\alpha^2 
\partial_x^2)e^{-|x|/\alpha}=2 \alpha \delta 
(x)\,.
\end{equation}
For this case, and with vanishing boundary 
conditions at spatial
infinity, equations
(\ref{convect-b-eqn}) and 
(\ref{pkn-g}) imply the peakon equation on
the real 
line,
\begin{equation} 
\label{peakon-u-eqn}
u_t+uu_x
=
- \, \frac{1}{2 
\alpha}
\int_{-\infty}^\infty 
e^{-|x-y\,|/\alpha}
\Big(buu_y+(3-b)\alpha^2u_yu_{yy}\Big)
\,dy\,.
\end{equation}
Taking 
the $x-$derivative gives the equation for the slope 
$u_x(x,t)$
\begin{eqnarray}
u_{xt}+uu_{xx} 
+ u_x^2
&=&
- \, \frac{1}{2 \alpha}
\frac{\partial}{\partial x}
\int_{-\infty}^\infty 
e^{-|x-y\,|/\alpha}
\Big(buu_y+(3-b)\alpha^2u_yu_{yy}\Big)
\,dy
\nonumber\\
&&\hspace{-3cm}
=
\frac{1}{2 \alpha^2}
\int_{-\infty}^\infty e^{-|x-y\,|/\alpha}
\,{\rm 
sgn}\,(x-y)
\Big(buu_y+(3-b)\alpha^2u_yu_{yy}\Big)
\,dy
\,.\label{peakon-ux-eqn}
\end{eqnarray}
We shall use these expressions to prove the following.
\begin{proposition}[Peakon 
Steepening Lemma]\label{B-steepening}
For $b$ in the range $1< b\le3$ 
a sufficiently negative slope at an
inflection point of $u$ will 
become vertical in finite time under the
dynamics of the peakon 
equation 
(\ref{peakon-u-eqn}).
\end{proposition}

\paragraph{Proof.}
Following 
Camassa and Holm \cite{CH[1993]}, we shall consider
the evolution
of 
the slope $u_x$ at an inflection point $x=\bar{x}(t)$.
\rem{initially 
located to the right of the solution's maximum. }
Define the slope at 
the inflection point as $s(t)=u_x(\bar{x}(t),t)$ and
note that 
$u_{xx}(\bar{x}(t),t)=0$. Then equation (\ref{peakon-ux-eqn})
yields 
the following evolution equation for 
$s(t)$
\begin{eqnarray} 
\label{peakon-slope-eqn1}
\frac{ds}{dt} + s^2
=
    \frac{1}{2 
\alpha^2}
\int_{-\infty}^\infty
\!\!\!\!\!
e^{-|\,\bar{x}(t)-y\,|/\alpha}
\,{\rm 
sgn}\,(\bar{x}(t)-y)
\frac{1}{2}\frac{\partial}{\partial 
y}
\Big(bu^2+(3-b)\alpha^2u_y^2\Big)
\,dy
\end{eqnarray}
Integrating 
by parts using the definition $|y|=y\,{\rm sgn}\,(y)$, so 
that
$d|y|/dy=\,{\rm sgn}\,(y)+2y\delta(y)$, and recalling 
that
$y\delta(y)=0$, 
gives
\begin{eqnarray} 
\label{peakon-slope-eqn2}
\frac{ds}{dt}
=
-\,\Big(\frac{b-1}{2}\Big)s^2
+
\frac{b}{2\alpha^2}\,u^2
-
\frac{1}{2 
\alpha^2}
\int_{-\infty}^\infty
\!\!\!\!\!
e^{-|\,\bar{x}(t)-y\,|/\alpha}
\frac{1}{2\alpha}
\Big(bu^2+(3-b)\alpha^2u_y^2\Big)
\,dy
\nonumber
\end{eqnarray}
Hence,
in the range $0\le b\le3$ the last term is negative and we have 
the
slope 
inequality,
\begin{eqnarray} 
\label{peakon-slope-ineq}
\frac{ds}{dt}
\le
-\,\Big(\frac{b-1}{2}\Big)s^2
+
\frac{b}{2\alpha^2}\,u^2
\quad\hbox{for}\quad
0\le 
b\le3
\,.
\end{eqnarray}
We 
suppose the solution satisfies $(bu^2/\alpha^2)<M$ for
some constant 
$M$.%
\footnote{If 
this inequality is violated, we have another type of 
singularity.
However, for $b=2$, the constant $M$ can be estimated 
by 
using a
Sobolev inequality. In
fact, $M=4H_1(u)/\alpha^2$ because 
for 
this case we have
\[
\max_{x\in{\bf 
R}}[u^2(x,t)]
\le
\frac{1}{\alpha}\int_{-\infty}^\infty 
(u^2+\alpha^2u_x^2)\,dx
=
2H_1=const
\quad\hbox{for}\quad
b=2
\,.\]
}
Then,
\begin{eqnarray} 
\label{peakon-slope-ineq-M}
\frac{ds}{dt}
\le
-\,\Big(\frac{b-1}{2}\Big)s^2
+
\frac{M}{2}
\quad\hbox{for}\quad
0\le 
b\le3
\,.
\end{eqnarray}
Consequently, 
if 
$b>1$,
\begin{eqnarray} 
\label{peakon-slope-ineq-d}
\frac{dX}{1-X^2}
=
d\,{\rm 
coth}^{-1}(X)
\le
\sqrt{M}
\quad\hbox{for}\quad
X=\sqrt{\frac{b-1}{M}}\,s
\,.
\end{eqnarray}
This 
implies, for $s\le -\sqrt{M}$ initially negative, 
that
\begin{eqnarray} 
\label{peakon-slope-ineq-sol}
s
\le
\sqrt{M}\,{\rm coth}\Big(\sigma 
+ 
\sqrt{\frac{b-1}{M}}\frac{M}{2}\,t\Big)
\quad\hbox{for}\quad
1< 
b\le3
\,,
\end{eqnarray}
where 
the dimensionless integration constant $\sigma<0$ determines the 
initial
slope, which is negative. Under these circumstances, the 
slope at the
inflection
point must become vertical by time 
$t=-2\sigma/\sqrt{M(b-1)}$.

\hfill 
$\Box$

\paragraph{Remarks 
for $1< b\le3$.}
\begin{itemize}
\item
If the initial condition is 
antisymmetric, for $1< b\le3$, then the
inflection point at $u=0$ is 
fixed and $d\bar{x}/dt=0$, due to the
mirror reflection symmetry 
$(u,x)\rightarrow(-u,-x)$ admitted by equation
(\ref{peakon-u-eqn}). 
In this case, $M=0$ and equation
(\ref{peakon-slope-ineq-M}) 
implies
\begin{equation}\label{unused-eee}
ds /dt\le 
-\,\Big(\frac{b-1}{2}\Big)s^2
\quad\Rightarrow\quad
s(t)\le 
\frac{-2}{b-1}\Big(\frac{1}{t_0-t}\Big)\,,
\end{equation}
so 
verticality $s=-\infty$ will develop in finite time, regardless of 
how
small the initial slope $|s(0)|$, provided it is negative, 
$s(0)<0$, as in
figure \ref{pkn-anti-pkn}. If the initial slope is 
positive, then under this
evolution it will relax to zero from 
above.
\item
Consequently, traveling wave solutions of 
(\ref{peakon-u-eqn}) cannot
have the usual sech-like shape for solitons because inflection points with
sufficiently negative slope can produce unsteady changes in the shape of the
solution profile.
\end{itemize}

\subsection{Cases $0\le{b}\le1$}

In the range $0\le{b}\le1$, we have from (\ref{|M|cons}) that
\begin{equation}
\int_{-\infty}^\infty |m|^{1/b}\,dx
=
\int_{-\infty}^\infty |m_0|^{1/b}\,dx
\,, \quad\hbox{where}\quad
m_0(x)=m(x,0)
\,. \label{|M|cons1}
\end{equation}
This conservation law implies an elliptic regularity estimate
showing that the slope $s=u_x$ is always bounded under the
dynamics of the peakon equation (\ref{peakon-u-eqn}).
See Holm and Titi [2002] for a proof of this result and more discussion of
its implications.

\section{Adding viscosity to peakon dynamics}\label{Viscosity}

In the remainder of this paper, we shall restrict our attention to the
peakon case $g(x)=e^{-|x|/\alpha}$ with length scale $\alpha$, and
investigate the fate of the peakon solutions when
viscosity is introduced for given values of $b$ and $\alpha$. For purposes
of comparison with previous results in the literature, we shall also extend
equation (\ref{b-family}) to a new family of equations that includes the
Burgers equation by introducing two additional real parameters. These are the
viscosity $\nu$ and a multiplier $\beta$ for the stress, or pressure
gradient.

First, we shall introduce constant viscosity $\nu>0$ into
(\ref{b-family}) to form the viscous b-family of equations for the peakon case
$g(x)=e^{-|x|/\alpha}$, as follows,
\begin{equation}\label{viscous-b-family-1}
m_t\
+\
\underbrace{\ \ um_x\ \
}
_{\hspace{-2mm}\hbox{convection}\hspace{-2mm}}\
+\
\underbrace{\ \ b\,u_xm\ \
}
_{\hspace{-2mm}\hbox{stretching}\hspace{-2mm}}\
=\
\underbrace{\ \
\nu\,m_{xx}\
}
_{\hspace{-2mm}\hbox{viscosity}
}\,, \quad\hbox{with}\quad
m=u-\alpha^2 u_{xx}
\,.
\end{equation}
As in equation (\ref{b-family-u}), this equation with viscosity may be
expressed solely in terms of the velocity $u(x,t)$ as
\begin{eqnarray}\label{b-family-u-visc}
u_t + (b+1)uu_x - \nu u_{xx}
&=&
\alpha^2(u_{xxt}+uu_{xxx}+bu_x u_{xx}-\nu u_{xxxx} )
\\
&=&
\alpha^2\partial_x
\Big(u_{xt} + u u_{xx} - \nu u_{xxx} + \frac{b-1}{2} u_x^2 \Big)
\nonumber\\
&=&
\alpha^2\partial_x^2
\Big(u_{t} + u u_{x} - \nu u_{xx} + \frac{b-3}{2} u_x^2 \Big)
\,.
\nonumber
\end{eqnarray}
Thus, the nonlinear steepening term increases with $b$ as $(b+1)uu_x$.
When $\alpha\to0$ the previous equation reduces to
\begin{equation}\label{unused-fff}
u_t + (b+1)uu_x - \nu u_{xx} = 0
\,,
\end{equation}
and one then recovers the usual {\bf Burgers equation} either by rescaling
dimensions, or by setting $b=0$. For $b=2$, equation
(\ref{viscous-b-family-1}) is the one-dimensional version of the
three-dimensional Navier-Stokes-alpha model for turbulence
\cite{Chen-etal[1998]}.

The viscous b-family of peakon equations (\ref{viscous-b-family-1}) may
be rearranged into two other equivalent forms that are convenient for
making the second extension of a stress multiplier. These are either its
equivalent {\bf conservative form},
\begin{equation}\label{conserv-viscous-b-family}
m_t
=-\, \frac{\partial}{\partial x}
\Big(
mu
+ \frac{b-1}{2} u^2
- \frac{b-1}{2} \alpha^2 u_x^2\Big)
+
\nu \,m_{xx}
\,,
\end{equation}
or its equivalent {\bf convective form},
\begin{equation} \label{convect-viscous-b-family}
(1-\alpha^2\partial_x^2)\big(u_t + u u_x - \nu u_{xx}\big)
=
-\, \partial_x
\Big(\frac{b}{2}u^2+\frac{3-b}{2}\alpha^2u_x^2\Big)
\,.
\end{equation}
\paragraph{Stress
multiplier $\beta$.}
Next, we shall introduce a stress multiplier
$\beta$ as a second
parameter that for $\beta\ne1$ deforms the
convective form of the viscous
b-family of equations
(\ref{convect-viscous-b-family}) into the following
family of
Burgers-like equations with four parameters $b$, $\alpha$, $\nu$
and
$\beta$,
\begin{equation}
\label{convect-viscous-Bab-eqn}
u_t + u u_x - \nu u_{xx}
=
-
\,\beta\,\tau_x
\quad\hbox{with}\quad
(1-\alpha^2\partial_x^2)\tau
=
\frac{b}{2}u^2+\frac{3-b}{2}\alpha^2u_x^2
\,.
\end{equation}
When
$\beta=0$, the Burgers$-\alpha\beta$
equation
(\ref{convect-viscous-Bab-eqn})
recovers the usual Burgers
equation. When $\beta=1$, equation
(\ref{convect-viscous-Bab-eqn})
recovers the viscous b-family of peakon
equations
(\ref{viscous-b-family-1}).

We shall seek solutions of the
Burgers$-\alpha\beta$ equation
(\ref{convect-viscous-Bab-eqn}),
either on the real line and
vanishing at spatial
infinity, or in a
periodic domain, for various values of its four parameters
$b$,
$\alpha$, $\nu$ and $\beta$.  Under these boundary conditions,
when
$\beta\to1$, equation (\ref{convect-viscous-Bab-eqn}) recovers 
the 
convective
form (\ref{convect-viscous-b-family}) of the viscous 
b-family for peakons with
$g(x)=e^{-|x|/\alpha}$.
Thus, the viscous 
b-family of 
equations
(\ref{viscous-b-family-1}-\ref{convect-viscous-b-family}) 
deforms into the
Burgers$-\alpha\beta$ equation 
(\ref{convect-viscous-Bab-eqn}) when
$\beta\ne1$
and the 
Burgers$-\alpha\beta$ equation
(\ref{convect-viscous-Bab-eqn}) 
reduces to
the usual Burgers equation when $\beta=0$. We shall be
interested in
the effects
of the four parameters $b$, $\alpha$, $\nu$
and $\beta$ on the solutions of the
Burgers$-\alpha\beta$ equation
(\ref{convect-viscous-Bab-eqn}). We shall be
interested especially in
the fate of the peakon solutions upon introducing the
parameters
$\nu$ and $\beta$ so as to retain $H_\alpha^1$ control of
the
velocity. As we shall see, such control requires a special
relation between the
parameters $b$ and
$\beta$, namely,
$(3-b)\beta=1$.

\subsection{Burgers$-\alpha\beta$ equation: analytical estimates}

\begin{proposition}[$H_\alpha^1$
control of the velocity]\label{H1Control-Prop}
The
Burgers$-\alpha\beta$ equation (\ref{convect-viscous-Bab-eqn})
controls
the $\alpha-$weighted $H^1$ norm of the
velocity,
\[
\|u\|_{H^1_\alpha}^2 =
\int_{-\infty}^{\infty}\!\!
\big(u^2+\alpha^2
u_x^2\big) \, dx
\]
for $\alpha^2\ne0$,
provided
$(3-b)\beta=1$.
\end{proposition}

\paragraph{Proof.}
The
spatial derivative of the Burgers$-\alpha\beta$
equation
(\ref{convect-viscous-Bab-eqn}) yields the dynamics for the
slope $s=u_x$
as
\begin{eqnarray*}
u_t+uu_x
- \nu u_{xx}
&=&
-\, \beta \tau_x
\,,\\
s_t+us_x+s^2 - \nu
s_{xx}
&=&
    -\,\beta\,
\tau_{xx}
\,,\quad\hbox{with}\quad
s=u_x
\,,\\
-\alpha^2
\tau_{xx}
&=&
\frac{b}{2}\,u^2
+\frac{3-b}{2}\alpha^2\,u_x^2 -
\tau
\,.
\end{eqnarray*}
In turn, these slope dynamics equations
imply the following  evolution of the
$\alpha-$weighted $H^1$
density, cf.~equation
(\ref{peakon-slope-eqn1}),
\begin{eqnarray*}
\frac{\partial}{\partial t}
\Big(\frac{1}{2}u^2+ \frac{\alpha^2}{2}
s^2\Big)
&=&
\frac{\partial}{\partial
x}
\bigg(\frac{1}{3}
\Big(1-\frac{b \beta}{2}\Big)u^3
+ \beta u
\tau
+ \frac{\alpha^2}{2}u s^2
+ \nu uu_x
+ \nu \alpha^2 s
s_x
\bigg)
\\&&
-\
\nu u_x^2\
-\, \nu \alpha^2
s_x^2\
+\,\frac{\alpha^2}{2}
\Big( (3-b) \beta-1\Big) s^3
\,.
\end{eqnarray*}
Thus, provided
     \[(3-b) \beta=1\,,\]
the last term will vanish. Under this condition,
for periodic or vanishing boundary conditions the
$\alpha-$weighted $H^1$ norm
\[
\|u\|_{H^1_\alpha}^2 =
\int_{-\infty}^{\infty}\!\!
\big(u^2+\alpha^2 u_x^2\big) \, dx
\]
will decay monotonically under the Burgers$-\alpha\beta$ dynamics for
$\alpha^2\ne0$.

\hfill $\Box$

\paragraph{Remarks.}
\begin{itemize}
\item
When $\nu\to0$ in the Burgers$-\alpha\beta$ equation, the
$\alpha-$weighted $H^1$ norm is conserved for $(3-b)\beta=1$.
This relation cannot be satisfied for $b=3$. Thus, the proof of decay of the
$\alpha-$weighted $H^1$ norm under the Burgers$-\alpha\beta$ dynamics is
inconclusive for $\nu\ne0$ when $b=3$. However, one can expect on
physical grounds that this norm will also decay for $b=3$ if $\nu$ is
sufficiently large.

\item
We shall restrict our remaining considerations to those values of $b$ and
$\beta$  for which the $\alpha-$weighted $H^1$ norm is bounded, or decays
monotonically.  In one dimension, this control of the $\alpha-$weighted
$H^1$ norm  implies the solution for the velocity will be {\it continuous}.
\item
Namely, we shall consider the following cases with $(3-b)\beta=1$
         \begin{description}
         \item
         (b=0, $\beta=1/3$),
         (b=1, $\beta=1/2$) and
         (b=2, $\beta=1$).
         \end{description}
\end{itemize}

\begin{proposition}[Burgers$-\alpha\beta$
Steepening
Lemma]\label{B-alphabeta-steepening}
For $b$ and $\beta$
in the range
$(3-b)\beta\le2$ a sufficiently negative slope at
an
inflection point of velocity  $u$ will become vertical in finite
time under the
dynamics of the Burgers$-\alpha\beta$ equation
(\ref{convect-viscous-Bab-eqn})
with
$\nu=0$.
\end{proposition}

\paragraph{Proof.}The
proof follows that for the Peakon Steepening
Lemma \ref{B-steepening}
and uses the slope equation following from
Burgers$-\alpha\beta$
equation (\ref{convect-viscous-Bab-eqn}) with
$\nu=0$
that
corresponds to  (\ref{peakon-ux-eqn}) for the Peakons, modified
to include
$\beta$,
\begin{eqnarray}%
u_{xt}+uu_{xx} + u_x^2
&=&
- \,
\frac{\beta}{2 \alpha}
\frac{\partial}{\partial
x}
\int_{-\infty}^\infty
e^{-|x-y\,|/\alpha}
\Big(buu_y+(3-b)\alpha^2u_yu_{yy}\Big)
\,dy
\nonumber\\
&&\hspace{-3cm}
=
\frac{\beta}{2 \alpha^2}
\int_{-\infty}^\infty
e^{-|x-y\,|/\alpha}
\,{\rm
sgn}\,(x-y)
\Big(buu_y+(3-b)\alpha^2u_yu_{yy}\Big)
\,dy
\,.\label{B-alphabeta-ux-eqn}
\end{eqnarray}
Equation
(\ref{B-alphabeta-ux-eqn}) yields the inviscid
Burgers$-\alpha\beta$
evolution of the slope $s(t)=u_x(\bar{x}(t),t)$
at an inflection point
$x=\bar{x}(t)$
as
\begin{eqnarray}
\label{peakon-slope-ineq-M-2}
\frac{ds}{dt}
\le
-\Big(2-(3-b)\beta\Big)\frac{s^2}{2}
+
\frac{\beta
M}{2}
\quad\hbox{for}\quad
0\le
b\le3
\,,
\end{eqnarray}
This
holds provided we assume the solution satisfies $(bu^2/\alpha^2)<M$
for
some constant $M$. Consequently, if $2-(3-b)\beta>0$, we
have
\begin{eqnarray}
\label{peakon-slope-ineq-d-2}
\frac{dX}{1-X^2}
=
d\,{\rm
coth}^{-1}(X)
\le
\sqrt{M}
\quad\hbox{for}\quad
X=\sqrt{\frac{2-(3-b)\beta}{M\beta}}\,s
\,.
\end{eqnarray}
For
$s\le -\sqrt{M}$ initially negative and $\beta>0$, this
implies,
\begin{eqnarray}
\label{peakon-slope-ineq-sol-2}
s
\le
\sqrt{M}\,{\rm
coth}\Big(\sigma
+
\sqrt{\frac{2-(3-b)\beta}{M\beta}}\,\frac{M}{2}\,t\Big)
\quad\hbox{for}\quad
0\le
3-\frac{2}{\beta}<
b\le3
\,,
\end{eqnarray}
where
the dimensionless integration constant $\sigma<0$ determines
the
initial slope, which is negative. Under these circumstances,
provided
the inflection point continues to exist, its negative slope
must
become vertical by
time
$t=\frac{-2\sigma}{M}\sqrt{\frac{M\beta}{2-(3-b)\beta}}$.

\hfill
$\Box$

\begin{corollary}[Inviscid
Burgers$-\alpha\beta$ shocks]
Solutions of the inviscid
Burgers$-\alpha\beta$ equation
(\ref{convect-viscous-Bab-eqn}) with
$\nu=0$ that remain continuous in
velocity must develop negative
vertical slope in finite
time.
\end{corollary}

\paragraph{Proof.}
According
to Proposition \ref{H1Control-Prop}, continuity of the velocity
and,
hence,  control of the $H^1$ norm $\|u\|_{H^1_\alpha}$ requires
that
$(3-b)\beta=1$. This is in the parameter range where
Proposition
\ref{B-alphabeta-steepening} applies. Consequently,
verticality will form
at an inflection  point of negative slope under
the dynamics of the
inviscid Burgers$-\alpha\beta$ equation
(\ref{convect-viscous-Bab-eqn})
with $\nu=0$ for
$(3-b)\beta\le2$.
\hfill $\Box$

\begin{remark}
Hence, to remain
continuous without viscosity, the solution of the
inviscid
Burgers$-\alpha\beta$ equation must either develop verticality at
an
inflection point of negative slope, or it must evolve to eliminate
such
points
entirely.
\end{remark}

\subsection{Burgers$-\alpha\beta$ traveling waves for $\beta(3-b)=1$
\& $\nu=0$}

For  $\nu=0$, the Burgers$-\alpha\beta$ equation
(\ref{convect-viscous-Bab-eqn}) has traveling
waves given by
\begin{equation}\label{Bab-tw}
     (u-c)u\,'+ \beta \tau\,'=0
\quad\hbox{and}\quad
\tau-\alpha^2\tau\,''
=
\frac{b}{2}u^2+\frac{3-b}{2}\alpha^2(u\,')^2
\end{equation}
which yields
after one integration
\begin{equation}\label{1stConstant-bab}
\frac{u^2}{2}-c\,u + \beta \tau=K\,,
\end{equation}
where $K$ is the first integral.  Consequently, we find
\begin{equation}\label{tau-K}
\tau-\alpha^2\tau\,''
=
\frac{1}{\beta}
\Big(K+c\,u-\frac{u^2}{2}
+
\alpha^2\big((u-c)\,u\,''+(u\,')^2\big)
\Big)
\,.
\end{equation}
The second equation in (\ref{Bab-tw}) integrates
for the special case of $\beta(3-b)=1$,
\begin{equation}\label{2ndConstant-bab}
     2Ku+cu^2-\beta u^3 + \alpha^2(u-c)(u\,')^2=2H
\,.
\end{equation}
For the special case $K=0=H$ this becomes
\begin{equation}
\alpha^2(u-c)(u\,')^2
=
(\beta u-c)u^2
\quad\hbox{for}\quad
\beta(3-b)=1
\,,
\end{equation}
and we recover the peakon solution $u(z)=ce^{-|z|/\alpha}$ for $\beta=1$. In
the general case that $K\ne0$ and $H\ne0$, we rearrange equation
(\ref{2ndConstant-bab}) into the following quadrature for
inviscid Burgers$-\alpha\beta$ traveling waves,
\begin{equation}\label{tw-soln/Bab}
     \pm
\,\frac{dz}{\alpha} =
     \frac{(u-c)^{1/2}\,du}{\Big[2H - 2Ku - cu^2 + \beta u^3\Big]^{1/2}}
     \quad\hbox{for}\quad
\beta(3-b)=1\
\&\
\nu=0
     \,.
\end{equation}
In what follows, we shall consider the cases ($b=0$, $\beta=1/3$),
($b=1$, $\beta=1/2$) and ($b=2$, $\beta=1$) when $\nu\ne0$.

\section{The fate of the peakons under (1) adding viscosity and (2)
Burgers$-\alpha\beta$
evolution}\label{PeakonsViscousFate}

\subsection{The fate of peakons under adding viscosity}

The
following set of four figures shows the effects on the initial
value
problem for the viscous b-equation (\ref{viscous-b-family-1})
of
varying $\alpha$ and $b$ at fixed viscosity for an initial
velocity
distribution given by a peakon of width $w=5$ and initial
height
$U\simeq0.1$. The parameter $b$ takes the values
$b=0,1,2,3$.
In these four figures, the resolution is $2^{13}$ points
on a domain size
of 200 with viscosity $\nu=0.005$. This corresponds
to a grid-scale
Reynolds number of
$Re_{\Delta{x}}=U\Delta{x}/\nu=O(1)$ for velocity
$U\simeq0.1$.  The
pair of figures after these four then shows the effects
on the same
problem of increasing viscosity $\nu$ at fixed $\alpha$ for
$b=2$
and
$b=3$.

\begin{description}

\item
Figure \ref{pkn-ic-vb-eqn-b=0} shows three plots of the evolution of the
velocity profile under the viscous b-equation (\ref{viscous-b-family-1})
of an initial peakon of width five, as a function of increasing
$\alpha=1/4,1,4$ at fixed viscosity $\nu=0.005$ for $b=0$. The peakon
leans to the right and develops a  Burgers-like triangular shock, or ramp
and cliff, whose width increases and peak height decreases as $\alpha$
increases. These three plots show no discernable differences for $b=0$ as
the viscosity is decreased to $\nu=10^{-6}$. Hence the width of the
cliff in the ramp and cliff structure for $b=0$ is set by the value of
$\alpha$ in this range of parameters.

\remfigure{
\begin{figure}
\begin{center}
     \leavevmode {
        \hbox{\epsfig{figure=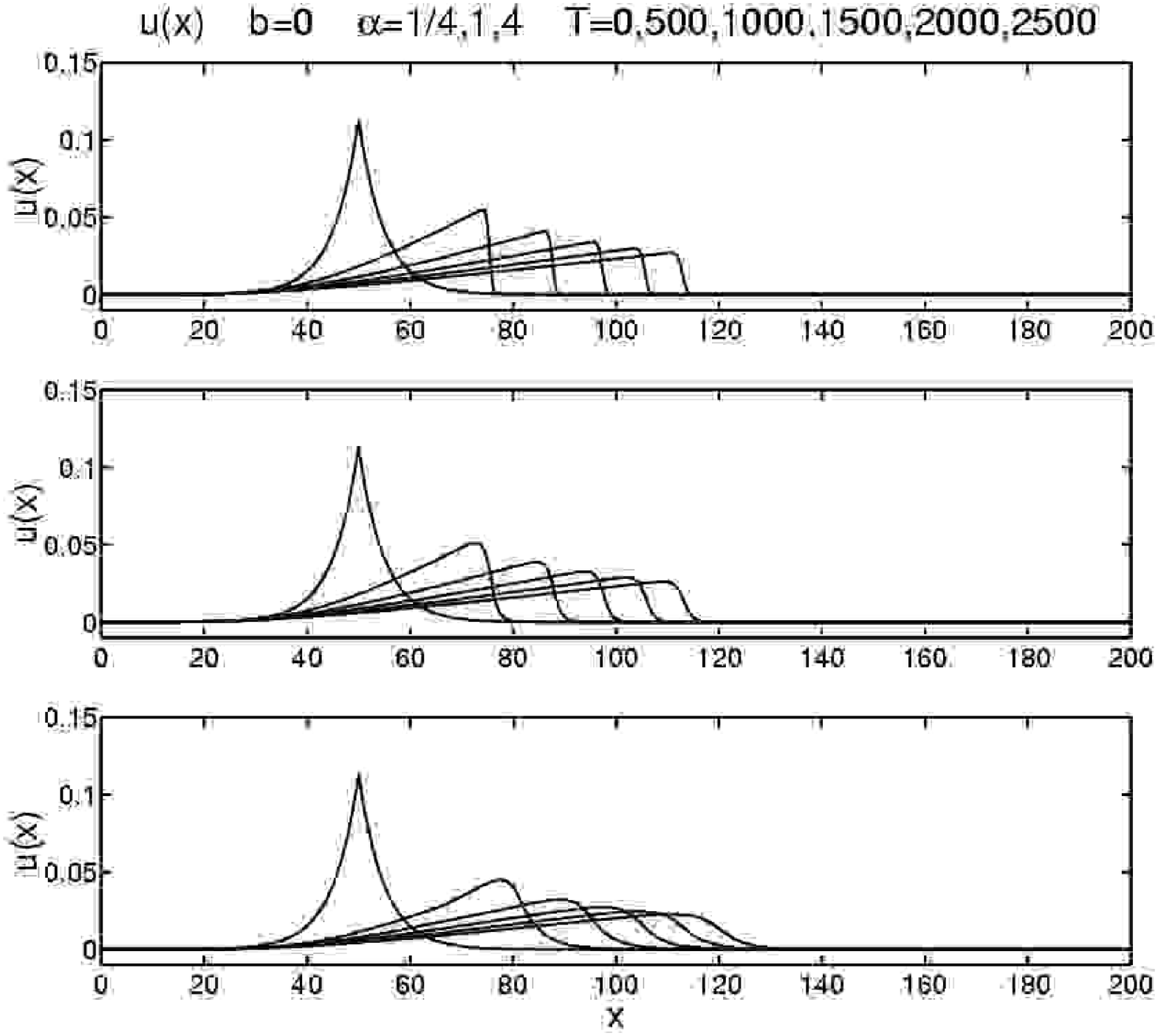, scale=0.4}}
     }
\caption{\label{pkn-ic-vb-eqn-b=0}
   {\bf Effect of increasing $\alpha$ for $b=0$.}
   Viscous $b$-family,
   $b=0$,
   $\alpha=1/4,1,4$,
   $\nu=0.005$,
   initial width $w=5$.
\rem{\sf
For $b=0$, evolution of the velocity profile under the viscous b-equation
(\ref{viscous-b-family-1}) of an initial peakon of width five, as a function
of increasing $\alpha=1/4,1,4$ at fixed viscosity $\nu=0.005$, for which
$Re_{\Delta{x}}=(b+1)U\Delta{x}/\nu\simeq4$.
}
}
\end{center}
\end{figure}
}

\item
Figure \ref{pkn-ic-vb-eqn-b=1} shows three plots of the same type of
evolution from a peakon initial condition of width $w=5$, as $\alpha$ is
varied for $b=1$. The front of the ramp and cliff structure propagates
faster and is sharper for $b=1$ than for $b=0$ when $\alpha=1/4$ and
$\alpha=1$. This increase in speed appears to occur because the
coefficient increases in the steepening term $(b+1)uu_x$ in equation
(\ref{b-family-u-visc}). A nascent peakon begins to form close behind
the front at the top of the ramp, then eventually gets absorbed into the
ramp and cliff. For $\alpha=4$, however, this nascent peakon forms more
completely and nearly escapes.

\remfigure{
\begin{figure}
\begin{center}
     \leavevmode {
        \hbox{\epsfig{figure=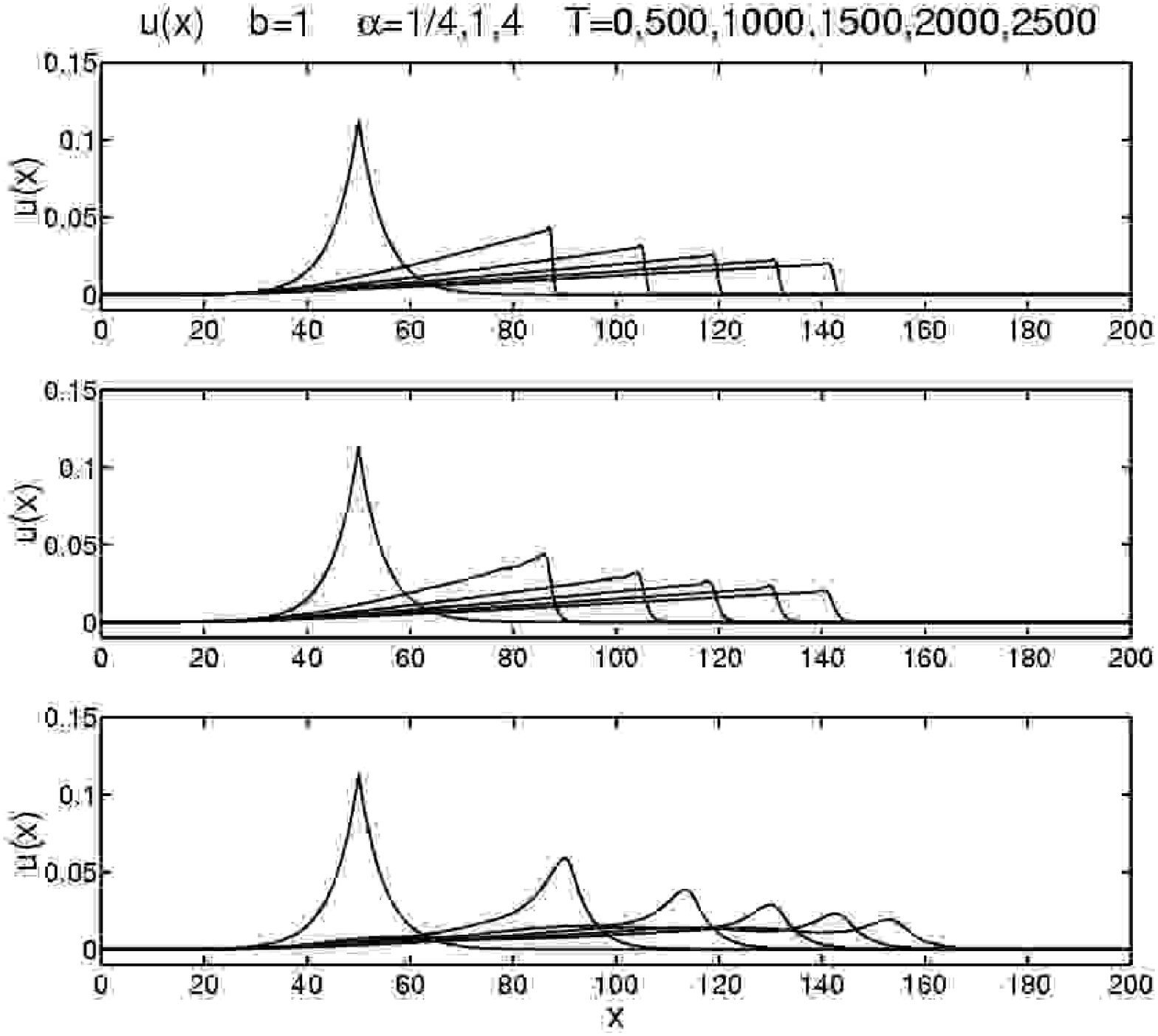, scale=0.4}}
     }
\caption{\label{pkn-ic-vb-eqn-b=1}
   {\bf Effect of increasing $\alpha$ for $b=1$.}
   Viscous $b$-family,
   $b=1$,
   $\alpha=1/4,1,4$,
   $\nu=0.005$,
   initial width $w=5$.
\rem{\sf
For $b=1$, evolution of the velocity profile under the viscous b-equation
(\ref{viscous-b-family-1}) of an initial peakon of width five, as a function
of increasing $\alpha=1/4,1,4$ at fixed viscosity $\nu=0.005$, for which
$Re_{\Delta{x}}=(b+1)U\Delta{x}/\nu\simeq8$.
}
}
\end{center}
\end{figure}
}

\item
Figure \ref{pkn-ic-vb-eqn-b=2} again shows three plots of
the evolution from a peakon initial condition of width $w=5$, as $\alpha$ is
varied, this time for $b=2$. The ramp and cliff structure is faster for
$b=2$ than for $b=1$ when $\alpha=1/4$. When $\alpha=1$ a series of
three nascent peakons forms close behind the front, then overtakes the
ramp and cliff structure and slightly affects its propagation before
eventually being absorbed. For $\alpha=4$, however, the initial peakon
simply propagates and decays under viscosity, although it is slightly
rounded at the top.

\remfigure{
\begin{figure}
\begin{center}
     \leavevmode {
        \hbox{\epsfig{figure=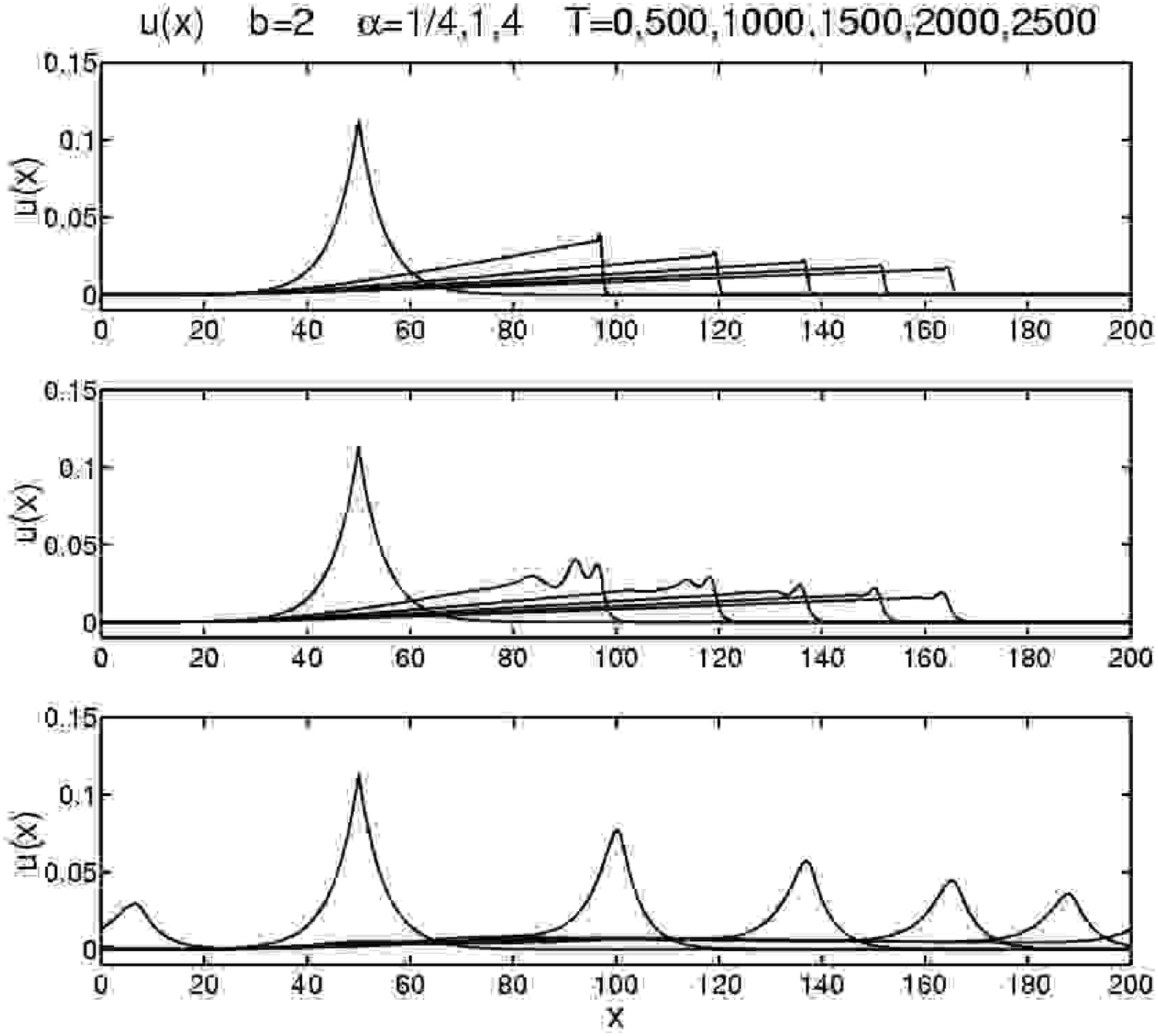, scale=0.4}}
     }
\caption{\label{pkn-ic-vb-eqn-b=2}
   {\bf Effect of increasing $\alpha$ for $b=2$.}
   Viscous $b$-family,
   $b=2$,
   $\alpha=1/4,1,4$,
   $\nu=0.005$,
   initial width $w=5$.
\rem{\sf
For $b=2$, evolution of the velocity profile under the viscous b-equation
(\ref{viscous-b-family-1}) of an initial peakon of width five, as a function
of increasing $\alpha=1/4,1,4$ at fixed viscosity $\nu=0.005$, for which
$Re_{\Delta{x}}=(b+1)U\Delta{x}/\nu\simeq12$.
}
}
\end{center}
\end{figure}
}

\item
Figure \ref{pkn-ic-vb-eqn-b=3} also shows three plots of
the evolution from a peakon initial condition of width $w=5$, as $\alpha$ is
varied, this time for $b=3$. The ramp and cliff structure moves faster
yet, and a single nascent peakon appears just behind the front already for
$\alpha=1/4$. When $\alpha=1$, a series of three nascent peakons
forms initially close behind the front and they nearly escape before being
slowed by viscosity. The leading peakon decays and slows due to
viscosity. Then the following ones overtake and collide with the
ones ahead as the ramp and cliff structure forms. These collisions occur at
higher relative velocity for $b=3$ than for $b=2$ and they significantly
affect the  propagation and eventual formation of the ramp and cliff. In
contrast, for $\alpha=4$, the initial peakon keeps its integrity and simply
propagates rightward and decays under viscosity. The propagating peakon
for $\alpha=4$ at this viscosity decays more slowly and is much sharper at
the top for $b=3$ than for $b=2$.

\remfigure{
\begin{figure}
\begin{center}
     \leavevmode {
        \hbox{\epsfig{figure=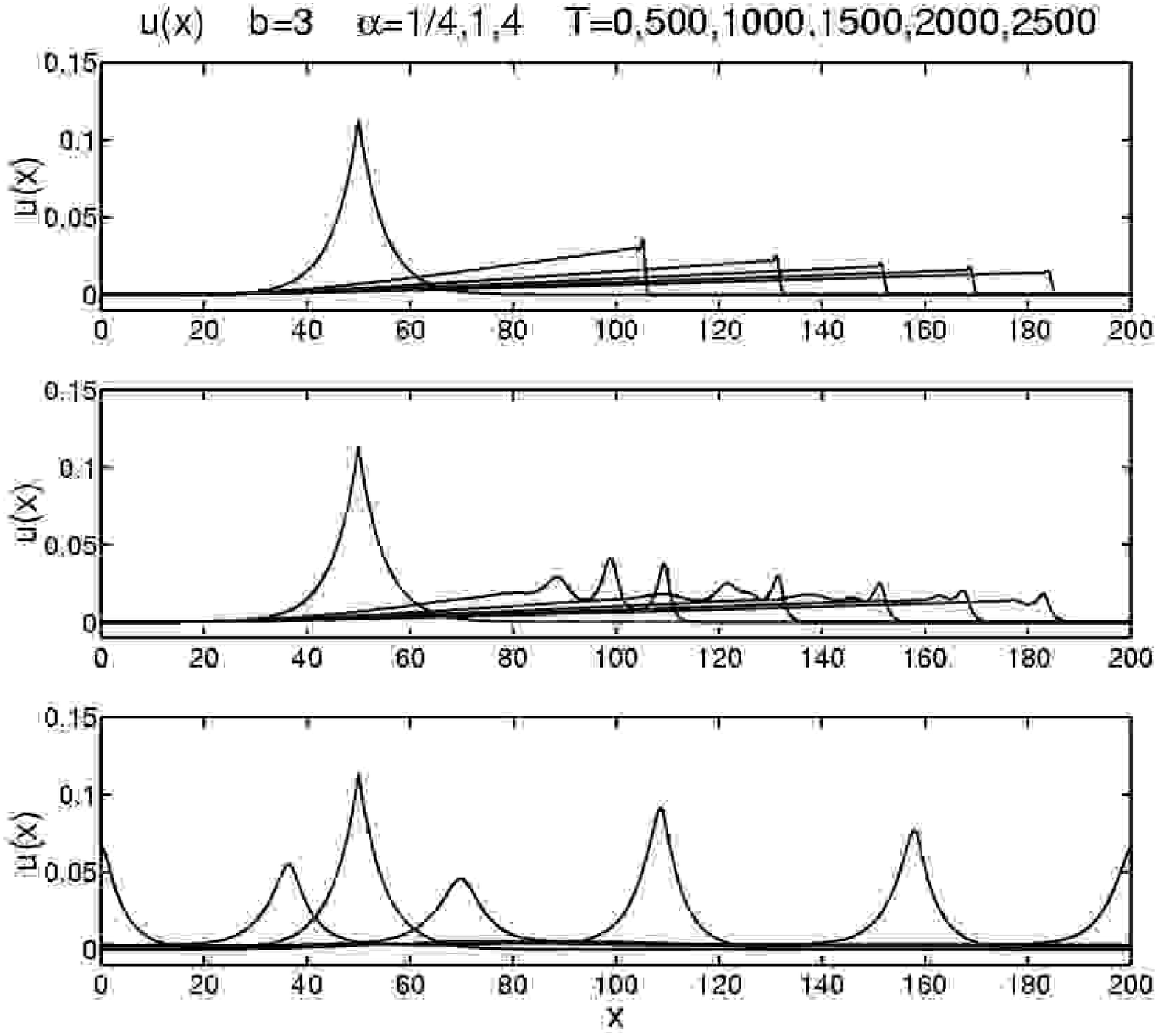, scale=0.4}}
     }
\caption{\label{pkn-ic-vb-eqn-b=3}
   {\bf Effect of increasing $\alpha$ for $b=3$.}
   Viscous $b$-family,
   $b=3$,
   $\alpha=1/4,1,4$,
   $\nu=0.005$,
   initial width $w=5$.
\rem{\sf
For $b=3$, evolution of the velocity profile under the viscous b-equation
(\ref{viscous-b-family-1}) of an initial peakon of width five, as a function
of increasing $\alpha=1/4,1,4$ at fixed viscosity $\nu=0.005$, for which
$Re_{\Delta{x}}=(b+1)U\Delta{x}/\nu\simeq16$.
}
}
\end{center}
\end{figure}
}

\end{description}

\begin{remark}[Exchange of stability]
To see the exchange of stability between the ramp/cliff structure and 
the peakon
as $b$ changes, we perform the following numerical experiment.  First, we run
the viscous b-equation (\ref{viscous-b-family-1}) with $b=0$, $\alpha=1$,
$\nu=10^{-5}$, and an initial peakon of width $w=5$.  As we see in Figures
\ref{ramp/cliff-ic-b-eqn-b=+2,+3-alpha1} and
\ref{ramp/cliff-ic-b-eqn-b=+2,+3-alpha1-profile},
this evolves into the ramp and cliff formation
even for nearly zero viscosity. Once the final ramp/cliff state is formed, we
then use it as the new initial condition for equation 
(\ref{viscous-b-family-1})
with either $b=2$ or $b=3$.  The new evolution breaks the ramp/cliff structure
into peakons and the new final state is a rightward moving train of peakons
ordered by height.

For Figures
\ref{ramp/cliff-ic-b-eqn-b=+2,+3-alpha5} and
\ref{ramp/cliff-ic-b-eqn-b=+2,+3-alpha5-profile},
we ran the same
numerical experiment, this time with a value $\alpha=5$ equal to the width of
the initial peakon.  The initial peakon ``borrows from the negative'' to form
a ramp, which is not quite antisymmetric because the total area of the initial
peakon must be preserved.  At time $T=150$ we switch to $b=2$ (top plot) or
$b=3$ (bottom plot), and again observe a train of stable peakons emerging from
the now-unstable ramp.

Finally, for Figures
\ref{ramp/cliff-ic-b-eqn-b=-2,-3-alpha1} and
\ref{ramp/cliff-ic-b-eqn-b=-2,-3-alpha1-profile},
we again run the numerical experiment with $\alpha=1$ and an initial peakon
width $w=5$, but this time changing to $b=-2$ or $b=-3$ after the ramp has
formed.  The new evolution breaks the ramp/cliff structure into leftons like
those in Figures \ref{Gauss-ic-pkn/a1,b-2/w10-20} and
\ref{Gauss-ic-pkn/a1,b-3/w10-20}.

\remfigure{
\begin{figure}
\begin{center}
     \leavevmode {
\hbox{\epsfig{figure=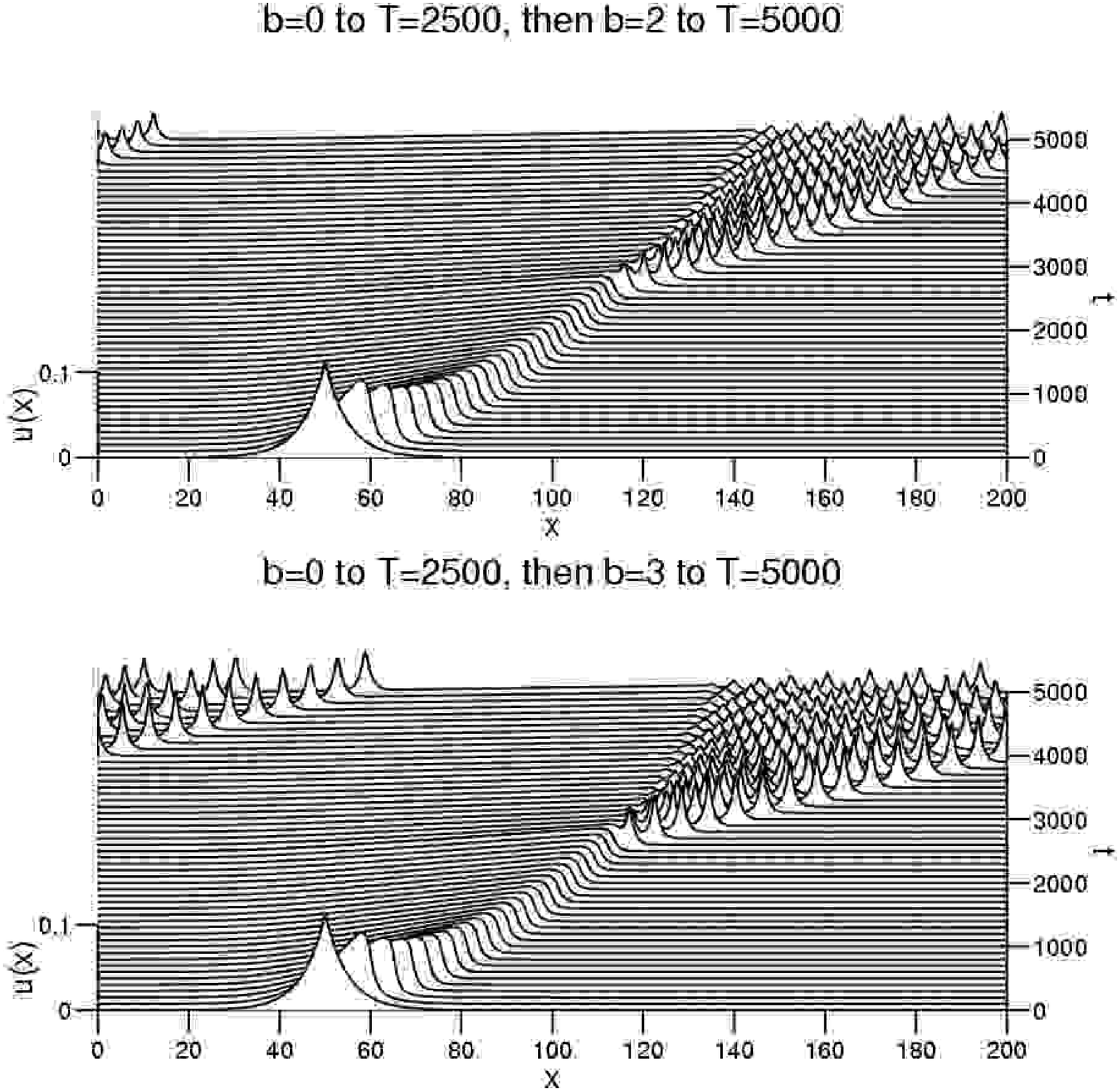, scale=0.45}}
     }
\caption{\label{ramp/cliff-ic-b-eqn-b=+2,+3-alpha1}
   {\bf Exchange of stability between ramps and peakons
      for $b=0,2,3$, when width $>\alpha$.}
   Viscous $b$-family,
   $b=0\rightarrow2,3$,
   $\alpha=1$,
   $\nu=10^{-5}$,
   initial width $w=5$.
\rem{\sf
Exchange of stability between ramps and peakons.  We ran
the viscous b-equation (\ref{viscous-b-family-1}) with the peakon initial
condition with width $w=5$, for $b=0$, $\alpha=1$, and $\nu=10^{-5}$.  At
time $T=2500$ we switched to $b=2$ (top figure) or to $b=3$ (bottom figure).
The ramp and cliff structure, which was stable for $b=0$, splits into
peakons when we change to $b>1$.
}
}
\end{center}
\end{figure}
}

\remfigure{
\begin{figure}
\begin{center}
     \leavevmode {
\hbox{\epsfig{figure=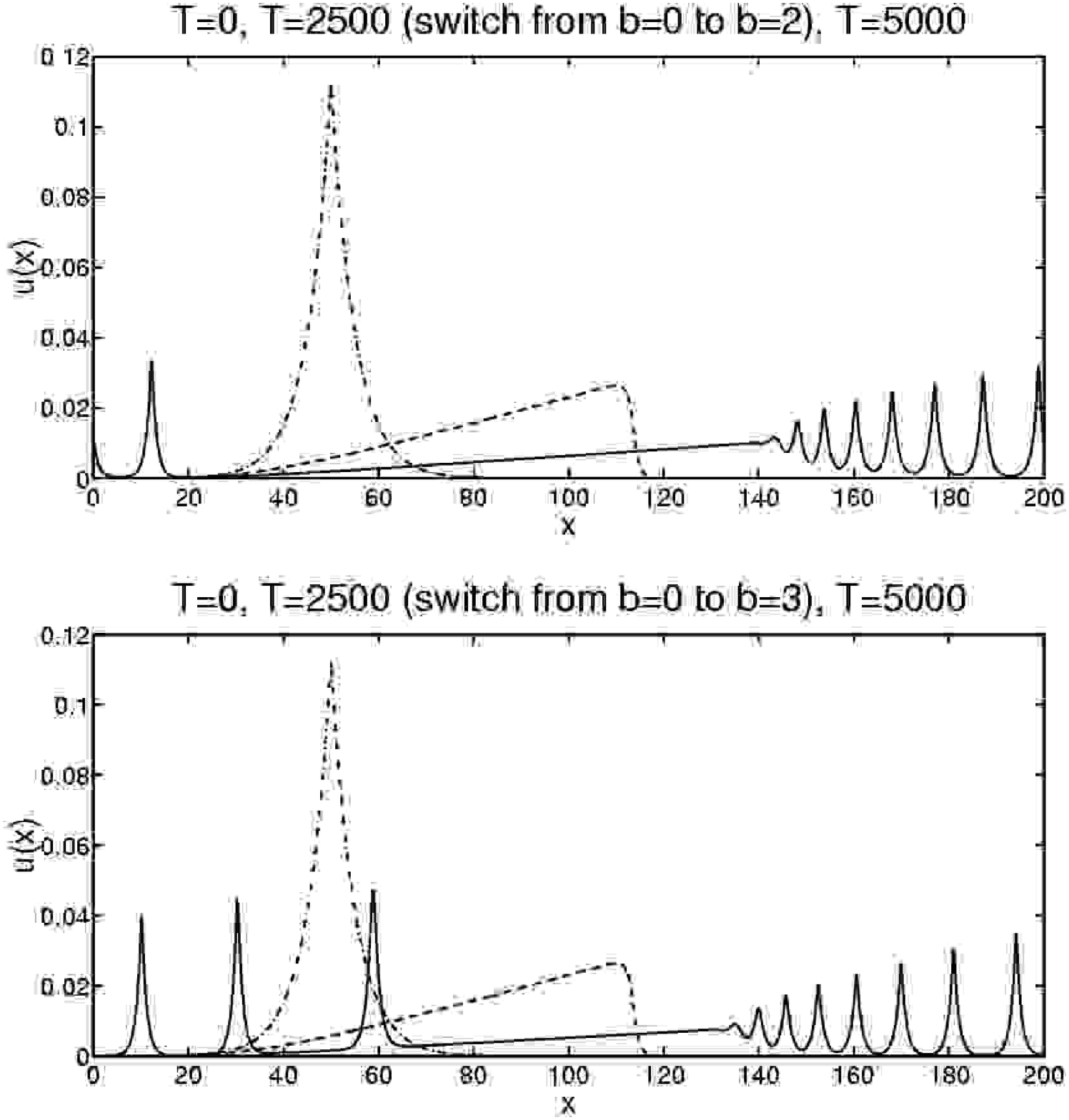, scale=0.45}}
     }
\caption{\label{ramp/cliff-ic-b-eqn-b=+2,+3-alpha1-profile}
   {\bf Exchange of stability between ramps and peakons
      for $b=0,2,3$, when width $>\alpha$: profiles.}
   Viscous $b$-family,
   $b=0\rightarrow2,3$,
   $\alpha=1$,
   $\nu=10^{-5}$,
   initial width $w=5$.
\rem{\sf
Exchange of stability between ramps and peakons.  The figures show profiles of
the data from Figure \ref{ramp/cliff-ic-b-eqn-b=+2,+3-alpha1}, at times T=0, T=2500 (at
which we change from b=0 to b=2 or b=3), and T=4000.
}
}
\end{center}
\end{figure}
}

\remfigure{
\begin{figure}
\begin{center}
     \leavevmode {
\hbox{\epsfig{figure=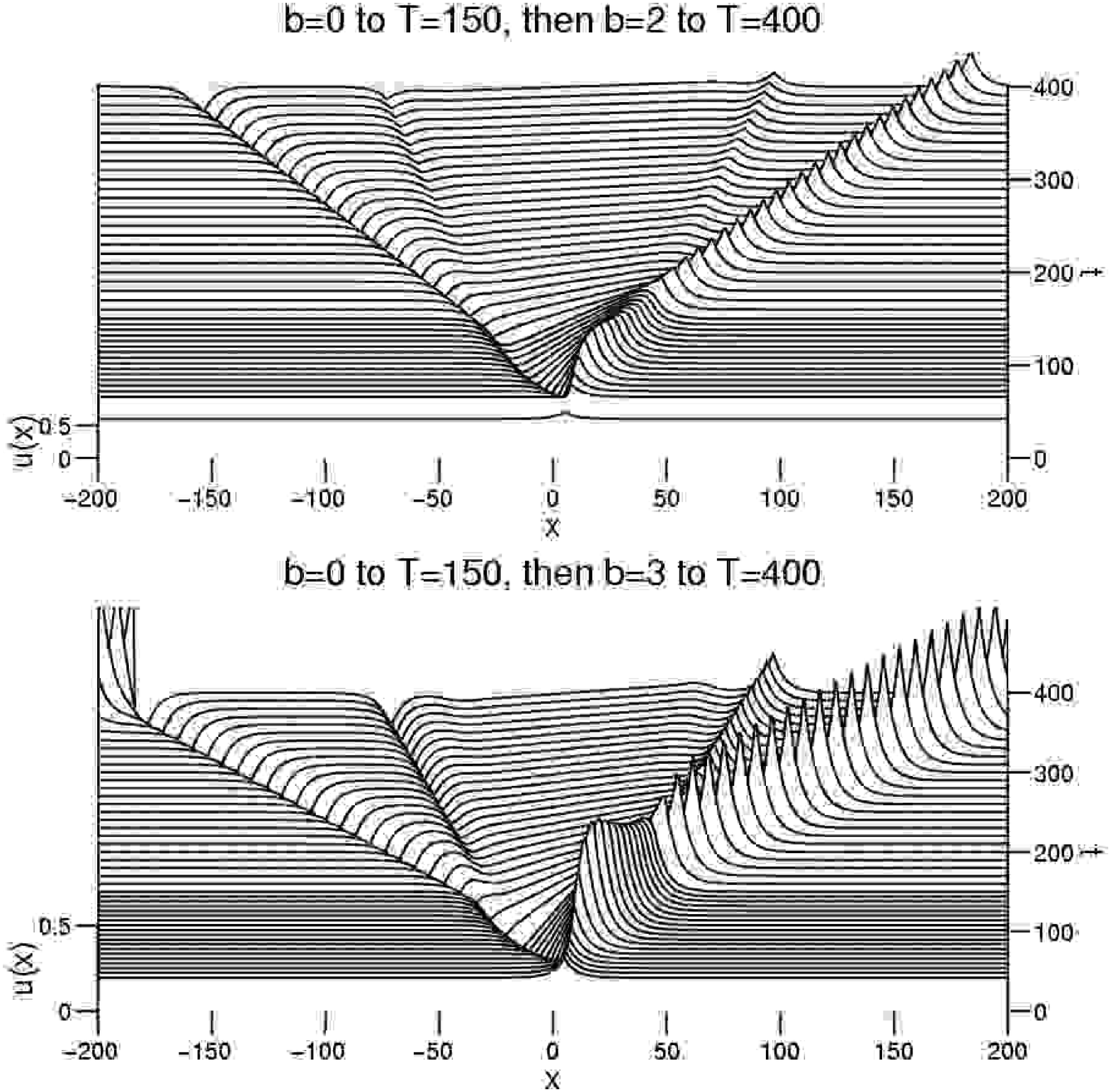, scale=0.45}}
     }
\caption{\label{ramp/cliff-ic-b-eqn-b=+2,+3-alpha5}
   {\bf Exchange of stability between ramps and peakons
      for $b=0,2,3$, when width $=\alpha$.}
   Viscous $b$-family,
   $b=0\rightarrow2,3$,
   $\alpha=5$,
   $\nu=10^{-5}$,
   initial width $w=5$.
\rem{\sf
Exchange of stability between ramps and peakons, beginning with an initial
peakon of width equal to $\alpha$.
}
}
\end{center}
\end{figure}
}

\remfigure{
\begin{figure}
\begin{center}
     \leavevmode {
\hbox{\epsfig{figure=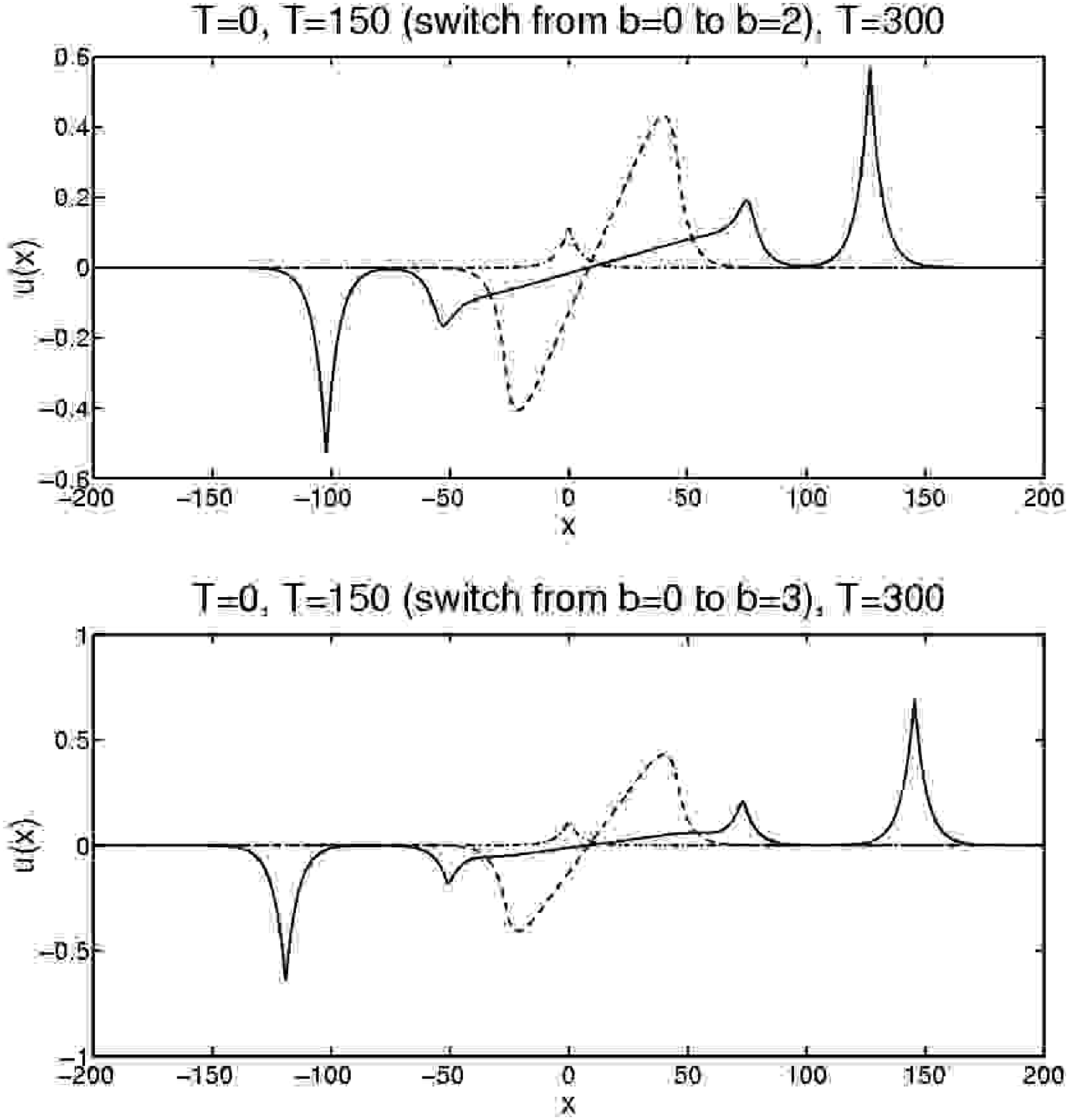, scale=0.45}}
     }
\caption{\label{ramp/cliff-ic-b-eqn-b=+2,+3-alpha5-profile}
   {\bf Exchange of stability between ramps and peakons
      for $b=0,2,3$, when width $=\alpha$: profiles.}
   Viscous $b$-family,
   $b=0\rightarrow2,3$,
   $\alpha=5$,
   $\nu=10^{-5}$,
   initial width $w=5$.
\rem{\sf
Exchange of stability between ramps and peakons, beginning with an initial
peakon of width equal to $\alpha$.
}
}
\end{center}
\end{figure}
}

\remfigure{
\begin{figure}
\begin{center}
     \leavevmode {
\hbox{\epsfig{figure=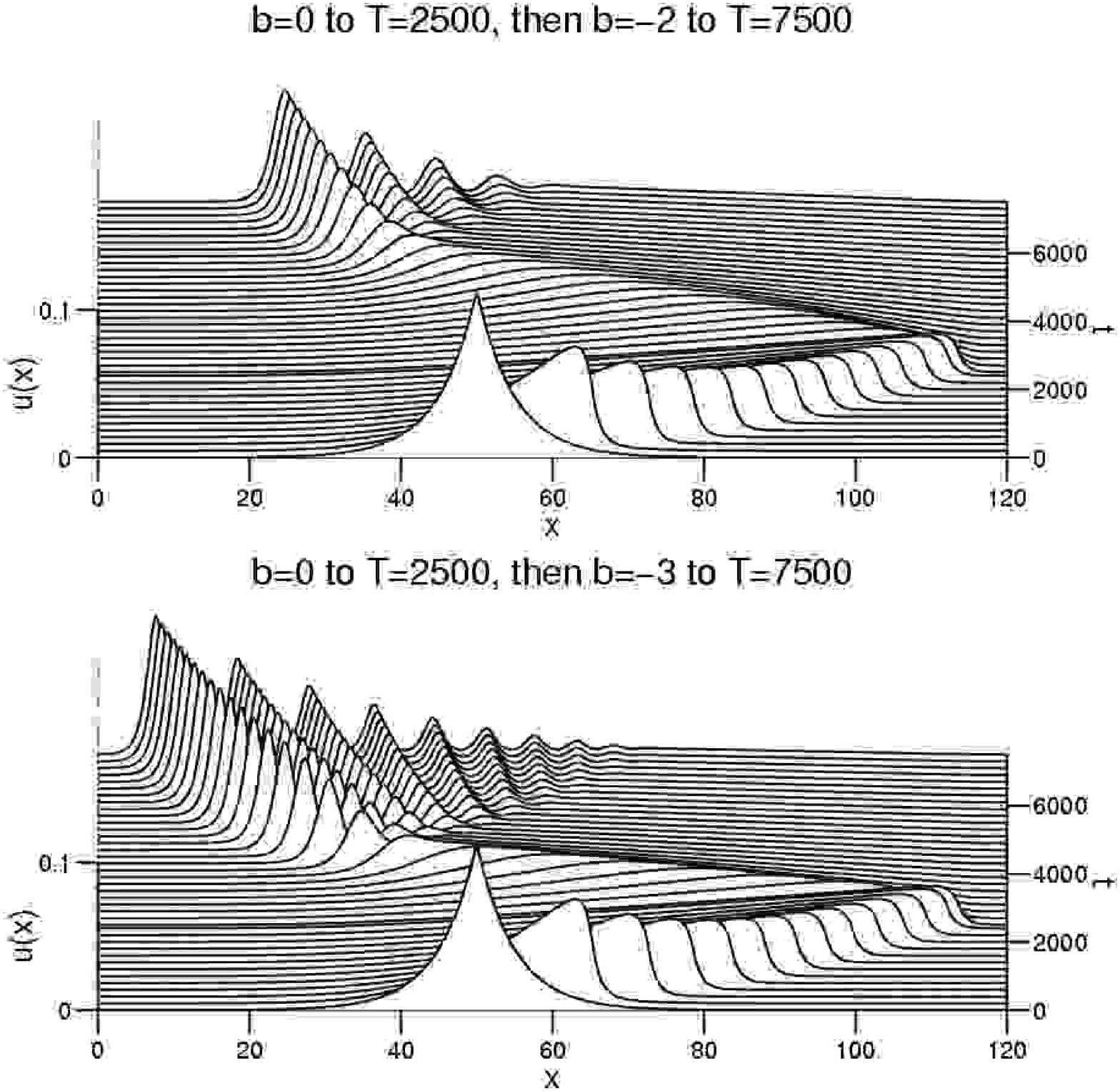, scale=0.45}}
     }
\caption{\label{ramp/cliff-ic-b-eqn-b=-2,-3-alpha1}
   {\bf Exchange of stability between ramps and leftons
      for $b=0,-2,-3$, when width $>\alpha$.}
   Viscous $b$-family,
   $b=0\rightarrow-2,-3$,
   $\alpha=1$,
   $\nu=10^{-5}$,
   initial width $w=5$.
\rem{\sf
Exchange of stability between ramps and peakons.  We ran
the viscous b-equation (\ref{viscous-b-family-1}) with the peakon initial
condition with width $w=5$, for $b=0$, $\alpha=1$, and $\nu=10^{-5}$.  At
time $T=2500$ we switched to $b=-2$ (top figure) or to $b=-3$ (bottom figure).
The ramp and cliff structure, which was stable for $b=0$, splits into leftons.
}
}
\end{center}
\end{figure}
}

\remfigure{
\begin{figure}
\begin{center}
     \leavevmode {
\hbox{\epsfig{figure=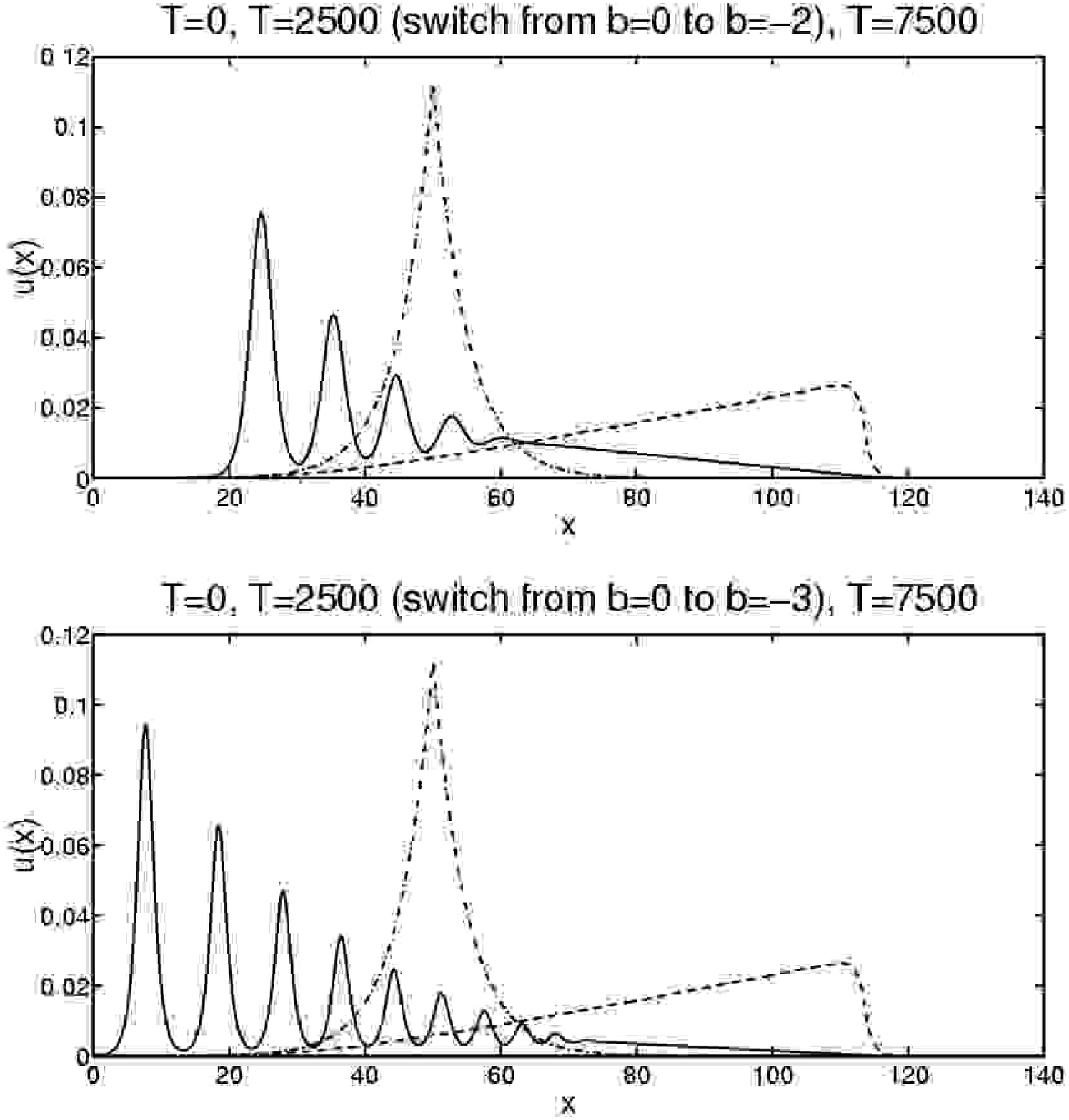, scale=0.45}}
     }
\caption{\label{ramp/cliff-ic-b-eqn-b=-2,-3-alpha1-profile}
   {\bf Exchange of stability between ramps and leftons
      for $b=0,-2,-3$, when width $>\alpha$: profiles.}
   Viscous $b$-family,
   $b=0\rightarrow-2,-3$,
   $\alpha=1$,
   $\nu=10^{-5}$,
   initial width $w=5$.
\rem{\sf
Exchange of stability between ramps and peakons.  The figures show profiles of
the data from Figure \ref{ramp/cliff-ic-b-eqn-b=-2,-3-alpha1}, at times T=0, T=2500 (at
which we change from b=0 to b=-2 or b=-3), and T=7500.
}
}
\end{center}
\end{figure}
}

\end{remark}

\begin{remark}[Increasing viscosity]
The effect of increasing viscosity on the evolution of the peakon initial
condition can be estimated from the $\alpha-$scale Reynolds number defined
by,
\[Re_\alpha=U\alpha/\nu=(\alpha/\Delta{x})Re_{\Delta{x}}\,.\] For
$(\alpha/\Delta{x})=40$, $U=0.1$ and increasing viscosity $\nu$, the
Reynolds numbers $Re_{\Delta{x}}$ and $Re_\alpha$ decrease as
\[Re_{\Delta{x}}\simeq2,0.2,0.02
\quad\hbox{and}\quad
Re_\alpha=\simeq80,8,0.8
\quad\hbox{for}\quad
\nu=0.01,0.1,1.0
\,.\]
Perhaps not surprisingly, when $Re_\alpha=O(1)$ the viscosity will diffuse
through the initial peakon before it can fully form.  Figures
\ref{b2_increase_nu} and \ref{b3_increase_nu} show that this effect increases
as $Re_\alpha$ decreases.
\end{remark}

\remfigure{
\begin{figure}
\begin{center}
     \leavevmode {
        \hbox{\epsfig{
figure=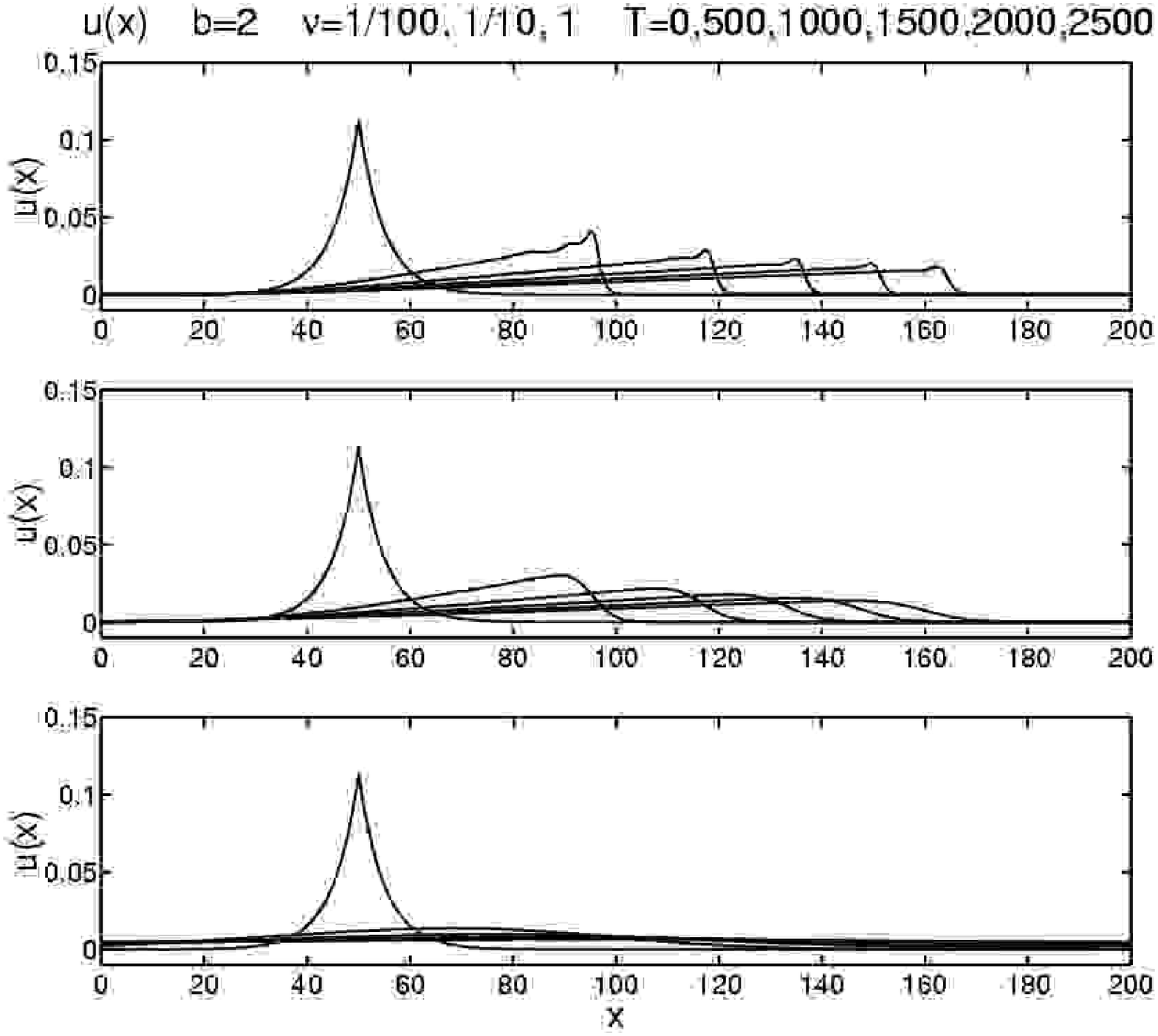, scale=0.4}}
     }
\caption{\label{b2_increase_nu}
   {\bf Effect of increasing viscosity for $b=2$.}
   Viscous $b$-family,
   $b=2$,
   $\alpha=1$,
   $\nu=1/100,1/10,1$,
   initial width $w=5$.
\rem{\sf
For $b=2$, evolution of the velocity profile under the viscous b-equation
(\ref{viscous-b-family-1}) of an initial peakon of width five, as a function
of increasing $\nu$ at fixed $\alpha=1$.
}
}
\end{center}
\end{figure}
}

\remfigure{
\begin{figure}
\begin{center}
     \leavevmode {
        \hbox{\epsfig{
figure=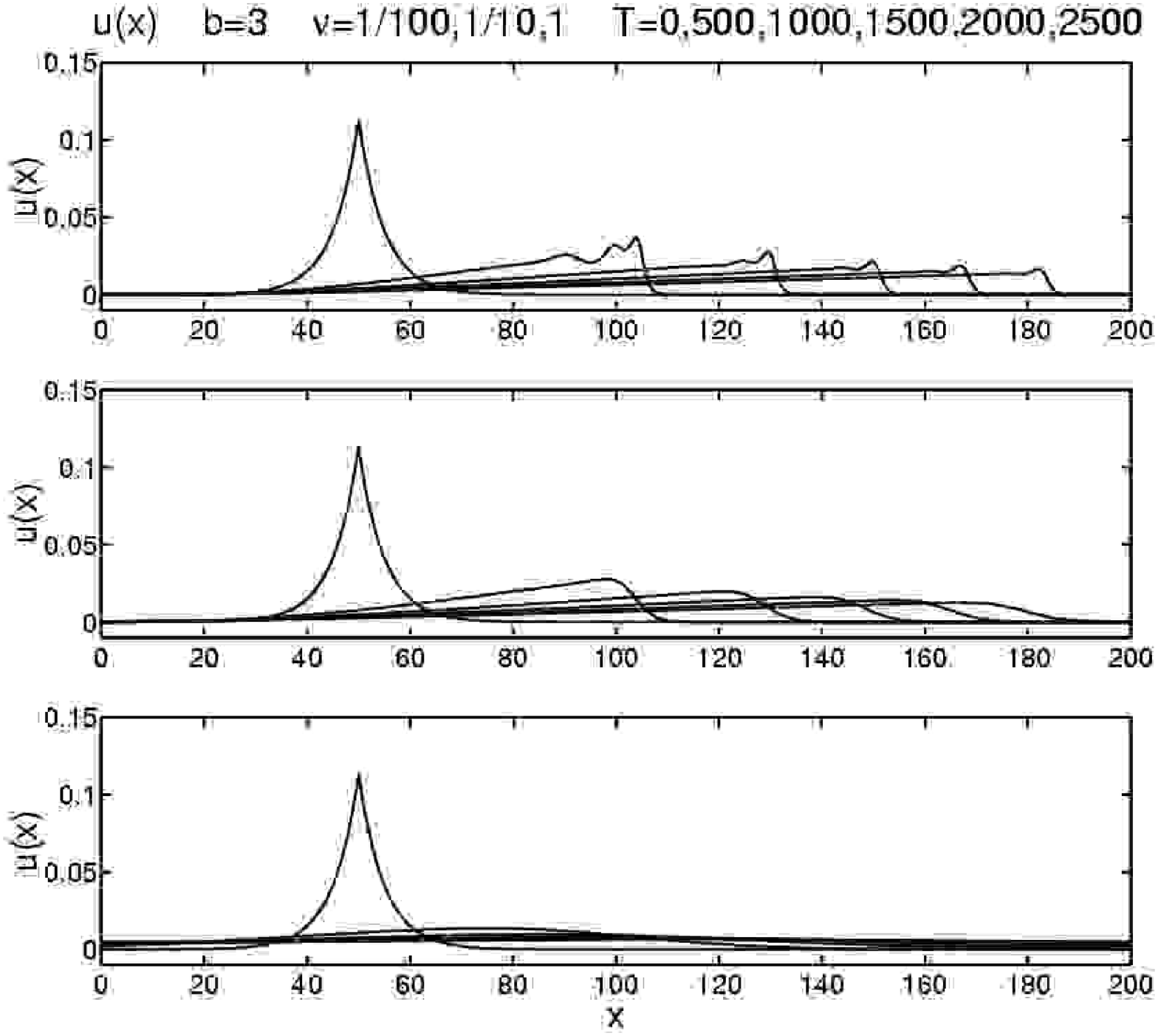, scale=0.4}}
     }
\caption{\label{b3_increase_nu}
   {\bf Effect of increasing viscosity for $b=3$.}
   Viscous $b$-family,
   $b=3$,
   $\alpha=1$,
   $\nu=1/100,1/10,1$,
   initial width $w=5$.
\rem{\sf
For $b=3$, evolution of the velocity profile under the viscous b-equation
(\ref{viscous-b-family-1}) of an initial peakon of width five, as a function
of increasing $\nu$ at fixed $\alpha=1$.
}
}
\end{center}
\end{figure}
}

\subsection{The fate of peakons under Burgers$-\alpha\beta$ evolution}

Figures \ref{beta1div3_vary_alpha} and \ref{beta1div2_vary_alpha}
show the effects on the peakon initial value
problem for the Burgers$-\alpha\beta$ evolution of varying $\alpha$
and $b$ with $(3-b)\beta=1$ at constant viscosity.
We shall consider the following cases with $(3-b)\beta=1$:
         \begin{description}
         \item
         $b=0$, $\beta=1/3$, $\nu=0.005$, $\alpha=1/4,1,4$, and
         \item
         $b=1$, $\beta=1/2$, $\nu=0.005$, $\alpha=1/4,1,4$.
         \end{description}
\begin{remark} [Lowering $\beta$ has little effect on the ramp/cliff]
Lowering $\beta$ to follow $(3-b)\beta=1$ instead of keeping $\beta=1$
has little effect on the development of the ramp/cliff solution for $b=0$
and $b=1$. Lowering $\beta$ for these cases only makes the activity
slightly less lively at the front for ($b=0$, $\beta=1/3$) and ($b=1$,
$\beta=1/2$) than for the corresponding cases of $b=0$ and $b=1$ with
$\beta=1$ in Figures \ref{pkn-ic-vb-eqn-b=0} and
\ref{pkn-ic-vb-eqn-b=1}. This lessened activity at lower
$\beta$ can only be discerned in the solution for the largest value
$\alpha=4$. The remaining case (b=2, $\beta=1$) recovers the
viscous b-equation (\ref{viscous-b-family-1}) for $b=2$ in Figure
\ref{pkn-ic-vb-eqn-b=2}, in which the larger $b$
produces much livelier steepening and, hence, more activity at the front
of the rightward moving pulses.
\end{remark}

\remfigure{
\begin{figure}
\begin{center}
     \leavevmode {
        \hbox{\epsfig{
figure=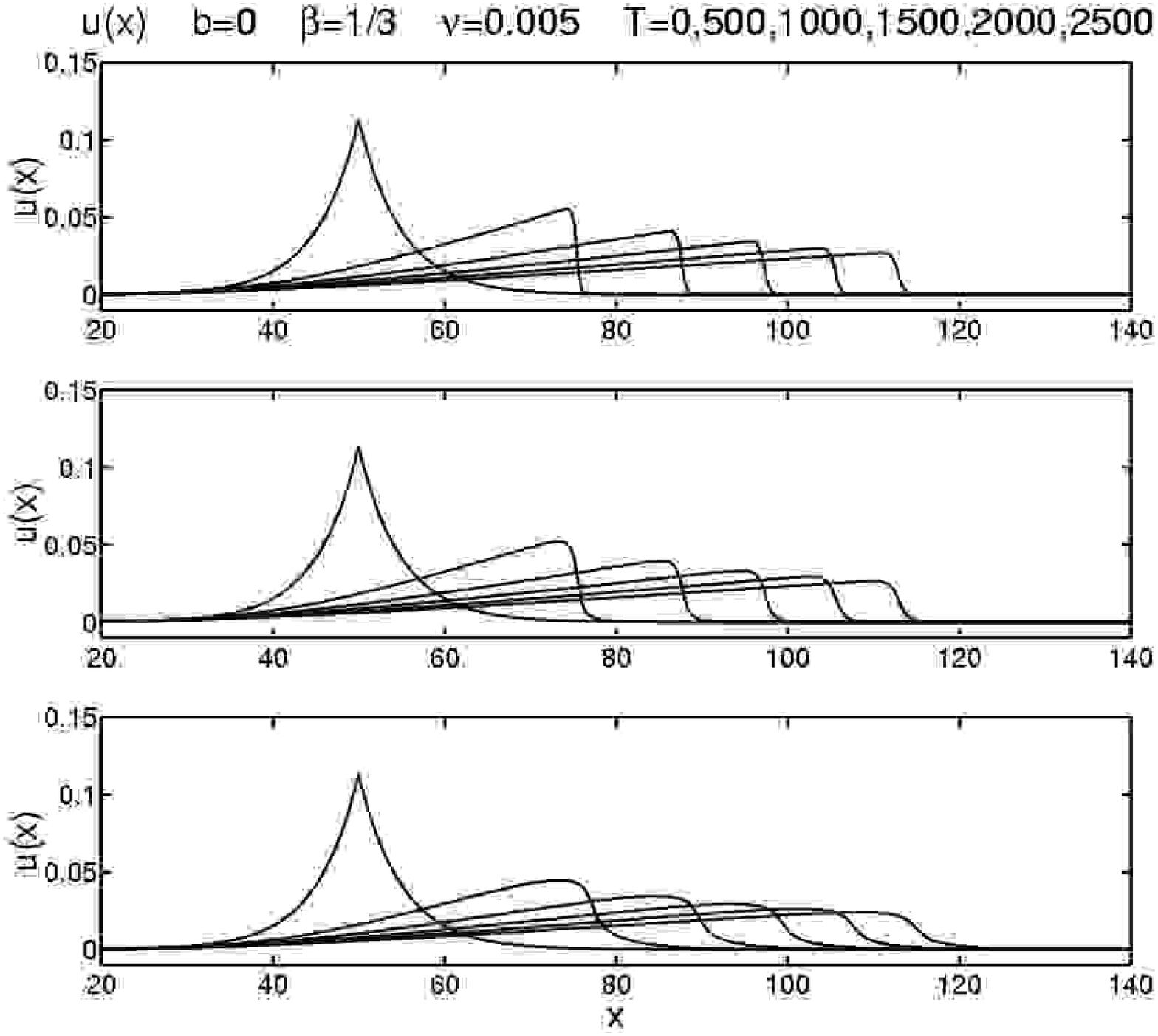, scale=0.4}}
     }
\caption{\label{beta1div3_vary_alpha}
   {\bf Effect of increasing $\alpha$ when $(3-b)\beta=1$,
      for $b=0$ and $\beta=1/3$.}
   Burgers-$\alpha\beta$,
   $b=0$,
   $\alpha=1/4,1,4$,
   $\beta=1/3$,
   $\nu=0.005$,
   initial width $w=5$.
\rem{\sf
For $b=0$, $\beta=1/3$ and $\nu=0.005$, evolution of the velocity profile
under the Burgers$-\alpha\beta$ equation
(\ref{convect-viscous-Bab-eqn}) of an initial
peakon of width five, as a function of increasing $\alpha=1/4,1,4$.
}
}
\end{center}
\end{figure}
}

\remfigure{
\begin{figure}
\begin{center}
     \leavevmode {
        \hbox{\epsfig{
figure=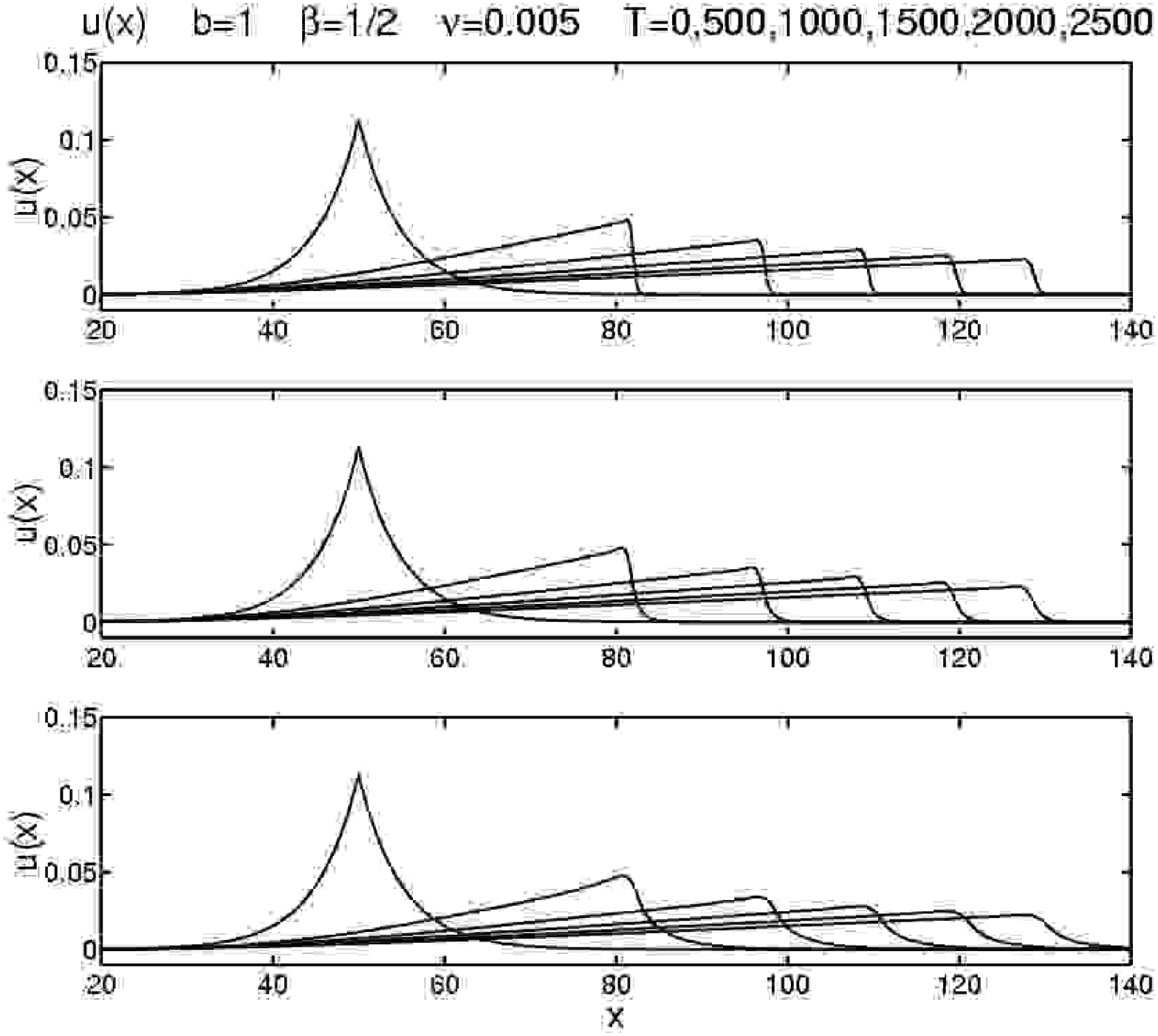, scale=0.4}}
     }
\caption{\label{beta1div2_vary_alpha}
   {\bf Effect of increasing $\alpha$ when $(3-b)\beta=1$,
      for $b=1$ and $\beta=1/2$.}
   Burgers-$\alpha\beta$,
   $b=1$,
   $\alpha=1/4,1,4$,
   $\beta=1/2$,
   $\nu=0.005$,
   initial width $w=5$.
\rem{\sf
For $b=1$, $\beta=1/2$ and $\nu=0.005$, evolution of the velocity profile
under the Burgers$-\alpha\beta$ equation
(\ref{convect-viscous-Bab-eqn}) of an initial
peakon of width five, as a function of increasing $\alpha=1/4,1,4$.
}
}
\end{center}
\end{figure}
}

\section{Numerical results for peakon scattering and initial
          value problems}\label{IVP}

We shall begin by summarizing the results in the figures given earlier, and
then describe the numerical methods used in producing them and discuss some
of the ways we verified and validated the results.

\subsection{Peakon initial value problems}

\subsubsection{Inviscid b-family of equations}
\begin{description}

\item{\bf Ramps and cliffs for $b=0$.}
Figure \ref{Gauss-ic-pkn/a1,b0/w2.5-10} shows the formation of a ramp and cliff
pattern for $b=0$, $\alpha=1$, and a set of Gaussian initial conditions of
increasing width $w=2.5,5,10$.

\item{\bf Peakons for $b=2,3$.}
Figures \ref{Gauss-ic-pkn/a1,b2/w2.5-10} and \ref{Gauss-ic-pkn/a1,b3/w2.5-10}
show the formation of peakons for $b=2$ and $b=3$, for $\alpha=1$ and a set of
Gaussian initial conditions of increasing width $w=2.5,5,10$.
\rem{
Peakons develop
and travel faster when $b=3$ than when $b=2$.
}

\item{\bf Ramps and cliffs for $b=-1/2$.}
Figure \ref{Gauss-ic-pkn/a1,b-0.5/w10-20} shows the formation of a ramp and
cliff pattern for $b=-1/2$, $\alpha=1$, and a set of Gaussian initial 
conditions
of increasing width $w=10,15,20$.

\item{\bf Stationary solutions for $b\le-1$.}
Figure \ref{Gauss-ic-pkn/a1,b-1/w10-20} shows an essentially 
stationary solution
with a slight rightward drift and leaning slightly to the right due 
to nonlinear
curvature terms with higher order derivatives in equation (\ref{b-family-u}),
for $\alpha=1$ and a set of Gaussian initial conditions of increasing width
$w=10,15,20$.  For the same $\alpha$ and same set of initial 
conditions, Figures
\ref{Gauss-ic-pkn/a1,b-2/w10-20} and \ref{Gauss-ic-pkn/a1,b-3/w10-20} show the
emergence of leftons.

Figure \ref{Gauss-ic-pkn/a1,b-2/w10} shows the leftons at time $T=2500$ for
the $b=-2$ case, versus the analytical $u(x) \simeq {\rm sech}^2(x/(2\alpha))$
from equation (\ref{hyper-limit-b-2}),
and for the $b=-3$ case, versus the analytical
$u(x) \simeq {\rm sech}(x/\alpha)$.

\item{\bf Peakons of width $\alpha$ for $b=2,3$.}
Figures \ref{2-3-pkn-ic/a1,b2} and \ref{2-3-pkn-ic/a1,b3} show 2-peakon and
3-peakon interactions for $b=2$ and for $b=3$, beginning with initial peakons
of width $w=\alpha=5$.

\item{\bf Peakons of width $>\alpha$ for $b=2,3$.}
Figure \ref{2-pkn/a1,b2-3/w5} shows the emergence of peakons of width 
$\alpha=1$
when we begin with peakons of width $w=5$ greater than $\alpha$, for $b=2$ and
$b=3$.

\item{\bf Peakon-antipeakon collisions for $b=1,2,3$.}
Figure \ref{pkn-anti-pkn} shows the dynamics of a peakon-antipeakon collision
for $b=1$, $b=2$, and $b=3$, for $\alpha=1$, at four successive times.

\end{description}

\subsubsection{Viscous b-family of equations}
\begin{description}

\item{\bf Effect of $\alpha$ for $b=0,1,2,3$.}
Figures \ref{pkn-ic-vb-eqn-b=0} -- \ref{pkn-ic-vb-eqn-b=3} show the evolution
of an initial peakon of width $w=5$ as a function of increasing 
$\alpha=1/4,1,4$
at fixed viscosity $\nu=0.005$, for $b=0$, $b=1$, $b=2$, and $b=3$.

\item{\bf Exchange of stability between ramps and peakons.}
Figures
\ref{ramp/cliff-ic-b-eqn-b=+2,+3-alpha1} and
\ref{ramp/cliff-ic-b-eqn-b=+2,+3-alpha1-profile}
show the exchange of stability between ramps and peakons suggested in the
previous four figures, with $\alpha=1$ and an initial peakon of width 
$w=5$, but
this time with a very small viscosity $\nu=10^{-5}$ so that the peakons, when
stable, do not noticeably decay.  The exchange of stability occurs when we
switch from $b=0$ to $b=2$ or $b=3$.  Figures
\ref{ramp/cliff-ic-b-eqn-b=+2,+3-alpha5} and
\ref{ramp/cliff-ic-b-eqn-b=+2,+3-alpha5-profile}
again show the exchange of
stability, this time using $\alpha=5$ so that the initial peakon has width
$\alpha$.

\item{\bf Exchange of stability between ramps and leftons.}
Figures
\ref{ramp/cliff-ic-b-eqn-b=-2,-3-alpha1} and
\ref{ramp/cliff-ic-b-eqn-b=-2,-3-alpha1-profile}
show the exchange of stability as in Figures
\ref{ramp/cliff-ic-b-eqn-b=+2,+3-alpha1} and
\ref{ramp/cliff-ic-b-eqn-b=+2,+3-alpha1-profile},
but we switch to $b=-2$ or $b=-3$ instead, and see the emergence of stable
leftons from the ramp.

\item{\bf Effect of viscosity for $b=2,3$.}
Figures \ref{b2_increase_nu} and \ref{b3_increase_nu} show the effect of
increasing viscosity $\nu=0.01,0.1,1$ on the evolution of an initial peakon
of width $w=5$, with $\alpha=1$ and $b=2$ or $b=3$.

\end{description}

\subsubsection{Burgers-$\alpha\beta$ equation}
\begin{description}

\item{\bf Effect of $\alpha$ when $(3-b)\beta=1$.}
Figures \ref{beta1div3_vary_alpha} and \ref{beta1div2_vary_alpha} show the
effect of increasing $\alpha=1/4,1,4$ on the evolution of an initial peakon
of width $w=5$, for fixed $\nu=0.005$ and two sets of values for $b$ 
and $\beta$
for which $(3-b)\beta=1$:
$b=0$, $\beta=1/3$ for the first figure, and
$b=1$, $\beta=1/2$ for the second figure.

\end{description}

\rem{
\item
{\bf Rear-end peakon-peakon collisions:}
Figures 
\ref{2-3-pkn-ic/a1,b2} and \ref{2-3-pkn-ic/a1,b3} show the 
time
evolution of the velocity profile under the inviscid b-equation 
(\ref{b-family})
for rear-end peakon-peakon collision interactions 
when $b=2$ and $b=3$,
respectively.

\item
{\bf  Emission peakon trains from Gaussian initial conditions:}
Figures \ref{Gauss-ic-pkn/a1,b2/w2.5-10} and \ref{Gauss-ic-pkn/a1,b3/w2.5-10}
show the evolution of the velocity profile under the inviscid b-equation
(\ref{b-family}) of an initial Gaussian of increasing width $w=2.5,5,10$ for
$b=2$ and $b=3$, respectively.  The initial Gaussian, which has unit area for
width $w=5$, leans to the right and emits a series of peakons that are each of
width $\alpha=1$ and form a train ordered by height.  Because the domain is
periodic, the fastest peakons leave at the right and re-enter from the left,
thereby experiencing many elastic rear-end collision interactions with the
slower peakons.  The figures show that more peakons are emitted as the width of
the initial Gaussian increases relative to the peakon width $\alpha$.
}

\subsection{Description of our numerical methods}\label{NumMethod}

For our numerical runs we advanced equations (\ref{convect-b-eqn}),
(\ref{convect-viscous-b-family}), and (\ref{convect-viscous-Bab-eqn})
with an explicit, variable timestep fourth/fifth order Runge-Kutta-Fehlberg
(RKF45) predictor/corrector.  We selected the timestep for numerical stability
by trial and error, while our code selected the timestep for numerical accuracy
(not to exceed the timestep for numerical stability) according to the 
well-known
formula from numerical analysis,
\begin{equation}\label{vartimestep-control}
h_i = \gamma h_{i-1}\left(
\frac{\epsilon|h_{i-1}|}{||\bar u_i - \hat u_i||}
\right)^{1/p}
\,.
\end{equation}
This is used in the following way.  At step $i$ of the calculation, we know the
predicted solution $\bar u_i$, the corrected solution $\hat u_i$, and the
previous timestep $h_{i-1}$.  The predictor's order of accuracy is $p=4$, while
the corrector's order of accuracy is $p+1$.  A new timestep $h_i$ is 
chosen from
(\ref{vartimestep-control}) based on the old timestep $h_{i-1}$ and the norm of
the difference between the current predicted and corrected
solutions.  We used a very strict relative error tolerance per timestep,
$\epsilon=10^{-8}$, a safety factor $\gamma=0.9$, and an $L_2$ norm
$||\cdot||_2$.

We computed spatial derivatives using 4th order finite differences, 
generally
at resolutions of $2^{13}=8192$ or $2^{14}=16384$ zones.
To 
invert the Helmholtz operator
in transforming between $m(x,t)$ and 
$u(x,t)$, we convolved $m(x,t)$ with the
Green's function in Fourier 
space.  When the numerical approximation of the
nonlinear terms had 
aliasing errors in the high wavenumbers, we applied the
following 
high pass filtered artificial 
viscosity,
\begin{equation}
\nu(k)=
\begin{cases}
0
   &\hbox{if\ \ } 
0\le k\le\frac{N}{3},\cr
\frac{3\delta}{N}\left(k-\frac{N}{3}\right)
&\hbox{if\ \ } \frac{N}{3}<k<\frac{2N}{3},\cr
\delta
   &\hbox{if\ \ 
} \frac{2N}{3}\le k\le N,\cr
\end{cases}
\end{equation}
where 
$\delta=0.01$ for the present simulations.  $N$ is one-half the 
number
of zones, because for each zone we have both a Fourier sine 
coefficient and
a Fourier cosine coefficient.

The quality of the 
numerical convergence may be checked analytically in the
case of rear-end two-pulson collisions, for which equation (\ref{q-min}) in
Corollary \ref{cor-q-min} yields
\begin{equation}\label{pulson-qmin}
g(q_{min}) = g(q)\Big|_{p=0} = 1-
\Big(\,\frac{4c_1c_2}{(c_1+c_2)^2}\Big)^{{1}/(b-1)}.
\end{equation}
For peakons with $b=2$ and $g(x)=e^{-|x|/\alpha}$, this formula gives the
minimum separation,
\begin{equation}\label{peakon-qmin1}
q_{min} = -2\alpha\,{\rm ln} \Big(\,\frac{c_1-c_2}{c_1+c_2}\Big) >0
\,.
\end{equation}
When $c_1=1$, $c_2=1/2$, and $\alpha=5$, as in figure \ref{2-3-pkn-ic/a1,b2},
this formula implies $q_{min}=10\,{\rm ln}\,3=10.9861$.  Our numerical
results with the resolution of $2^{14}$ zones yield $q_{min}=11.0049$.  The
very small discrepancy, less than $0.2\%$, occurs largely because our numerical
measurement of $q_{min}$ is obtained by examining the peakon positions at each
internal timestep in the code, while the code's time discretization effectively
means we're unlikely to land exactly on the time at which the minimum 
separation
occurs.  The code's true accuracy is thus better than the above measure
indicates, because the intermediate steps involved in advancing the solution
from one discrete time to the next with an RKF45 method cancel the
higher-order discretization errors.

Likewise, for peakons with $b=3$ and $g(x)=e^{-|x|/\alpha}$, formula
(\ref{pulson-qmin}) gives the minimum separation,
\begin{equation}\label{peakon-qmin2}
q_{min} = -\alpha\,{\rm ln}
\Big(\,1-\frac{\sqrt{c_1c_2}}{(c_1+c_2)/2}\Big)
>0
\,.
\end{equation}
When $c_1=1$, $c_2=1/2$, and $\alpha=5$, as in figure \ref{2-3-pkn-ic/a1,b3},
this formula implies $q_{min}=5\,{\rm ln}\,(3/(3-\sqrt{8}))=14.3068$.
This time our numerical results yield $q_{min}=14.2924$, a discrepancy of only
$0.1\%$.

Of course, the two-body collision is rather simple compared to the  plethora
of other multi-wave dynamics that occurs in this problem.  For this reason, we
also checked the convergence of our numerical algorithms
by verifying that the relative phases of the peakons in the various
figures remained invariant under grid refinement.  Moreover, the integrity of
the waveforms in our figures attests to the convergence of the numerical
algorithm -- after scores of collisions, the waveforms given by the Green's
function for each case are still extremely well preserved.  The preservation
of these soliton waveforms after so many collisions would not have occurred
unless the numerics had converged well.

\section{Conclusions}\label{Conclusions}

Equation (\ref{b-family}) introduced a new family of
reversible, parity invariant,
evolutionary 1+1 PDEs describing motion by active transport
\begin{equation}\label{b-family-1}
m_t\
+\
\underbrace{\ \ um_x\ \
}
_{\hspace{-2mm}\hbox{convection}\hspace{-2mm}}\
+\
\underbrace{\ \ b\,u_xm\ \
}
_{\hspace{-2mm}\hbox{stretching}\hspace{-2mm}}\
=\
0
\,, \quad\hbox{with}\quad
u=g*m
\,.
\end{equation}
We 
analyzed the transformation properties and conservation laws of 
this
family of equations, which led us to choose $g$ to be an even 
function. Then
we classified its traveling waves, identified the 
bifurcations of its
traveling wave solutions as a function of the 
balance parameter $b$ and
for some choices of the convolution kernel 
$g(x)$ we studied
its  particle-like solutions and their interactions 
when $b>1$. These were
obtained by superposing $N$ traveling  wave 
solutions
$u(x,t)=cg(x-ct)$ 
as
\begin{equation}\label{pulson-soln-1}
u(x,t)=\sum_{i=1}^Np_i(t)g(x-q_i(t))
\quad\hbox{and}\quad
m(x,t)=\sum_{i=1}^Np_i(t)\delta(x-q_i(t))\,,
\end{equation}
for any real constant $b$ and $u=g*m$, in which the function $g$
is even $g(-x)=g(x)$, so that $g\,'(0)=0$, and is bounded, so we may
set $g(0)=1$.

Following \cite{FH[2001]}, we call these solutions ``pulsons.'' We have
shown that for any $b>1$, once they are initialized on their invariant
manifold (which may be finite  dimensional), the pulsons
undergo particle-like dynamics in terms of the moduli variables
$p_i(t)$ and $q_i(t)$, with $i=1,\dots,N$. The pulson dynamics we studied
for $b>1$ in this  framework on a finite-dimensional
invariant manifold displayed all of the classical soliton interaction
behavior for pulsons found in \cite{FH[2001]} for the case $b=2$. This
behavior included pairwise elastic scattering of pulsons,
dominance of the initial value problem by confined pulses and
asymptotic sorting according to
height  -- all without requiring complete integrability.  Thus, the
``emergent pattern'' for $b>1$
in the nonlinear evolution governed by the active transport equation
(\ref{b-family}) was the rightward moving
pulson train, ordered by height. Thus, the moduli variables $p_i(t)$ and
$q_i(t)$ are collective coordinates on an invariant
manifold for the PDE motion governed by equation (\ref{b-family}). Once
initialized for $b>1$, these  collective degrees of freedom persist
and emerge as a train of stable pulses, arranged in order of their
heights, that then undergo particle-like collisions.

In contrast, the emergent pattern in the Burgers  parameter region
$0\le{b}<1$ is the classic ramp/cliff structure as in Figure
\ref{pkn-ic-vb-eqn-b=0}. That the behavior should depend on the value of
$b$ is clear from the velocity form of equation (\ref{b-family}) written
in (\ref{b-family-u-visc}),
\begin{eqnarray}\label{b-family-u-visc-1}
u_t + (b+1)uu_x - \nu u_{xx}
&=&
\alpha^2(u_{xxt}+uu_{xxx}+bu_x u_{xx}-\nu u_{xxxx} )
\\
&=&
\alpha^2\partial_x
\Big(u_{xt} + u u_{xx} - \nu u_{xxx} + \frac{b-1}{2} u_x^2 \Big)
\nonumber\\
&=&
\alpha^2\partial_x^2
\Big(u_{t} + u u_{x} - \nu u_{xx} + 
\frac{b-3}{2} u_x^2 \Big)
\,.
\nonumber
\end{eqnarray}
Thus, 
nonlinear terms in this equation change sign at four integer 
values
of the parameter $b$. Nonlinear $\alpha^2-$terms change sign 
when
$b=0,1,3$. Also, the nonlinear steepening term increases with 
$b$ as
$(b+1)uu_x$. So this term changes sign when $b=-1$. In the 
parameter regime
$b>-1$ (resp. $b<-1$) the solutions of equation 
(\ref{b-family}) move
rightward (resp. leftward), provided the terms 
on the right hand side of
equation (\ref{b-family-u-visc-1}) are 
sufficiently small.

\paragraph{Three regions of $b$.} We found that 
the solution behavior for
equation (\ref{b-family})  changes its 
character near  the boundaries of
the following three regions in the 
balance parameter $b$.
\begin{description}
\item (B1)
In the stable 
pulson region $b>1$, the Steepening Lemma for peakons
proven for 
$1<b\le3$ in Proposition \ref{B-steepening} allows inflection
points 
with negative  slopes to escape verticality by producing a jump
in 
spatial derivative at the peak of a  traveling wave that eliminates 
the
inflection points altogether. Pulson behavior dominates this 
region,
although ramps of positive slope are also seen to coexist 
with the
pulsons. When $b\le1$ we found the solution behavior of the 
active
transport equation (\ref{b-family}) changed its character and 
excluded
the pulsons entirely.

\item (B2)
In the Burgers region 
$0\le{b}\le1$, the $L^{1/b}$ norm of the
variable $m$ is 
controlled%
\footnote{For $b=0$, this is a maximum principle for 
$|m|$.}
and the solution behavior is dominated by ramps and cliffs, 
as for
the usual Burgers equation.
Similar ramp/cliff solution 
properties hold for the region
$-1\le{b}\le0$, for which the 
$L^{1/b}$ norm
of the variable $1/|m|$ is controlled.
At the boundary 
of the latter region, for
$b=-1$, the active transport
equation 
(\ref{b-family}) admits stationary plane waves as exact
nonlinear 
solutions.

\item (B3)
In the steady pulse region $b<-1$, pulse 
trains form that move
leftward from a positive velocity
initial 
condition (instead of moving rightward, as for $b>-1$). These 
pulse
trains seem to approach a steady 
state.

\end{description}

\paragraph{Effects of viscosity.}
Almost 
any numerical investigation will introduce some viscosity or 
other
dissipation. Consequently, we studied the fate of the peakons 
when
viscosity was added  to the b-family in 
equation
(\ref{viscous-b-family-1}). Viscous solutions of 
equation
(\ref{viscous-b-family-1}) for the peakon 
case
$g(x)=e^{-|x|/\alpha}$ with $\alpha=1$ were studied in each of 
the three
solution regions (B1)-(B3). In the Burgers region (B2) near 
$b=0$ we
focused on the shock-capturing  properties of the solutions 
of
equation (\ref{b-family}) and this family of equations was 
extended
for $\beta\ne1$ to the Burgers$-\alpha\beta$ 
equation
(\ref{convect-viscous-Bab-eqn}),
\begin{equation} 
\label{convect-viscous-Bab-eqn-1}
u_t + u u_x - \nu u_{xx}
=
- 
\,\beta\,\tau_x
\quad\hbox{with}\quad
(1-\alpha^2\partial_x^2)\tau
=
\frac{b}{2}u^2+\frac{3-b}{2}\alpha^2u_x^2
\,.
\end{equation}
According 
to Proposition \ref{H1Control-Prop}, the
Burgers$-\alpha\beta$ 
equation
(\ref{convect-viscous-Bab-eqn-1}) controls the 
$\alpha-$weighted
$H^1$ norm of the velocity for
$\alpha^2\ne0$, 
provided  $(3-b)\beta=1$. This analytical property
guided our study 
of this new equation by
identifying a class of equations for which a 
priori estimates
guarantee continuity of the solution $u(x,t)$.
The 
shock-capturing properties of the Burgers$-\alpha\beta$ 
equation
(\ref{convect-viscous-Bab-eqn-1}) and its $\alpha\to\infty$ 
limit
will be reported in a later paper \cite{HLS[2002]}.

\rem{
Finally, we studied the $\alpha\to\infty$ limit of the
Burgers$-\alpha\beta$ equation, which integrates once to yield the
slope dynamics equation for $s=u_x$, cf.~equation (\ref{b-family-u-visc}),
\begin{equation} \label{Slope-Bab-eqn-alpha2infty}
s_t + u s_x - \nu s_{xx} + \frac{b-1}{2} s^2
=
0\,.
\end{equation}
}

\section{Acknowledgements}\label{Acknowledgements}

We are grateful to A.~Degasperis, A.~N.~W.~Hone, J.~M.~Hyman, S.~Kurien,
C.~D.~Levermore, R.~Lowrie and E.~S.~Titi for their thoughtful remarks,
careful reading and attentive discussions that provided enormous help
and encouragement during the course of writing this paper.


\end{document}